\documentclass[iop]{emulateapj}
\pdfoutput=1

\submitted{To be submitted to ApJ}

\usepackage{epsfig}
\usepackage{natbib}
\usepackage{color}
\usepackage{datetime}
\def\simgt{\lower.5ex\hbox{$\; \buildrel > \over \sim \;$}}
\def\simlt{\lower.5ex\hbox{$\; \buildrel < \over \sim \;$}}

\def\amin{\ifmmode^{\prime}\else$^{\prime}$\fi}
\def\asec{\ifmmode^{\prime\prime}\else$^{\prime\prime}$\fi}

\def\simgt{\lower.5ex\hbox{$\; \buildrel > \over \sim \;$}}
\def\simlt{\lower.5ex\hbox{$\; \buildrel < \over \sim \;$}}

\newcommand\chandra{{\it Chandra}}

\newcommand\xmm{{\it XMM-Newton}}

\newcommand\suzaku{{\it Suzaku}}
\newcommand\integral{{\it INTEGRAL}}

\newcommand\nustar{\hbox{\it NuSTAR\/}}
\newcommand\fermi{{\it Fermi\/}}

\newcommand\eflux{erg\,cm$^{-2}$\,s$^{-1}$}

\def\pwncen{G359.95$-$0.04}
\def\fila{G359.97$-$0.038}
\def\filb{G359.964$-$0.052}
\def\filc{G0.13$-$0.11}
\def\aeknot{G359.89$-$0.08}

\def\gevsrc{2FGL~J1745.6$-$2858}
\def\hesssrc{HESS~J1745$-$290}
\def\xrb1743{1E1743.1$-$2843}

\newcommand\intsrc{IGR~J17456$-$2901}

\slugcomment{Accepted to Astrophysical Journal}

\shorttitle{{\it NuSTAR} observation of Galactic Center diffuse emission}
\shortauthors{K.~Mori et al.}

\begin{document}

%\vspace{-0.5in}

\title{NuSTAR Hard X-ray Survey of the Galactic Center region I: Hard X-ray morphology and spectroscopy of the diffuse emission}

\author{Kaya~Mori\altaffilmark{1}, Charles~J.~Hailey\altaffilmark{1}, Roman~Krivonos\altaffilmark{2,3}, Jaesub~Hong\altaffilmark{4}, 
Gabriele~Ponti\altaffilmark{5}, Franz~Bauer\altaffilmark{6,7,8}, Kerstin~Perez\altaffilmark{1,9}, 
Melania~Nynka\altaffilmark{1}, Shuo~Zhang\altaffilmark{1}, John~A.~Tomsick\altaffilmark{2}, David~M.~Alexander\altaffilmark{10}, Frederick~K.~Baganoff\altaffilmark{11}, Didier~Barret\altaffilmark{12,13}, 
Nicolas~Barri\`{e}re\altaffilmark{2}, Steven~E.~Boggs\altaffilmark{2}, 
Alicia~M.~Canipe\altaffilmark{1}, Finn~E.~Christensen\altaffilmark{14}, William~W.~Craig\altaffilmark{2,15}, Karl~Forster\altaffilmark{16}, Paolo~Giommi\altaffilmark{17}, Brian~W.~Grefenstette\altaffilmark{16}, Jonathan~E.~Grindlay\altaffilmark{4}, 
Fiona~A.~Harrison\altaffilmark{16}, Allan~Hornstrup\altaffilmark{14}, Takao~Kitaguchi\altaffilmark{18, 19}, Jason~E.~Koglin\altaffilmark{20}, 
Vy~Luu\altaffilmark{1}, 
Kristen~K.~Madsen\altaffilmark{16}, Peter~H.~Mao\altaffilmark{16}, 
Hiromasa~Miyasaka\altaffilmark{16}, Matteo~Perri\altaffilmark{17, 21}, 
Michael~J.~Pivovaroff\altaffilmark{15}, Simonetta~Puccetti\altaffilmark{17, 21}, Vikram~Rana\altaffilmark{16}, Daniel~Stern\altaffilmark{22}, Niels~J.~Westergaard\altaffilmark{14}, William~W.~Zhang\altaffilmark{23} and 
Andreas~Zoglauer\altaffilmark{2}}

\altaffiltext{1}{Columbia Astrophysics Laboratory, Columbia University, New York, NY 10027, USA; kaya@astro.columbia.edu}
\altaffiltext{2}{Space Sciences Laboratory, University of California, Berkeley, CA 94720, USA} 
\altaffiltext{3}{Space Research Institute, Russian Academy of Sciences, Profsoyuznaya 84/32, 117997 Moscow, Russia}      
\altaffiltext{4}{Harvard-Smithsonian Center for Astrophysics, Cambridge, MA 02138, USA}     
\altaffiltext{5}{Max-Planck-Institut~f.~extraterrestrische~Physik, HEG, Garching, Germany}
\altaffiltext{6}{Instituto de Astrof\'{\i}sica, Facultad de F\'{i}sica, Pontificia Universidad Cat\'{o}lica de Chile, 306, Santiago 22, Chile} 
\altaffiltext{7}{Millennium Institute of Astrophysics, Santiago, Chile} 
\altaffiltext{8}{Space Science Institute, 4750 Walnut Street, Suite 205, Boulder, Colorado 80301}
\altaffiltext{9}{Haverford College, 370 Lancaster Avenue, KINSC L109, Haverford, PA 19041, USA}
\altaffiltext{10}{Department of Physics, Durham University, Durham DH1 3LE, UK}
\altaffiltext{11}{Kavli Institute for Astrophysics and Space Research, Massachusets Institute of Technology, Cambridge, MA 02139, USA}
\altaffiltext{12}{Universit\'{e} de Toulouse, UPS-OMP, IRAP, Toulouse, France}
\altaffiltext{13}{CNRS, Institut de Recherche en Astrophysique et Plan\'{e}tologie, 9Av. colonel Roche, BP 44346, F-31028 Toulouse Cedex 4, France}
\altaffiltext{14}{DTU Space - National Space Institute, Technical University of Denmark, Elektrovej 327, 2800 Lyngby, Denmark}
\altaffiltext{15}{Lawrence Livermore National Laboratory, Livermore, CA 94550, USA}
\altaffiltext{16}{Cahill Center for Astronomy and Astrophysics, California Institute of Technology, Pasadena, CA 91125, USA}
\altaffiltext{17}{ASI Science Data Center, Via del Politecnico snc I-00133, Roma, Italy}
\altaffiltext{18}{Department of Physical Science, Hiroshima University, Higashi-Hiroshima, Hiroshima 739-8526, Japan}
\altaffiltext{19}{Core of Research for the Energetic Universe, Hiroshima University, Higashi-Hiroshima, Hiroshima 739-8526, Japan}
\altaffiltext{20}{Kavli Institute for Particle Astrophysics and Cosmology, SLAC National Accelerator Laboratory, Menlo Park, CA 94025, USA}
\altaffiltext{21}{INAF - Astronomico di Roma, via di Frascati 33, I-00040 Monteporzio, Italy}
\altaffiltext{22}{Jet Propulsion Laboratory, California Institute of Technology, Pasadena, CA 91109, USA}
\altaffiltext{23}{NASA Goddard Space Flight Center, Greenbelt, MD 20771, USA}

\begin{abstract}

We present the first sub-arcminute images of the Galactic Center above 10~keV, obtained with {\it NuSTAR}. {\it NuSTAR} resolves the hard X-ray source \intsrc\ into non-thermal X-ray filaments, molecular clouds, point sources and a previously unknown central component of hard X-ray emission (CHXE). 
{\it NuSTAR} detects four non-thermal X-ray filaments, extending the detection of their power-law spectra 
with $\Gamma\sim1.3$-2.3 up to $\sim50$ keV.
A morphological and spectral study of the filaments suggests that their origin 
may be heterogeneous, where previous studies suggested a common origin in young pulsar wind nebulae (PWNe).
\nustar\ detects non-thermal X-ray continuum emission spatially correlated with the 6.4 keV Fe 
K$\alpha$ fluorescence line emission associated with two Sgr A molecular clouds: MC1 and the Bridge. 
Broad-band X-ray spectral analysis with a Monte-Carlo based X-ray reflection model self-consistently 
determined their intrinsic column density ($\sim10^{23}$~cm$^{-2}$), primary X-ray spectra (power-laws with $\Gamma\sim2$) and 
set a lower limit of the X-ray luminosity of Sgr A* flare illuminating the Sgr A clouds to $L_X \ga 10^{38}$~erg\,s$^{-1}$.   
Above $\sim20$~keV, hard X-ray emission in the central 10 pc region around Sgr A* consists of the candidate PWN 
\pwncen\ and the CHXE, possibly resulting from an 
unresolved population of massive CVs with white dwarf masses $M_{WD} \sim 0.9 M_{\odot}$.   
Spectral energy distribution analysis suggests that \pwncen\ is likely the hard X-ray counterpart of the ultra-high 
gamma-ray source \hesssrc, strongly favoring a leptonic origin of the GC TeV emission.  

\end{abstract}

\section{Introduction}

The relative proximity of the Galactic Center (GC), at $\sim8$~kpc, allows for sensitive, high-resolution observations that are not possible for more distant galactic nuclei. 
Over the last two decades, the GC and Galactic Ridge have been extensively surveyed by X-ray telescopes, which have revealed various diffuse X-ray components including  
the Galactic Ridge X-ray emission \citep[GRXE; ][]{Revnivtsev2006, Krivonos2007, Yuasa2012}, an X-ray haze that extends out from the GC for $\sim60$ degrees in longitude and a few degrees in 
latitude \citep{Worrall1982}, and large-scale diffuse Fe line emission \citep{Koyama1989, Koyama1996}. In the inner 20\amin\ around the GC, a separate, unresolved $\sim8$~keV thermal component   
has been observed by \chandra\ and \xmm\ \citep{MunoDiffuse2004, Heard2013}. \chandra\ has resolved thousands of point sources in the $2^{\circ}\times0.8^{\circ}$ GC field, suggesting the  $kT\sim8$ keV thermal emission represents a population of unresolved magnetic CVs \citep{Wang2002, Revnivtsev2009, Muno2009}. 
In addition, \chandra\ has performed arcsecond-scale mapping in the crowded soft X-ray (2-10~keV) band \citep{Baganoff2003},
identifying emission from the central supermassive black hole Sgr A*, hot gas from winds of the surrounding central stellar cluster, the supernova remnant (SNR) Sgr A East, 
non-thermal filamentary structures, molecular clouds and thousands of X-ray point sources \citep{Muno2008}. 

In the GC region, \chandra\ has detected nearly two dozen X-ray filaments, most of which exhibit ``cometary'' or ``filamentary''
shapes and featureless
non-thermal spectra with $\Gamma\sim$1.5-2.5 \citep{Johnson2009}.
\citet{Muno2008} and \citet{Lu2008} speculated that the X-ray filaments are young pulsar wind nebulae (PWNe) 
since they possess similar spectral
and morphological properties, although there is as yet no direct evidence for a PWN. 
Whether X-ray filaments are PWNe or not, if synchrotron 
radiation is responsible for their non-thermal X-ray emission, hard X-ray spectroscopy probes the highest energy ($\sim10$-100~TeV) 
electrons that are accelerated  since the 
synchrotron photon energy  $E_\gamma \sim 40 (E_e/10 \rm{TeV})^2 (B/1 \rm{mG})$ keV where $E_e$ is the electron energy and 
$B$ is the magnetic field strength typically $\sim$0.1-1 mG inside radio filaments \citep{Yusef1987, Ferriere2009}. 

Many of the Galactic Center molecular clouds (GCMCs) in the Sgr A, B and C regions are known to produce  
diffuse Fe K$\alpha$ fluorescence emission at 6.4 keV \citep{Ponti2014}.
Two models, the so-called X-ray reflection nebula (XRN) model \citep{Sunyaev1993} and the low-energy cosmic-ray (LECR) model 
\citep{Yusef2002}, have been proposed to account for Fe K$\alpha$ line emission by photo-ionization by an external X-ray source and collisional ionization by low-energy cosmic-rays, 
repectively \citep[see][for a review]{Ponti2013}. 
The XRN scenario seems more plausible since the \xmm\ and \chandra\ surveys of the GC region over the last decade have revealed the
year-scale time variation of strong Fe K$\alpha$ line with equivalent width $\sim1$ keV in the Sgr A 
clouds \citep{Ponti2010, Capelli2012, Clavel2013} and Sgr B2 \citep{Terrier2010}. 
It has been proposed that the X-ray emission of GCMCs is associated with Sgr A* past flares, or nearby
 X-ray transients \citep{Sunyaev1993, Koyama1996}.
However, it is still possible that the LECR emission contributes
as an additional component given that a large population of cosmic rays are expected in the GC \citep{Capelli2012}. In either case, 
there has been no clear detection of X-ray continuum emission intrinsic to the Sgr A clouds. 
In the soft X-ray band, thermal diffuse emission as well as point sources heavily contaminate X-ray emission from the Sgr A clouds, 
while hard X-ray telescopes such as \integral\ were not able to resolve X-ray continuum 
emission from the Sgr A clouds. 

In the gamma-ray band, the CANGAROO-II and HESS arrays of Cherenkov telescopes discovered the ultra-high energy gamma-ray source
\hesssrc\ \citep{Tsuchiya2004, Aharonian2004} and later its 0.1-10~TeV spectrum was well measured
by different TeV telescopes such as HESS \citep{Aharonian2009}, VERITAS \citep{Archer2014} and MAGIC \citep{Albert2006}.
Both Sgr A* and the cometary PWN candidate \pwncen\ have been proposed as counterparts of
 the TeV source \hesssrc, as both lie within its 13\asec\ error radius \citep{Acero2010}.
This has led to two possible interpretations: a leptonic and a hadronic origin. In the leptonic scenario, high-energy electrons are
accelerated by Sgr A* flares,
PWNe, supernova remnants (SNRs) interacting with molecular clouds and stellar winds. These TeV electrons emit
synchrotron radiation in the X-ray band in an ambient interstellar medium (ISM) magnetic field of $\sim10$ $\mu$G
and also emit inverse-Compton radiation in the gamma-ray band by up-scattering
 ultraviolet and far-infrared photons in the high radiation density field of the
GC \citep{Hinton2007}. In the hadronic scenario, relativistic protons accelerated from Sgr A*
or SNRs interacting with the
surrounding medium emit gamma-rays via pion decay, then secondary electrons emit X-rays via synchrotron or non-thermal
bremsstrahlung radiation \citep{Chernyakova2011}.
Both hadronic and leptonic models can also explain the 0.1-10~TeV spectrum of the GC, either via pion decay from protons
injected into the diffuse interstellar medium by past Sgr A* outflows
\citep{Aharonian2005b, Ballantyne2007, Chernyakova2011} or via inverse-Compton emission from a population of electrons
ejected from Sgr A* \citep{Kusunose2012} or the PWN candidate \pwncen\ \citep{Hinton2007}.
As a more exotic scenario, dark matter annihilation
at the GC has been also proposed \citep{Cembranos2013}. 
Since the soft X-ray emission of the GC is dominated by diffuse thermal emission and point sources that are mostly unrelated
to the GeV to TeV emission, it is extremely important to identify a hard X-ray counterpart of \hesssrc. 
However, the central 10 pc region of the GC has been difficult to localize due to the $\ga10'$ angular resolution of hard X-ray instruments 
\citep{Winkler2003, Gehrels2004}, leaving the origin and nature of \hesssrc\ a subject of controversy. 

Above 20 keV, the \integral\ observatory discovered a persistent hard X-ray source \intsrc, which is particularly 
bright in the 20-40~keV range, 
within $1'$ of the GC \citep{Belanger2006}. 
The emission at energies above 40 keV, however, seems to shift several arcminutes to the east of both Sgr A* and Sgr A East.
This variation in the position of the emission combined with the $12'$ spatial resolution of the \integral/IBIS coded aperture mask has led to speculation that the
emission results not from a single object, but from a collection of the many surrounding diffuse and point-like 
X-ray sources \citep{Krivonos2007}. 
However, without high-resolution, high-energy images of the region available, the existence of a new source of high-energy X-ray 
emission could not be ruled out. 

The \nustar\ hard X-ray telescope \citep{Harrison2013}, with its arcminute angular resolution and effective area extending from 
3 to 79 keV, can make unique contributions to understanding the emission mechanisms of X-ray filaments, GCMCs and the gamma-ray 
source \hesssrc. Broadband X-ray spectroscopy with \nustar\  provides a powerful diagnostic that can distinguish 
between different models of GCMC X-ray emission and tightly constrain parameters when combined with self-consistent X-ray emission models.
In addition, \nustar\ is the key to filling 
the gap between the well-studied soft X-ray populations and the persistent gamma-ray emission in the central parsec region of the GC. 

In this paper, we report \nustar\ hard X-ray observations of diffuse emission in the GC region, while our companion paper \citep{Hong2015} 
focuses on the hard X-ray point sources. \S\ref{sec:obs} outlines the \nustar\ and \xmm\ observations adopted  
for studying GC diffuse emission, followed by \S\ref{sec:analysis} describing our imaging and spectral analysis methods. \S\ref{sec:morphology} presents the hard X-ray morphology of the GC 
region above 10 keV. For three hard X-ray diffuse source categories, namely non-thermal 
X-ray filaments (\S\ref{sec:filaments}), molecular clouds (\S\ref{sec:mc}) and the central 10 parsec region around Sgr A* 
(\S\ref{sec:10pc}), we present our 
\nustar\ spectral and morphological analysis jointly with archived \chandra\ and \xmm\ data, 
and discuss implications for their hard X-ray emission mechanisms. \S\ref{sec:summary} summarizes our results from the \nustar\ 
GC survey. The Appendix describes \nustar\ background components and background subtraction 
methods, some of which are peculiar to the \nustar\ GC observations, as well as X-ray reflection models for GCMCs particularly on a Monte-Carlo based self-consistent MYTorus model. Throughout the paper, we assume a distance to the GC of 8~kpc 
\citep{Reid1993}. 

%%%%%%%%%%%%%%%%%%%%%%%%%%%%%%%%%%%%%%%%%%%%%%%%%%%%%%%%%%%%

\section{\nustar\ Observations}
\label{sec:obs}

{\it NuSTAR} consists of coaligned X-ray telescopes with corresponding focal plane modules (FPMA and FPMB) with an angular resolution of $58^{\prime\prime}$ Half Power Diameter 
(HPD) 
and $18^{\prime\prime}$ Full Width Half Maximum \citep[FWHM; ][]{Harrison2013}. \nustar\ operates in the 3-79 keV band with $\sim400$ eV (FWHM) energy resolution below $\sim50$ keV and 
$\sim900$ eV at 68~keV.  
Soon after its launch in June 2012, \nustar\ initiated a large GC survey to study both point sources and diffuse emission in the hard X-ray band. A number of 
single pointing observations, occasionally coordinated with other telescopes, were performed to study Sgr A* flaring \citep{Barriere2014}, the newly discovered magnetar SGR~1745$-$29 
\citep{Mori2013, Kaspi2014}, the Cannonball \citep{Nynka2013}, Sgr A-E \citep{Zhang2014}, X-ray transients in outbursts 
\citep{koch2014, Barriere2015}, the Arches cluster \citep{Krivonos2014} and Sgr B2 molecular cloud \citep{Zhang2015}. 

For the analysis of GC diffuse emission presented in this paper, we used the nine \nustar\ observations listed in Table \ref{tab:obslog}. 
The first three observations were pointed at 
Sgr A*, while the next six observations, each with $\sim25$~ks depth, 
covered the $\sim0.4^{\circ} \times 0.3^{\circ}$ area between Sgr A* and the low mass X-ray binary (LMXB) \xrb1743\ (hereafter referred to as the ``mini-survey''). The \nustar\ mini-survey was originally motivated to study the 
{\it INTEGRAL} source \intsrc.  
We did not use \nustar\ observations of Sgr A* in 2013 and 2014 since the data were heavily 
contaminated by outbursting X-ray transients. The other observations surveying larger regions,  $\ga10'$ away from the GC,    
were primarily aimed at studying point sources, and so they are not included in this analysis. 
In addition, four \xmm\ observations in 2012 (Table \ref{tab:obslog}) were obtained in the Full Frame mode 
with the medium filter and their data are used for joint spectral analysis.

\begin{deluxetable*}{lcccc}
\tablecaption{{\it NuSTAR} and \xmm\ Galactic Center Observations in 2012}
\tablewidth{0pt}
\tablecolumns{4}
\tablehead { \colhead{ObsID}    &   \colhead{Start Date}   &   \colhead{Exposure}   &  \colhead{Target} \\
 \colhead{     }  & \colhead{(UTC)}        & \colhead{(ks)}  & \colhead{  }    }
\startdata
 \multicolumn{4}{c}{\textit{NuSTAR}} \\ \hline
30001002001  & 2012 07 20 &  166.2    & Sgr~A*\\
30001002003  & 2012 08 04 &   83.8   & Sgr~A*\\
30001002004  & 2012 10 16 &   53.6   & Sgr~A*\\
40010001002  & 2012 10 13 &   23.9    & Mini-survey\\
40010002001  & 2012 10 13 &   24.2  & Mini-survey\\
40010003001  & 2012 10 14 &   24.0     & Mini-survey\\
40010004001  & 2012 10 15 &   24.0   & Mini-survey\\
40010005001  & 2012 10 15 &   25.7   & Mini-survey\\
40010006001  & 2012 10 16 &   23.5  & Mini-survey\\ \hline
 \multicolumn{4}{c}{\textit{XMM-Newton}\tablenotemark{a} } \\ \hline
0694640301  & 2012 08 31 &   35.5    & CMZ\tablenotemark{b} \\
0694640401  & 2012 09 02 &   43.9   & CMZ\tablenotemark{b}  \\
0694641001  & 2012 09 23 &   40.7   & CMZ\tablenotemark{b}  \\
0694641101  & 2012 09 24 &   35.5   & CMZ\tablenotemark{b}  
\enddata
\tablecomments {The exposure times listed are corrected for good time intervals.} 
\tablenotetext{a}{All \xmm\ observations were operated in Full Frame mode with the medium filter.} 
\tablenotetext{b}{Central Molecular Zone}
\label{tab:obslog}
\end{deluxetable*}     

%%%%%%%%%%%%%%%%%%%%%%%%%%%%%%%%%%%%%%%%%%%%%%%%%%%%%%%%%%%%%%%%%%%%%%%%%%%%

\section{Data reduction and analysis}
\label{sec:analysis}

\begin{figure*}[t]
\centerline{
\includegraphics[width=0.7\linewidth]{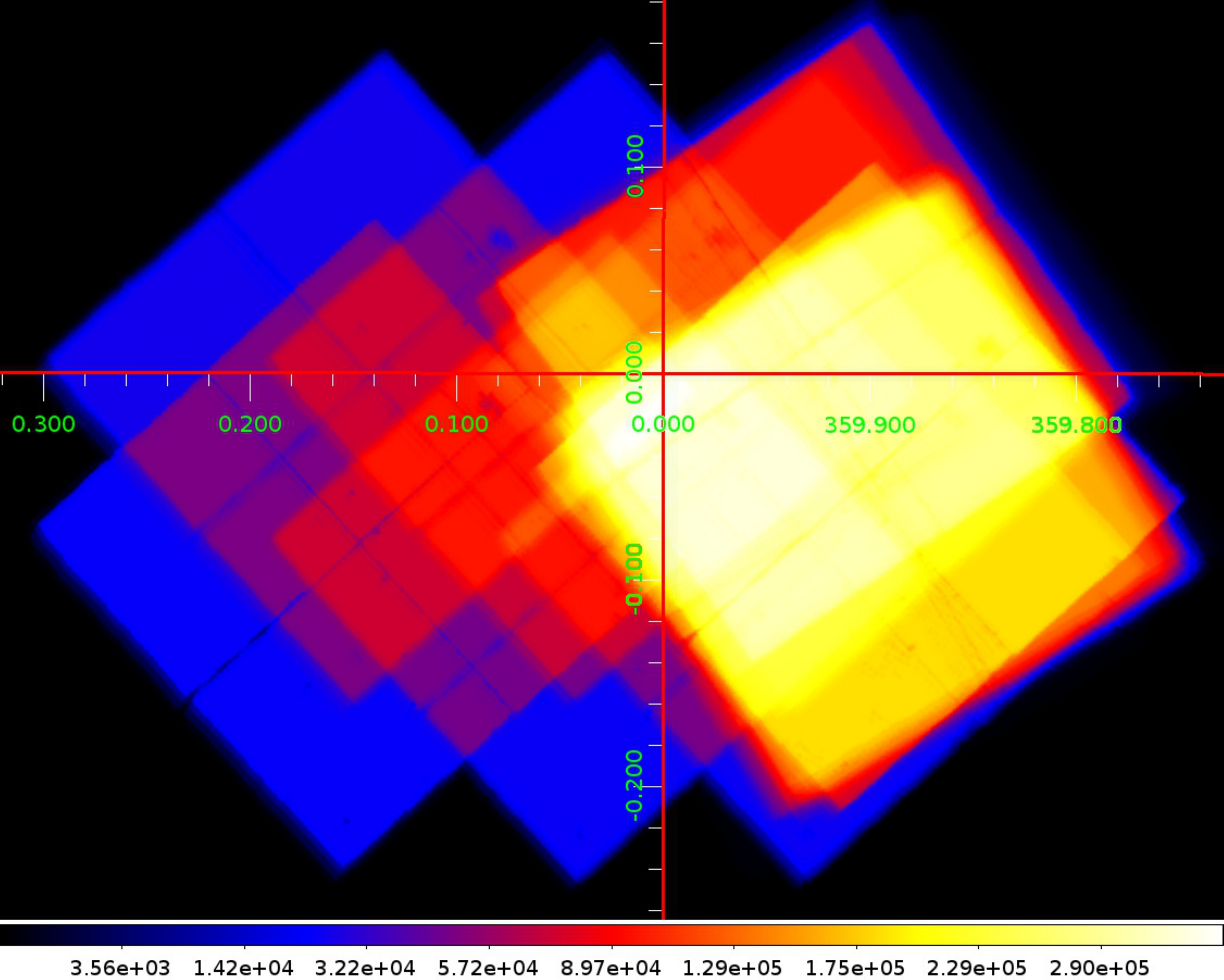}
}
\caption{Exposure map of the nine \nustar\ observations of the GC region combined before removing the high stray-light
background
regions contaminated by nearby bright point sources. Exposure time in seconds is plotted in the square root scale in the
Galactic coordinates [$^\circ$].}
\label{fig:nustar_expmap}
\end{figure*}

In this section, we describe our imaging and spectral analysis of the \nustar\ GC data.  
All the \nustar\ data were processed using the \nustar\ {\it Data Analysis Software (NuSTARDAS)} v1.3.1.
After filtering high background intervals during South Atlantic Anomaly (SAA) passages, we removed additional time
periods in which Sgr A* was in a flaring state,
as observed by \nustar\ or during coincident \chandra\ observations \citep{Barriere2014}.
An additional 70 ks was removed from the 2012 August observation (ObsID: 30001002003) due to a reduction in
the event rate of FPMA, possibly due to debris blocking
the detector.
After all quality cuts, the effective exposure time ranges from $\sim$25-100 ks (mini-survey) to $\sim300$ ks (Sgr A*)
(Table \ref{tab:obslog} and Figure \ref{fig:nustar_expmap}).

In most \nustar\ GC observations, the background below $\sim40$ keV is dominated by photons from outside the
field of view (FOV) entering through the
aperture stop \citep[so-called ``stray-light background'' or SLB hereafter][]{Harrison2013, Krivonos2014, Wik2014}.
In particular, SLB patterns from nearby bright point sources ($\ga 10^{-11}$~\eflux)
within $\sim5^\circ$ from the telescope's
pointing vector are visible at predictable locations on the FOV and completely dominate over other X-ray emission.
We filtered out events in the region of
heavy SLB contamination from the nearby bright source GX~3+1 (\citet{Seifina2012} see the Appendix for more details).
As a result, $\sim25$\% of FPMB events, mostly in detector chip 0, were removed from all observations, while
FPMA data do not have significant SLB from bright point sources.

\subsection{Imaging analysis} 

First, we applied astrometric corrections for individual \nustar\ event files by 
registering known soft X-ray sources to further improve our positioning accuracy for detailed morphological studies.   
These registration sources include bright \chandra\ point sources \citep{Muno2009}, the core of the Sgr A-E \citep{Sakano2003}, the 
Arches cluster 
\citep{Yusef2002b} and the Cannonball (a neutron 
star candidate located outside the Sgr A East shell) \citep{Park2005, Nynka2013}. 
They all have \chandra\ counterparts with known positions to better than $\sim0.5$\asec. 
Using the IDL centroiding routine {\verb=gcntrd=}, we determined the centroid of each registration source in the 3-10 keV band, 
matching \chandra's
sensitive energy band (2-8 keV) for highly absorbed GC sources.
Two of the Sgr A* observations (ObsID: 30001002001, 30001002004) contained several
bright flares from Sgr A* itself \citep{Barriere2014} that aided in determining the astrometric corrections.
For these two observations, an image file was created that contained the bright flares from Sgr A* and we defined the offset as 
the difference between the radio position 
of Sgr A* \citep{Reid1999} and the centroid of the \nustar\ emission. 
These event files, after removing Sgr A* flare intervals, were properly shifted using the Sgr A* flare position and used 
in the subsequent mosaic images presented here.
Translational shifts by as much as $\sim14''$ were required to place the target source at its known position, and
were applied to both event files and exposure maps.

Second, after each observation was corrected for its offset, we summed together all observations and 
normalized the resulting image by the effective exposure map. 
We neither subtracted background nor corrected for vignetting effects (so-called flat-fielding) in the subsequent
imaging analysis since the background is not spatially uniform (see Appendix).
Figure~\ref{fig:nustar_2figures} shows 
the exposure-corrected count rate images in the 10-79 keV band, after combining both FPMA and FPMB data from the nine \nustar\ observations.
For illustration purposes, we smoothed \nustar\ images with a Gaussian kernel of radius ${\sim}12''$ (5 pixels) unless otherwise instructed.
We verified the applied astrometric correction by determining the position of one or several additional sources in each individual 
observation and in the final mosaicked image. The \nustar\ positions are within 5\asec\ of the reported \chandra\ positions.

There are two particularly bright regions seen in the \nustar\ images. 
One of them is the Sgr A* complex containing thermal diffuse emission, hundreds of X-ray point sources, 
X-ray filaments and thermal emission from Sgr A East. Although these X-ray sources are unresolved by \nustar\ in the 3-10 keV band, 
most of them have soft X-ray spectra and fade out beyond 10 keV making a subset of X-ray sources more prominent (\S \ref{sec:morphology}). 
The other bright region near the left (east) side of the image is a persistent LMXB \xrb1743\ at 
RA = $17^{h}46^{m}21^{s}.094$ and DEC = $-28^{\circ}43^{\prime}42^{\prime\prime}.3$ \citep[J2000.0][]{Wijnands2006}. 
The LMXB looks ``extended'' because it is so bright, with a 3-79 keV flux of $2.2\times10^{-10}$ \eflux\ \citep{Lotti2015}, that its PSF wings extending beyond $\sim1$\amin\ are still dominant over other X-ray emission. Ghost-ray background (see its definition in the 
Appendix) from \xrb1743\ is not visible, and  our simulation confirmed that it is below the GC diffuse emission and SLB in most 
of the area covered by the \nustar\ mini-survey observations. 
The brightest X-ray filament is Sgr A-E (\aeknot) at RA = $17^{h}45^{m}40^{s}.4$ and 
DEC = $-29^{\circ}04^{\prime}29^{\prime\prime}.0$ \citep[J2000.0][]{Lu2003} and it is 
distinct from the Sgr A* complex. The molecular clouds in the region between the Sgr A*   
complex and \xrb1743\ \citep{Ponti2010} are also visible. 
The outer regions at $ b \ga 0.1^{\circ}$ or $b \la 0.2^\circ$  are dominated by SLB from the Galactic Ridge X-ray emission,
 and no GC emission is clearly visible there. 

\begin{figure*}[t]
\centerline{
\includegraphics[width=0.7\linewidth]{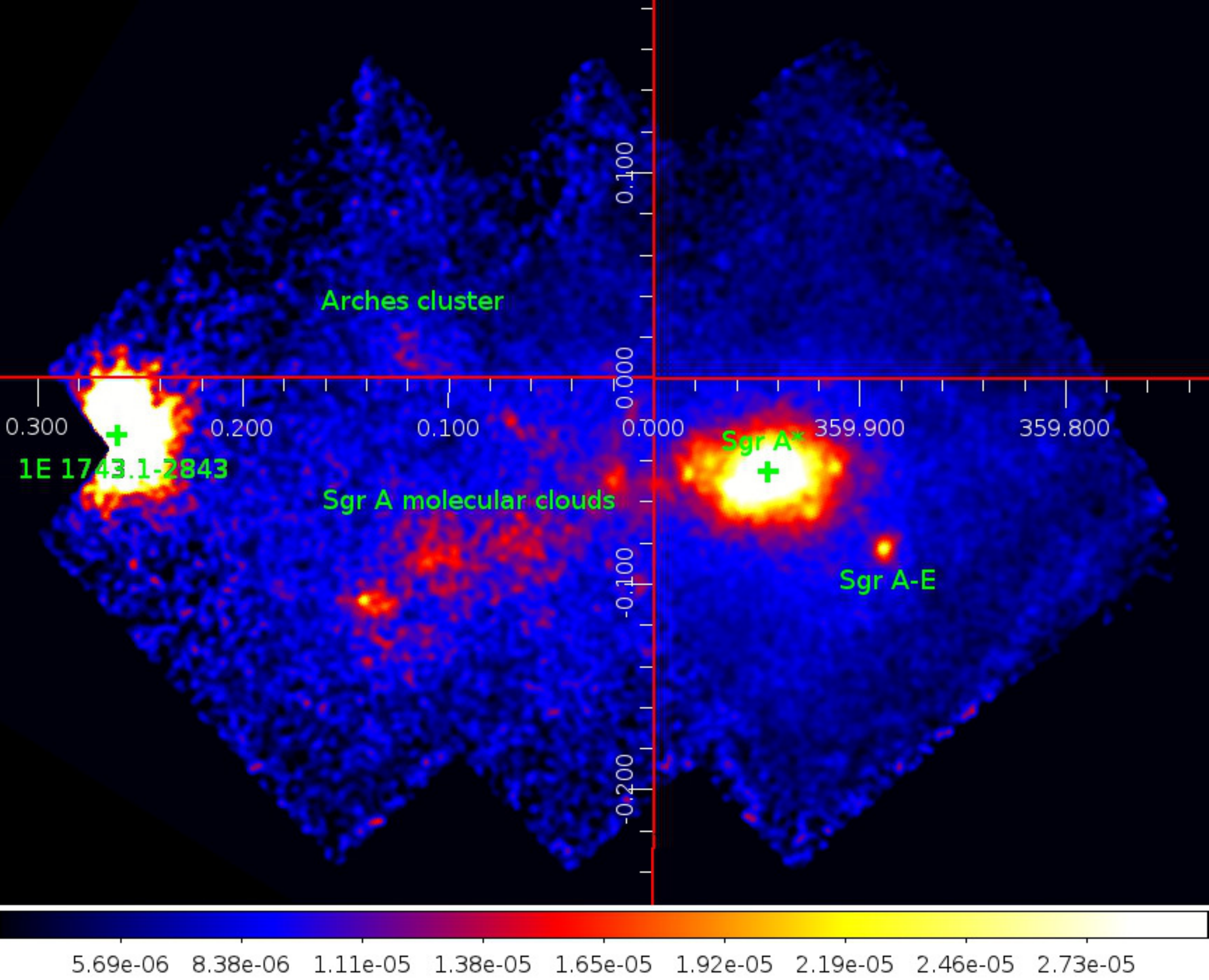}
}
\caption{\nustar\ 10-79 keV exposure-corrected smoothed image of the GC in the Galactic coordinates [$^\circ$]. The image was smoothed by a 5-pixel (12.3\asec) Gaussian kernel. The image scaling was adjusted to illustrate the X-ray features clearly. }
\label{fig:nustar_2figures}
\end{figure*}

\subsection{Trial probability map}

In addition to the exposure-corrected images, we present a sky map of detection significance, dubbed a `trial map', 
to illustrate detection significance of faint sources 
that are otherwise hidden in the count rate images (see \citet{Hong2015} for more details and applications to \nustar\ images). 
The value of the trial map at each sky position represents the number
of random trials required to produce the observed counts by purely random
Poisson fluctuations if no excess of X-ray sources relative to the
background is present at the location.  For every sky position, we first
define a source cell (e.g., 20\% encircled energy fraction of the PSF) and
a background cell (an annulus around the position), and then, using the
cells, we calculate the total observed counts ($S$) and their
background counts ($\lambda_B$). For each image pixel, the random trial number
is estimated to be a normalized incomplete gamma function of $S$ and
$\lambda_B$ \citep{Weisskopf2007, Kashyap2010}. We repeat the
procedure for other pixels to generate the map by sliding cell windows across the field.

Throughout the paper, the negative logarithm of the trial probability is plotted
in the trial map [e.g., $-\log(10^{-6}$) = 6 for a trial probability of $P_{\rm trial} = 10^{-6}$].
Thus, a brighter spot indicates higher detection significance.
When there are no significant systematic fluctuations,
the presence of an X-ray source is indicated by a position where
the trial probability is significantly smaller than the odds one can reach
with the total number of independent searches (i.e. the inverse of
the total number of searches).
The total number of independent searches can be
estimated as the maximal number of resolvable sources ($N_R$) in the field,
which is the ratio of the number of pixels in the image ($N_P=1.7\times10^5$~pixels in this example presented in
\S \ref{sec:image_40-79keV})
to the \nustar\ angular resolution in pixels (FWHM $\sim18$\asec\ diameter circle: $\sim40$ pixels).
For a given confidence level ($C$), one can claim a detection when $P_{\rm trial} < N_R * (1 - C)$.
For instance, to detect a source at 99.7\% confidence level (i.e. 3-$\sigma$ detection), $P_{\rm trial}$ should be less than
$10^{-6.1}$ or its trial map value should be greater than 6.1.
Similarly for 4- and 5-$\sigma$ detections, the trial map value should be greater than 7.8 and
9.9, respectively. Note that trial maps are not used to infer the actual
source brightness and they are presented without any smoothing.

\subsection{Spectral analysis}
\label{sec:spectra}

For some of the hard X-ray sources discussed below, we have jointly
analyzed the \nustar\ and \xmm\ spectra to investigate their X-ray
emission mechanisms by comparing with several existing models.
We extracted \nustar\ source spectra and generated response matrices and ancillary files using
{\it nuproducts}.
Background subtraction and modeling require extra caution due to the high background level and to
its complex multiple components (Appendix). Background substraction methods are specific to each source and they can be found
in later sections (e.g., \S \ref{sec:mc_spec} and \ref{sec:10pc_spec}) where we clarify our data selection and filtering.

We processed all the \xmm\ Observation Data Files (ODFs) with the \xmm\ Science Analysis System (SAS version 13.5.0) and 
the most recent calibration files. We restricted our analysis to 
the \xmm\ EPIC-PN data where photon pile-up effect is negligible for the sources we
analyzed. After filtering out time intervals with high soft proton flaring levels, we selected EPIC-PN events with 
FLAG = 0 and PATTERN$\le4$.
For each \xmm\ spectrum, the response matrix and effective area files are computed with the XMM-SAS tasks {\tt rmfgen} and 
{\tt arfgen}. For the background, we adopted the \xmm\ calibration observations closest in time to each of
the \xmm\ observations, and used their EPIC-PN data with the filter
wheel closed, thus blocking external X-rays and soft protons, and allowing us
to measure internal background components accurately.
First, we fit the so-called Filter Wheel Closed (FWC) spectra with several power-law continuum components and Gaussian 
lines to properly parameterize the background emission.
Since the ratio between the lines and continuum in background spectra is stable between
observations close in time, we scaled the overall normalization of the FWC model to match the
count rates from the same source-free region between the FWC model and actual \xmm\ science data,
while we froze all the other parameters.

We combined \nustar\ FPMA and FPMB spectra and response files using the {\tt FTOOL} {\it addascaspec}. \xmm\ EPIC-PN spectra 
and response files from individual observations were similarly combined, after ensuring that individual spectra were 
consistent with each other.
We grouped all spectra so that each bin had sufficient counts to 
ensure that it had a significance over background of at least $4\sigma$. 
All spectral fitting and flux derivations were performed in XSPEC \citep{Arnaud1996}, with photoionization cross
sections as defined in \citet{Verner1996}
and abundances for the interstellar absorption as defined in \citet{Wilms2000}. Chi-squared statistics were used for
spectral fitting, and all quoted errors are for 1-$\sigma$ level confidence.

%%%%%%%%%%%%%%%%%%%%%%%%%%%%%%%%%%%%%%%%%%%%%%%%%%%%%%%%%%%%%%%%%%%%%%%%%%%%%%%%%%%%%%%

\section{Hard X-ray morphology of the Galactic Center region}
\label{sec:morphology}

\subsection{10-20 keV band morphology}
\label{sec:image_10-20keV}

\begin{figure*}[t]
\centerline{
\includegraphics[height=0.6\linewidth]{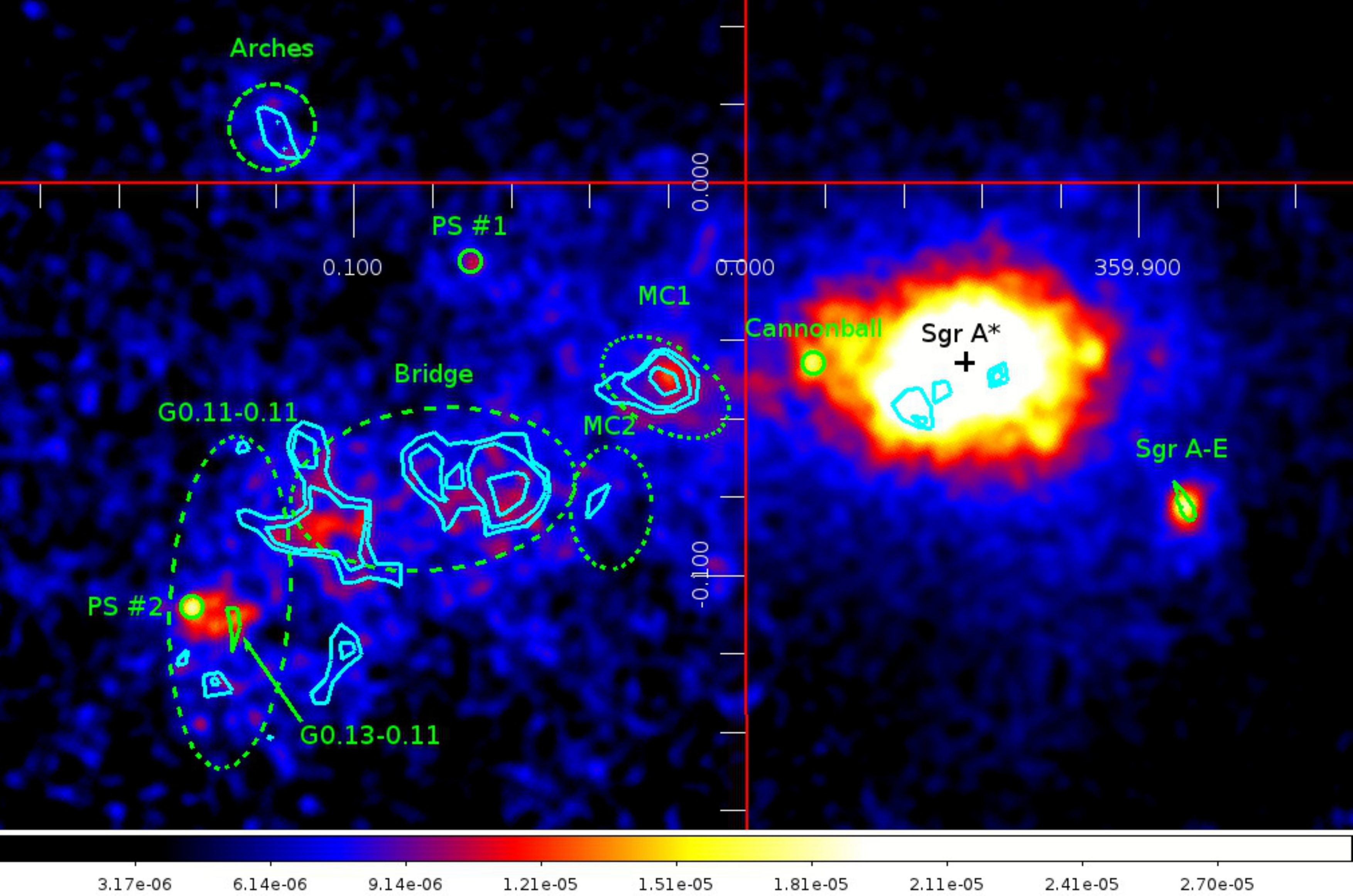}
}
\caption{\nustar\ 10-20 keV exposure-corrected smoothed image of the GC region in the Galactic coordinates [$^\circ$]. 
Green dashed ellipses: selected molecular clouds. Cyan contours: 
6.4 keV Fe K-$\alpha$ continuum-subtracted intensity contours from \xmm. 
Green polygons: \chandra\ morphologies of the two X-ray filaments detected above 10~keV.  
Green circles ($r=10$\asec): two hard X-ray point sources. 
PS \#1 and \#2 are known \chandra\ point sources, CXOUGC~J174551.9$-$285311 
and CXOUGC~J174622.7$-$285218, respectively \citep{Muno2009}.}
\label{fig:nustar_10-20keV}
\end{figure*}

Figure \ref{fig:nustar_10-20keV} shows the \nustar\ image in 
the 10-20 keV band. 
Some features are identified as point sources and marked with white circles \citep{Hong2015}.
While 8~keV thermal emission and SNR Sgr A East emission are still dominant, the Cannonball \citep{Nynka2013} is  
visible in the 10-20 keV image. Diffuse emission is present in the region east of the Sgr A* complex 
but west of the LMXB~\xrb1743. This diffuse emission is likely a mixture of 8~keV thermal emission and molecular cloud X-ray 
continuum. 
Three molecular clouds, namely MC1, the Bridge and the Arches cluster indicated by green dashed ellipses \citep[defined in][]{Ponti2010}, 
are clearly detected above 10~keV, while we do not detect diffuse 
emission from these clouds above 20~keV largely due to the high background level. 
The \nustar\ image is overlaid with Fe K$\alpha$ line intensity contours obtained from
the 2012 \xmm\ observations (cyan contours). The continuum emission in the 4.5-6.28 keV band was subtracted from the 6.28-6.53~keV 
\xmm\ image to emphasize just the Fe K$\alpha$ line 
emission \citep{Ponti2010}. The 10-20 keV hard X-ray emission is well correlated with the Fe K$\alpha$
line contours in these cloud regions. 

\begin{figure}[t]
\includegraphics[height=0.67\linewidth]{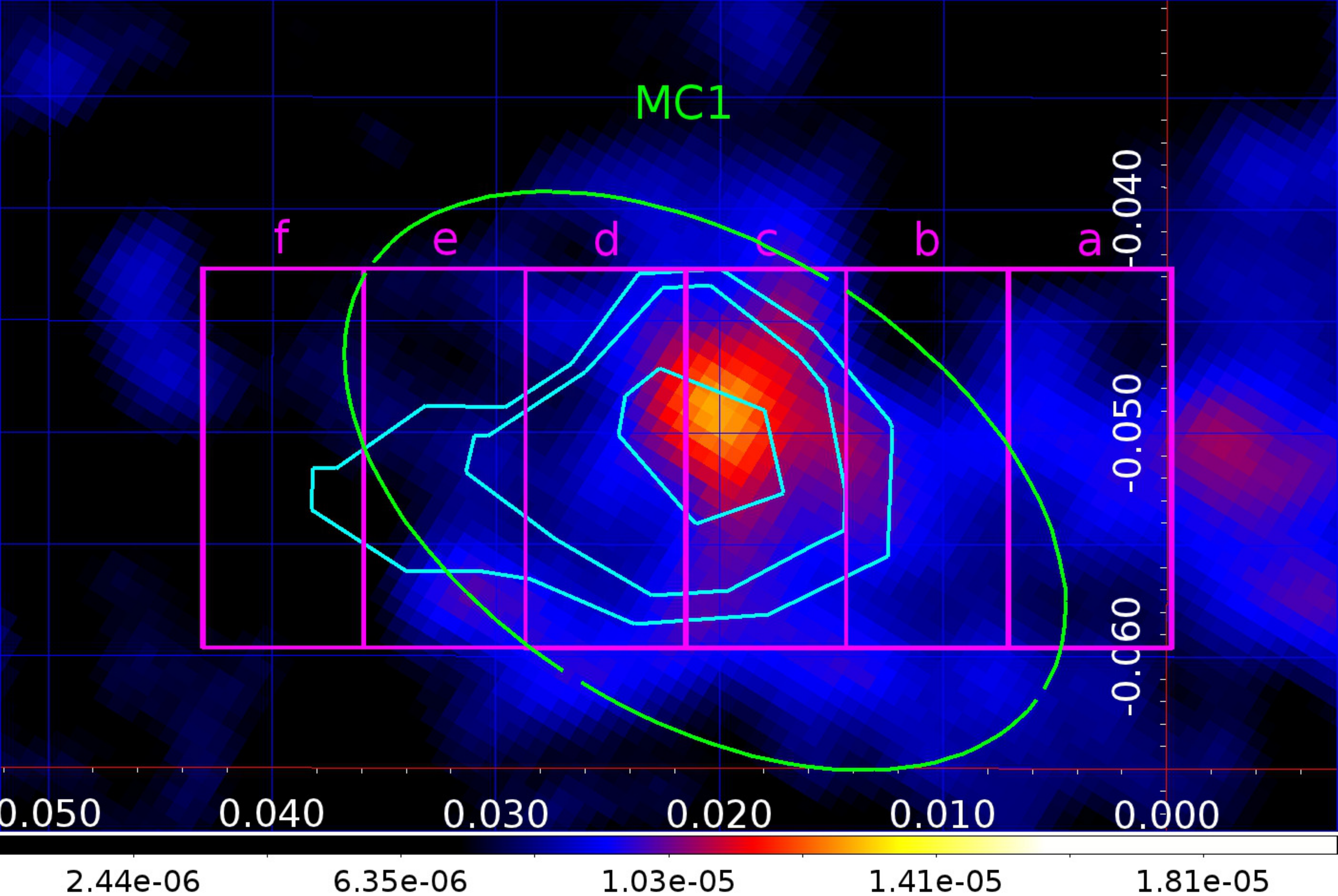}
\caption{\nustar\ 10-20 keV image zoomed around MC1 cloud region in the Galactic coordinates [$^\circ$] overlaid with a region used for 
extracting \xmm\ and \nustar\ spectra of MC1 (green), Fe~K$\alpha$ line intensity contours from 2012 \xmm\ observations (cyan), and the six 
$26''\times61''$ rectangular subregions (magenta) defined in \citet{Clavel2013}.}
\label{fig:mc1}
\end{figure}

With a separation of $\sim5$\amin, MC1 is the closest to Sgr A* in projection
among the molecular clouds emitting the Fe K$\alpha$ line. MC1 was one of the brightest clouds in 2012, 
and the Fe K$\alpha$  line flux of the overall cloud stayed nearly constant from 2000 to 2010 \citep{Ponti2010, Capelli2012}.
The bright 10-20 keV emission in MC1 coincides with the strong Fe K$\alpha$ line emission seen in 2012. 
Using \chandra\ data, \citet{Clavel2013} found different time
variations in sub-divided regions in the MC1 cloud between 2000 to 2010 --- the Fe~K$\alpha$ line flux decreased in the two regions dubbed
'a' and 'b' while it increased in the four regions dubbed 'c' to 'f' in Figure \ref{fig:mc1}. In 2012, \xmm\ and \nustar\ detected the 
brightest Fe~K$\alpha$ line and hard X-ray continuum emission coinciding with the 
'c' region, where Fe~K$\alpha$ line flux has been most prominent in MC1 since 2002 \citep{Clavel2013}.

The so-called Bridge is located on the east side of MC1. It contains multiple clouds exhibiting a range of Fe K$\alpha$
line flux light curves \citep{Ponti2010, Capelli2012, Clavel2013}.
In the Bridge, there are two bright regions both in the Fe K$\alpha$ line
and hard X-ray continuum emission. They correspond to the two distinct regions observed in the
N$_2$H$^+$ map at molecular line velocity $\sim$ +50~km\,s$^{-1}$  \citep{Jones2012}, dubbed the Br1 and Br2 regions
by \citet{Clavel2013}. Spectral analysis of MC1 and the Bridge is presented in \S \ref{sec:mc}, 
while detailed imaging and spectral analysis of the Arches cluster can be found in \citet{Krivonos2014}.

The east end of the Bridge is another molecular cloud, located at G0.11$-$0.11. In G0.11$-$0.11, there are two hard X-ray sources that do not
emit a strong Fe~K$\alpha$ line at 6.4 keV.
One is the X-ray filament \filc\ (\S \ref{sec:filc}), while the other is the bright magnetic CV
CXOUGC~J174622.7$-$285218 \citep{Muno2009}.
Unlike the GCMC, these two sources do not have strong Fe K$\alpha$ line emission and thus do not appear in the 6.4 keV Fe K$\alpha$
contours. Otherwise, we detected neither strong Fe K$\alpha$ nor hard X-ray continuum emission associated with the molecular cloud in
G0.11$-$0.11 probably because its X-ray flux has been decaying over the last decade \citep{Capelli2012}.
No hard X-ray emission above 10 keV was detected from other clouds such as
MC2, the 20 km\,s$^{-1}$ and  50 km\,s$^{-1}$ clouds.
Continuing the linearly decreasing trend of Fe K$\alpha$ line flux \citep{Clavel2013},
hard X-ray emission from MC2 may have slipped below the \nustar\ detection threshold.
Our results are consistent with no apparent detection of Fe K$\alpha$ emission from the
20 km\,s$^{-1}$ cloud and the 50 km\,s$^{-1}$ cloud \citep{Ponti2010}.
However, given their small offsets
($\la 10$~pc) relative to the GC where X-ray transients are highly concentrated \citep{Muno2005}, X-ray outbursts
lasting over a few years (e.g. SGR~J1745$-$29) may illuminate these clouds, and thus
X-ray reflection from there may be observed through time-varying Fe K$\alpha$ line and hard X-ray
continuum emission in the near future.

\subsection{20-40 keV band morphology} 
\label{sec:image_20-40keV}

\begin{figure*}[t]
\centerline{
\includegraphics[height=0.6\linewidth]{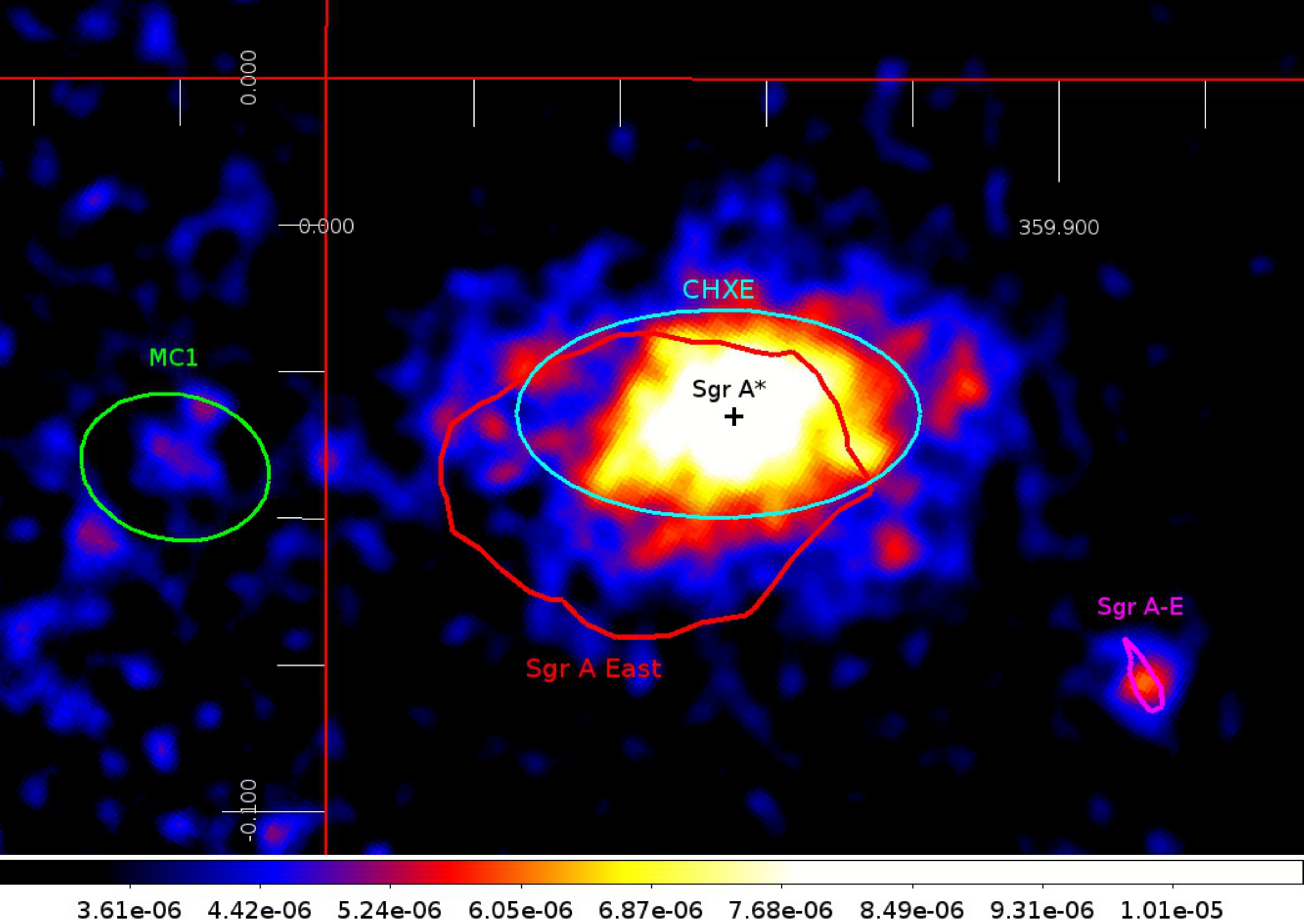}
}
\caption{\nustar\ 20-40 keV exposure-corrected smoothed image of the central 10\amin\ $\times$ 7\amin\ region around Sgr A* in the
Galactic coordinates [$^\circ$]. The image is overlaid with
the CHXE (cyan; the FWHM ellipse), Sgr A* (black), MC1 (green), Sgr A-E (magenta) and SNR Sgr A East non-thermal radio shell 
contours from 20 cm observation (red). Outside this region, only the LMXB 1E1743.1$-$2843 is visible above 20 keV.}
\label{fig:nustar_20-40keV}
\end{figure*}

Above 20 keV, besides the LMXB 1E1743.1$-$2843 and Sgr A-E, hard X-ray emission is observed within a $\sim3$\amin\ radius 
region around Sgr~A*.     
In order to investigate the central emission morphology precisely while avoiding image distortion of the Sgr~A* region
due to the off-axis PSF in the mini-survey data, we only used the three observations where Sgr~A* was on-axis.
Figure \ref{fig:nustar_20-40keV} shows the 20-40 keV \nustar\ image of the GC region. 
Diffuse emission from SNR Sgr~A~East (red contours in Figure \ref{fig:nustar_20-40keV}), which is comprised of a $kT \approx 1$~keV and $kT \approx 3$-6~keV
two-temperature thermal plasma \citep{Sakano2004, Park2005}, is no longer visible. 

Instead, the central $\sim10$ parsec of the persistent 20-40 keV emission is dominated by a point-like feature and 
a previously unknown diffuse X-ray component in the hard X-ray band \citep{Perez2015}. \citet{Perez2015} fit the raw count \nustar\ image in the 20-40 keV band with a two-dimensional model with 
two Gaussian profiles. The fitting range was restricted to the central $3'$ to minimize the effect of background variations between different detector chips 
and bias from the molecular cloud region to the northeast. 
The fitting procedure used the {\it Sherpa} package \citep{Freeman2001} to fully convolve the on-axis \nustar\ point spread function (PSF), telescope pointing fluctuation and vignetting function as well as fit a flat background component together. 

\begin{figure*}[t]
\centerline{
\includegraphics[height=0.6\linewidth]{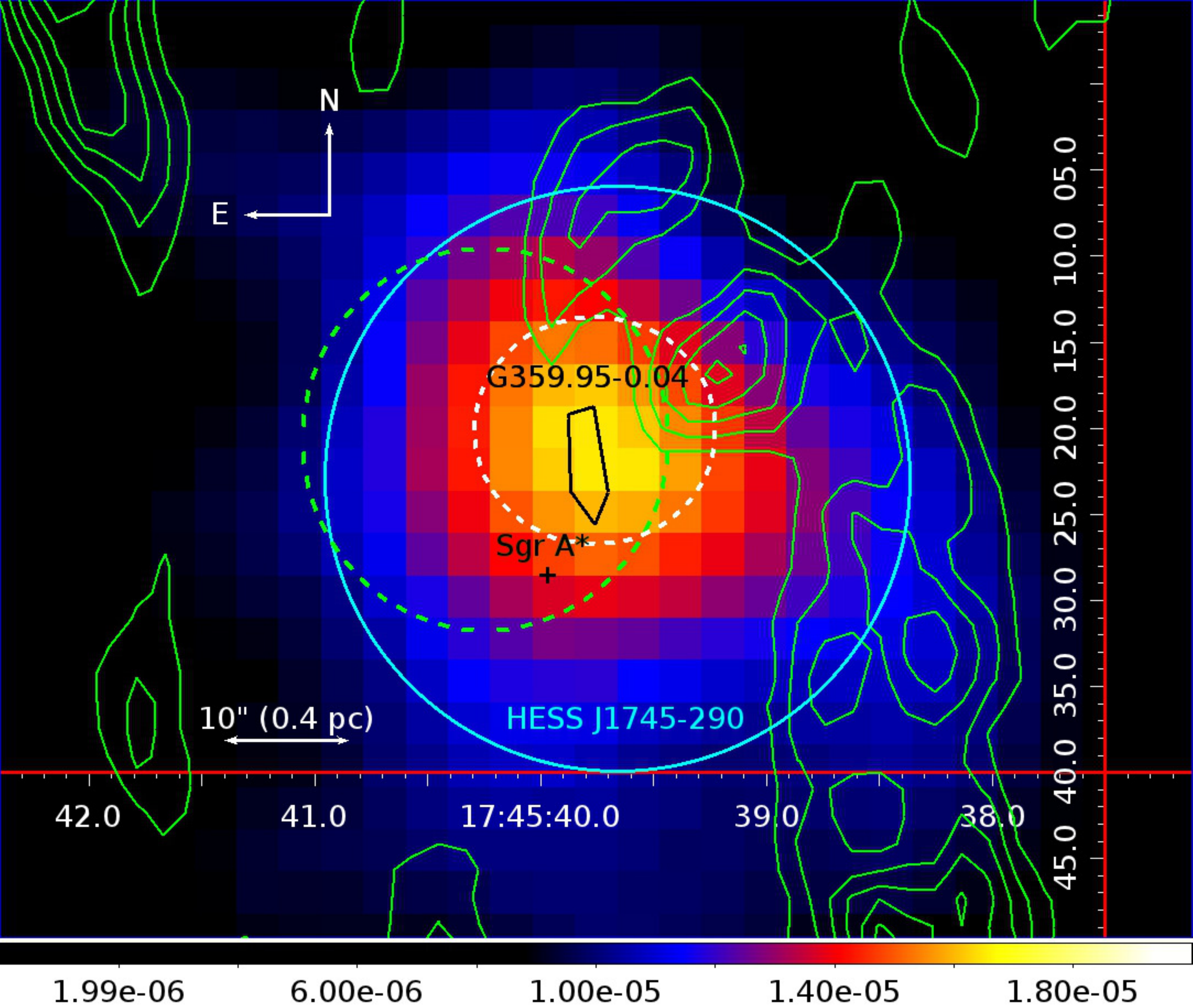}
}
\caption{\nustar\ 20-40 keV image zoomed in the central 3\amin\ region overlaid with Sgr A* (black cross), the centroid of the 
TeV source \hesssrc\ (cyan circle), PWN candidate \pwncen\ (black polygon) and circumnuclear disk (green contours). 
The centroid of the CHXE and point source detected in the 20-40 keV band are indicated by
green and white dashed circles with the 90\% c.l. circles including both statistical and systematic errors, respectively.}
\label{fig:central_20-40keV}
\end{figure*}

Figure \ref{fig:central_20-40keV} shows \nustar\ 20-40~keV image zoomed in the central 3\amin\ region around Sgr A*. 
The dominant feature is not resolved by \nustar\ and therefore consistent with a point source, and its centroid at 
RA = $17^{h}45^{m}39^{s}.76$ and DEC = $-29^{\circ}00^{\prime}20^{\prime\prime}.2$
(J2000; white dashed circle in Figure \ref{fig:central_20-40keV} indicating 90\% c.l position error of $7^{\prime\prime}$) aligns well with the head of the PWN candidate \pwncen. 
The compact size of \pwncen\ with $\sim6$\asec\ elongation as measured at $E<8$~keV band by \chandra\ \citep{Wang2006} (thus basically a point source with the $\sim1$\amin\ \nustar\ HPD) and spatial coincidence 
suggests that \pwncen\ is the likely counterpart to this point-like hard X-ray emission. This is further supported by our spectral analysis (\S \ref{sec:10pc}) and the fact that the head of \pwncen\ has the hardest power-law spectrum and the highest 2-8 keV 
flux in the filament \citep{Wang2006}.

On the other hand, the central hard X-ray emission (CHXE) is centered at 
RA = $17^{h}45^{m}40^{s}.24$ and DEC = $-29^{\circ}00^{\prime}20^{\prime\prime}.7$
(J2000; green dashed circle in Figure \ref{fig:central_20-40keV} indicating 90\% c.l position error of $11^{\prime\prime}$), and it has an extent (FWHM) of $l=3.3$\amin\  
 and $b=1.7$\amin\ or 8 pc and 4 pc assuming a GC distance of 8 kpc (cyan ellipse in Figure \ref{fig:nustar_20-40keV}). 
According to the detailed spectral study of two nearby intermediate polars and the CHXE 
by \citet{Hailey2015}, 
the CHXE emission is likely an unresolved population of massive magnetic CVs 
with white dwarf masses $M_{\rm WD}\sim 0.9 M_{\odot}$.  

\subsection{40-79 keV band morphology}
\label{sec:image_40-79keV}

\begin{figure*}[t]
\centerline{
\includegraphics[height=0.45\linewidth]{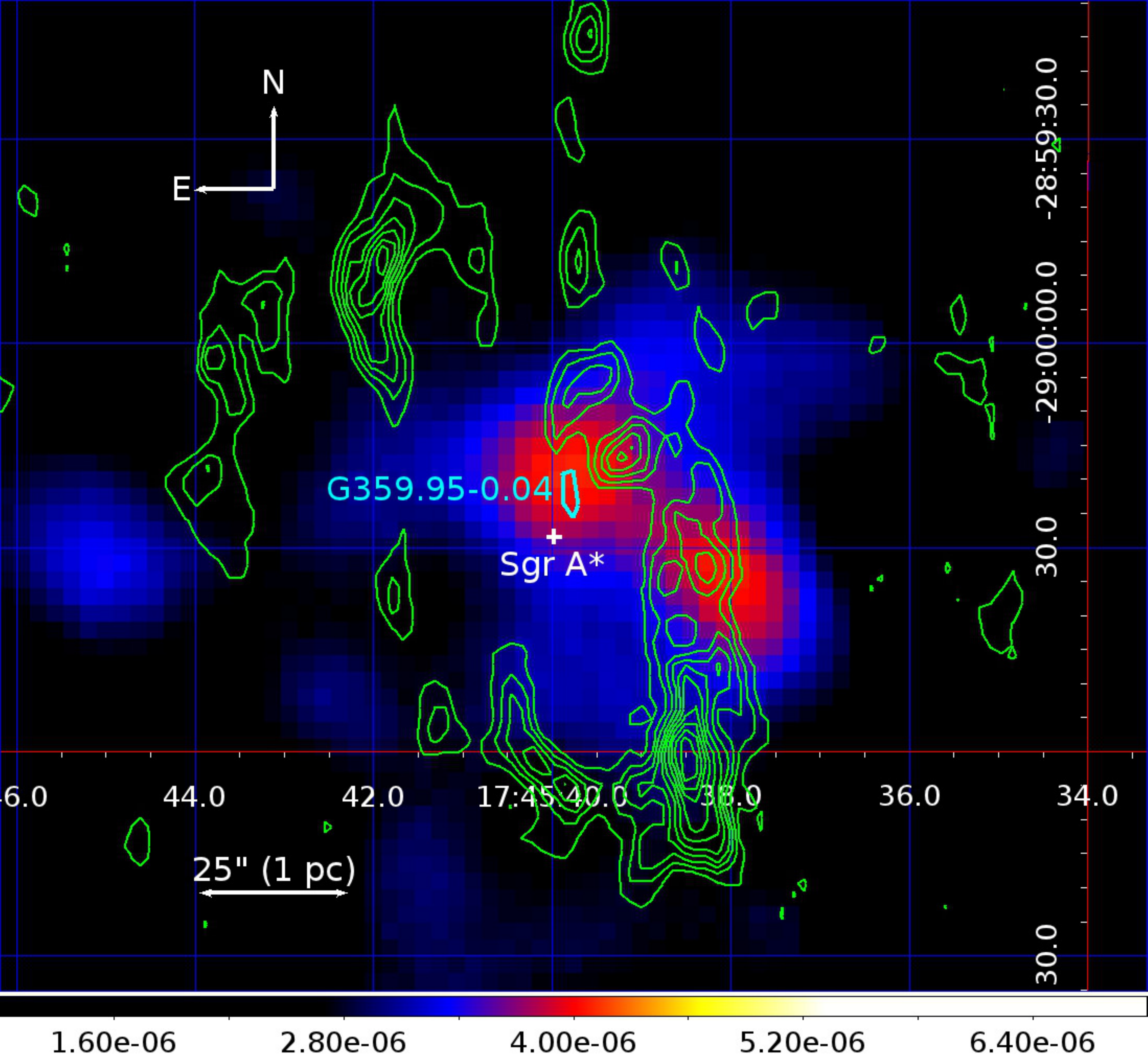}
\includegraphics[height=0.45\linewidth]{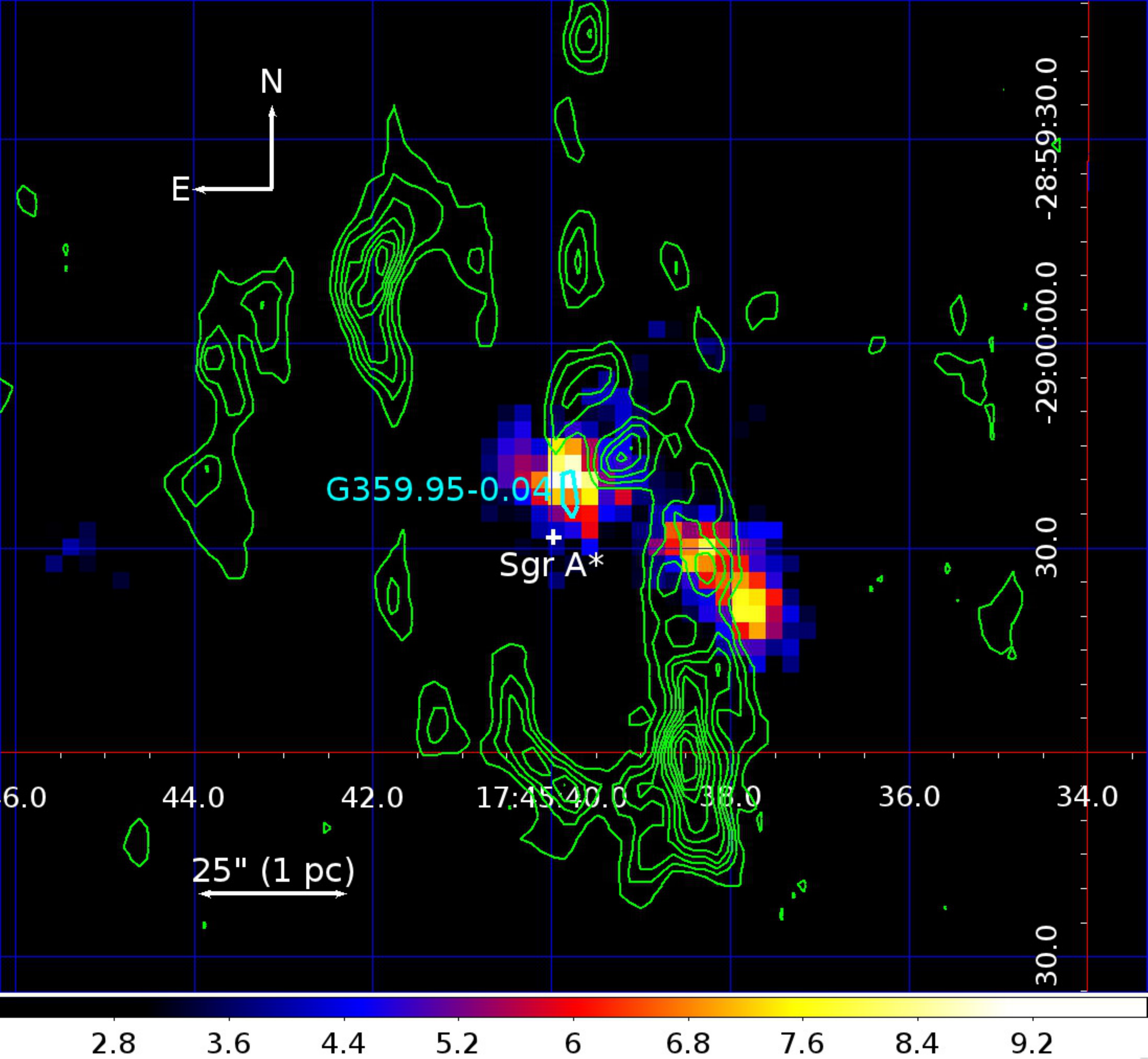} 
}
\caption{\nustar\ 40-79 keV exposure-corrected image (left) and trial probability map (right) of the central 1 arcmin region around 
the GC overlaid with PWN candidate \pwncen\ (cyan polygon), Sgr A* (white cross) and circumnuclear disk contours (green). The circumnuclear disk 
contours were obtained 
from OVRO HCN map \citep{Christopher2005}. The exposure-corrected image was smoothed by a Gaussian kernel with 7 pixel (17.5\asec) width, while the trial map is unsmoothed. In the trial map, 3-, 4- and 5-$\sigma$ detections correspond to values of 6.1 (orange), 7.8 (yellow) and 9.1 (white) 
respectively. }
\label{fig:nustar_40-79keV}
\end{figure*}

The only significant emission above 40 keV in this field is concentrated within the central 1\amin\ region of the GC, 
likely because the three Sgr A* observations 
have a longer combined exposure than the \nustar\ mini-survey (Figure \ref{fig:nustar_expmap}). 
Figure \ref{fig:nustar_40-79keV} shows a \nustar\ exposure-corrected smoothed image and the matching trial map of the central 1\amin\ region around Sgr A* 
in the 40-79 keV band, which is the only region with significant emission in this band. 
Given the fewer source counts, we smoothed the \nustar\ image with a larger Gaussian width of 17.5\asec\ (7 pixels) for better illustration. 
The view of the GC drastically simplifies above 40 keV --- the emission is centered around \pwncen\ with some potential substructures. 

The trial map clearly exhibits two distinct features above the 4-$\sigma$ level.  
One is a point-like feature centered at the head of \pwncen\ and also spacially coincident with the TeV 
source \hesssrc. 
This feature is persistent, observed in all individual observations.
The other is a protrusion elongated in the south-west direction. Its significance is highest in the FPMA data of one of the Sgr A* observations (ObsID: 30001002003), and thus the protrusion should be taken with some caution as a potential artifact.  
The protrusion is not spatially coincident with the radio \citep{Yusef2012} or X-ray jets \citep{Li2013}, 
but it intersects with the cooler molecular gas of the circumnuclear disk indicated by green contours 
\citep{Morris1996, Christopher2005}. There is no apparent counterpart in either the \chandra\ 2-8 keV image or the \xmm\ 6.4 keV Fe K$\alpha$ image. It is possible that 
soft X-ray emission from the protrusion may be heavily absorbed by  
the optically thick circumnuclear disk and also contaminated by 8 keV thermal emission.

%%%%%%%%%%%%%%%%%%%%%%%%%%%%%%%%%%%%%%%%%%%%%%%%%%%%%%%%%%%%%%%%%%%%%%%%%%%%%%%%%%%%%%%%%%%%%%%%%%%%%%%%

\section{Non-thermal X-ray filaments} 
\label{sec:filaments}

\begin{figure*}[t]
\centerline{
\includegraphics[height=0.6\linewidth]{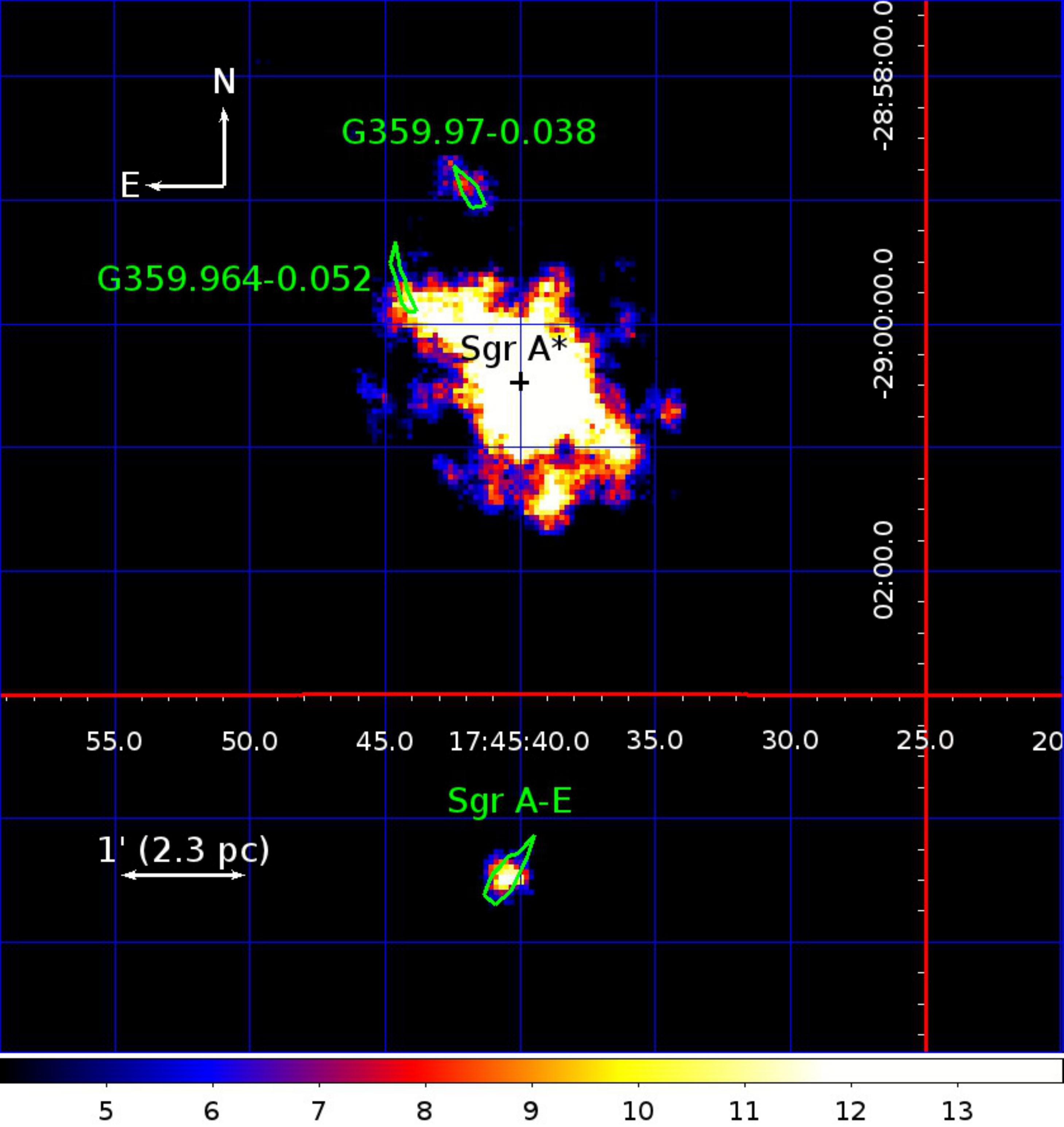} 
}
\caption{\nustar\ 20-40 keV trial map overlaid with three known X-ray filaments (green polygons). The polygons 
roughly traces the 2-8 keV morphologies of the filaments determined with \chandra. The image scale was chosen from 
log($P_{\rm trial}$) = 4 to 15 to illustrate the \nustar\ filaments. log($P_{\rm trial}$) $>9.1$ (orange color bar) indicates above 5-$\sigma$ detection.}
\label{fig:filament}
\end{figure*}

Throughout the Sgr A* and GC mini-survey observations, 
{\it NuSTAR} detected four non-thermal X-ray filaments (\aeknot\ or Sgr A-E, \fila, \filc\ and \pwncen) above 10 keV. 
The 20-40 keV trial map 
(Figure \ref{fig:filament}), where Sgr A East diffuse emission is no longer dominant, illustrates the filaments 
Sgr A-E and \fila. 
On the other hand, \filc\ is located in the molecular cloud G0.11$-$0.11 and it was detected by the mini-survey observation (shown as 
one of the green polygons in Figure \ref{fig:nustar_10-20keV}). \pwncen, which lies $9''$ away from Sgr A*, appears in the 
zoomed images around Sgr A* shown in \S\ref{sec:image_20-40keV} and \S\ref{sec:image_40-79keV}.  
 These hard X-ray filaments are 
among the brightest in the soft X-ray band with 2-8 keV fluxes above $1\times10^{-13}$ \eflux\ or an unabsorbed luminosity of 
$8\times10^{32}$ \,erg\,s$^{-1}$ at a distance of 8 kpc \citep{Johnson2009}. 
Although we detected hard X-ray emission from a part of \filb\ shown in Figure \ref{fig:filament}, its spectral 
identification as the known X-ray filament is unclear since it might be 
confused with a bright \chandra\ source CXO J174543.7$-$285947 with $F_{\gamma, \rm 2-8 keV} = 6.8\times10^{-6}$\,photons\,cm$^{-2}$
\,s$^{-1}$ 
within $\sim10$\asec\ of the filament \citep{Muno2009}, in addition to some contamination from 8 keV thermal emission and the CHXE. 
In the following sections, we individually discuss three out of the four hard X-ray filaments detected by \nustar\ above 10 keV. 
\pwncen\ will be later discussed in connection with the TeV source \hesssrc\ (\S\ref{sec:10pc}). 

\subsection{\aeknot\ (Sgr A-E) --- TeV electrons trapped in magnetic tubes} 

\aeknot\ is the brightest X-ray filament, with a 3-79 keV flux of $2.0\times10^{-12}$ \eflux\ or an unabsorbed X-ray luminosity of $2.6\times10^{34}$ 
\,erg\,s$^{-1}$ assuming a distance of 8 kpc \citep{Zhang2014}. The filament was detected up to $\sim50$ keV with a best-fit power-law index of $\Gamma=2.3\pm0.2$.
We do not detect another prominent radio filament Sgr A-F (G359.90$-$0.06), above 10 keV. This is 
consistent with the fact that Sgr A-F is significantly fainter than \aeknot\ \citep{Lu2008}.   
Based on the high-resolution radio and X-ray morphology of the filament as well as spectral 
analysis, \citet{Zhang2014} ruled out both a PWN scenario and a SNR-molecular cloud interaction. 
Instead, the most plausible scenario is that  magnetic flux tubes trap $\sim100$~TeV electrons, which emits synchrotron X-rays up to $\sim50$ keV. 
Since $\sim100$ TeV electrons have cooling times as short as $\sim1$ year for $B\sim0.1$~mG, 
electrons must be accelerated nearby before entering the filament. One possible external source of TeV electrons is 
relativistic protons accelerated from Sgr A* 
or SNRs interacting with the nearby 20 km\,s$^{-1}$ cloud which produces secondary electrons via pion decays. The electrons can diffuse out of the cloud before they cool significantly by synchrotron radiation, and become trapped in the magnetic flux tubes. Another (less likely) possibility is that 
a population of unresolved $\sim10^{5}$ year old PWNe accelerate  
electrons to TeV energies. 
\suzaku\ has detected extended X-ray emission from such $\sim10^5$ year-old PWNe elsewhere in the Galactic Plane \citep{Bamba2010}, but low brightness X-ray emission from old PWNe  
may be contaminated by the strong GC diffuse emission or be below the \nustar\ detection level. We therefore cannot completely rule 
out this scenario. 

\subsection{\fila\ - The Sgr A East shell interacting with the 50 km\,s$^{-1}$ cloud} 

\fila\ is located just outside the Sgr A East shell, and it is close to the ``Plume'' region. By jointly fitting the \nustar\ and \chandra\ spectra of the filament, \citet{Nynka2014NF} found that its non-thermal spectrum extends to $\sim50$ keV with 
the best-fit photon index $\Gamma =1.3_{-0.2}^{+0.3}$. The photon index of this filament is significantly harder than 
that of \aeknot\ ($\Gamma=2.3\pm0.2$).  
Using the high-resolution radio and \chandra\ image of the filament as well as SED model fitting including the \nustar\ results, \citet{Nynka2014NF} found that the PWN scenario is again highly unlikely. 
Instead, the filament is likely illuminated by the interaction between the shell of SNR Sgr A East and the 50 km\,s$^{-1}$ cloud 
\citep{Nynka2014NF}, as evidenced by the large width of the CS $J=1$-0 line, which exceeds the cloud bulk velocity of $\sim50$ km\,s$^{-1}$ \citep{Tsuboi2006}. The harder X-ray power-law spectrum of $\Gamma=1.3$ 
is typical of non-thermal bremsstrahlung or inverse Compton emission of electrons accelerated at the SNR-cloud interaction site \citep{Bykov2000, Bykov2005}. 
The lack of an apparent radio counterpart is also consistent with this picture.  
The GeV source 2FGL~J1745.6-2858 detected by \fermi\ is coincident with the position of \fila\ 
\citep{Nolan2012, Yusef2013}. If the GeV source is  
associated with \fila, it is additional evidence supporting the SNR-cloud interaction scenario since the SED 
model of \citet{Bykov2000}  
predicts a peak in the GeV band. 

\subsection{\filc\ - pulsar wind nebula? } 
\label{sec:filc}

The third hard X-ray filament \filc, shown in Figure \ref{fig:nustar_10-20keV}, is located near the Radio Arc region and is embedded in the molecular cloud 
G0.11$-$0.11. 
The filament is a candidate PWN due to its cometary shape and a point-like feature CXOGCS~J174621.5$-$285256 
\citep{Wang2002b}. It is one of the few X-ray filaments that 
has a radio counterpart. It is not possible to extract a clean \nustar\ spectrum of the filament due to the limited statistics, and contamination from the bright X-ray source CXOUGC~J174622.7$-$285218, $\sim40''$ away 
from the filament. CXOUGC~J174622.7$-$285218 is a magnetic CV with a 1745~s periodicity \citep{Muno2009} and \nustar\ detected its hard X-ray extension above 10 keV \citep{Hong2015}. A deeper \nustar\ observation with more than $\sim200$ ksec exposure will be 
required to perform useful spectral and timing analyses of this filament. 

\subsection{Heterogeneous origin of non-thermal X-ray filaments?}

Two of the three hard X-ray filaments (\aeknot\ and \fila) detected above 10 keV are unlikely to be PWNe, suggesting a heterogeneous origin for the X-ray filaments.  
\aeknot\ is likely powered by synchrotron radiation in 
magnetic flux tubes trapping TeV electrons, while \fila\ is illuminated by Sgr A East interacting with a 
50 km\,s$^{-1}$ 
cloud. Our results indicate a reservoir of relativistic electrons and protons in the central 10 pc region, rather than production and acceleration of  
particles locally inside the filaments as in the PWN scenario.  
Electrons may be accelerated to TeV energies by faint $\sim10^5$ year-old PWNe or they are by-products of  
hadronic interactions between relativistic protons and clouds. 

Alternatively, \citet{Linden2011} proposed dark matter annihilation as a potential source of GeV electrons that are trapped in magnetic 
flux tubes 
and emit synchrotron radiation. In this scenario, light neutralinos with $\sim5$-10~GeV mass annihilate directly to 
leptons that decay to GeV electrons. The four radio filaments (G0.2$-$0.0, G0.16$-$0.14, G0.08$+$0.15 and G359.1$-$0.2)
 investigated by \citet{Linden2011} using their model 
 are located outside the \nustar\ GC survey area or did not have sufficiently long exposure time in the \nustar\ mini-survey coverage to warrant study. Deep X-ray observations of these radio filaments could test the dark matter scenario    
since any X-ray detection of these radio filaments would indicate the presence of TeV electrons that cannot be produced in the annihilation of $\sim$5-10~GeV 
mass neutralinos.  
A more extensive hard X-ray survey of radio and X-ray filaments probe 
not only the spatial and energy distribution of cosmic-rays beyond the central 10 pc region but also dark matter physics. 

%%%%%%%%%%%%%%%%%%%%%%%%%%%%%%%%%%%%%%%%%%%%%%%%%%%%%%%%%%%%%%%%%%%%%%%%%%%%%%%%%%%%%%%%%%%%%%%%%%%%%%%%%%%%%

\section{Galactic center  molecular clouds}
\label{sec:mc}

All Sgr A clouds, including MC1, MC2, the Bridge and G0.11$-$0.11, were covered by the \nustar\ mini-survey as well as
\xmm\ observations in  2012. As Figure \ref{fig:nustar_10-20keV} in \S~\ref{sec:image_10-20keV} shows, we find that Fe
K$\alpha$ line emission (as measured by \xmm) and hard X-ray continuum (as measured by \nustar) emission are spatially
well-correlated in MC1, the Bridge and the
Arches cluster.
In October 2013, a 300 ks \nustar\ observation of the fading Sgr B2 spatially 
resolved hard X-ray emission from the Sgr B2 core and a newly discovered cloud feature G0.66$-$0.13 
\citep{Zhang2015}. 
Sgr C is not suitable for \nustar\ observations because of strong ghost-ray background from the bright persistent 
LMXB~1A1742$-$294. 

Two models, the so-called X-ray reflection nebula (XRN) model \citep{Sunyaev1993} and the low-energy cosmic-ray (LECR) model
\citep{Yusef2002}, predict distinct spectral and temporal properties for the X-ray emission from GCMCs.
In the XRN scenario, molecular clouds can reflect X-rays from an illuminating source by scattering continuum X-rays and producing 
fluorescence line emission
following photo-ionization of K-shell electrons.
The XRN model predicts (1) variability of Fe K$\alpha$ line and X-ray
continuum emission over the light-crossing time of a cloud ($\sim$1-10 years) or over the variability time scale of
an illuminating source, (2) a strong Fe K$\alpha$ line with equivalent width (EW) $\ga 1$ keV, (3) a Fe-K
photo-absorption edge at 7.1 keV and (4) a Compton reflection hump (i.e. curved power-law spectrum) in the hard X-ray band
if the cloud column density is high ($N_{\rm H} \gg 10^{24}$~cm$^{-2}$).
Alternatively, low-energy cosmic ray electrons (LECRe), protons and ions (LECRp) can eject K-shell electrons via
collisional ionization leading to fluorescence line emission.
The LECR model
predicts (1) a power-law spectrum orignating from non-thermal bremsstrahlung emission, (2) an Fe K$\alpha$ line with EW
$\la$~(0.25-0.4)~$Z_{Fe}$ keV where $Z_{Fe}$ is
the Fe abundance relative to solar \citep{Yusef2007, Yusef2013}, and
(3) time variability of Fe K$\alpha$ line and X-ray continuum emission over the
electron cooling/diffusion  time (LECRe) or long-term variability over $\ga 100$ years (LECRp).
The shape of the X-ray continuum is sensitive to the incident cosmic ray energy spectrum.

Previous soft X-ray observations have been mainly focused on tracking time evolution of Fe K-$\alpha$ line at 6.4 keV \citep[see][for a review]{Ponti2013}, due to their 
narrow bandpass (typically $\sim$4-8~keV) where different spectral components such as diffuse thermal emission,
X-ray continuum from the cloud, Fe 
K~edge and Fe K fluorescent lines are potentially all present and strongly degenerate.
Both the EW of the Fe K$\alpha$ line and the absorption depth ($\tau_{\rm FeK}$) of the Fe K edge are
highly sensitive to the underlying
X-ray continuum level. Diffuse thermal emission, if not properly subtracted, will enhance the underlying continuum level and thus decrease
both the Fe K$\alpha$ line EW and $\tau_{\rm FeK}$.
However, in the previous \xmm, \chandra\ and \suzaku\
analysis, intrinsic X-ray continuum
spectra either have been poorly constrained \citep{Inui2009, Ponti2010, Nobukawa2011} or have been assumed to be a power-law spectrum
with $\Gamma$ fixed to 1.9 \citep{Capelli2012}.
More importantly, previous X-ray studies determined the parameters of the GCMCs and illuminating X-ray sources
separately from individual components such as the Fe K$\alpha$ line
or absorption edge, therefore they lack self-consistency. In the XRN scenario, an Fe K$\alpha$ line flux measurement yields a 
luminosity of the illuminating primary souce 
only at $\sim8$ keV with some uncertainty associated with Fe abundance \citep{Sunyaev1993}. 

In constrast, a broad-band X-ray continuum measurement provides the most robust determination of the X-ray spectrum of 
the primary source in the XRN scenario and a cosmic ray energy spectrum in the LECR scenario. The hard X-ray continuum provides an 
excellent measurement of the intrinsic column density ($N_{\rm H}$) of the cloud 
\citep{Ponti2014}. 
In the subsequent sections,  we fit self-consistent spectral models to the broad-band 
X-ray spectra of the Sgr A clouds using the \nustar\ and \xmm\ data. This provides a powerful diagnostic that can distinguish between 
different models and tightly constrain parameters since it takes the full advantage of the broadband X-ray spectroscopy.  

\subsection{Spectral analysis of the Sgr A clouds: MC1 and the Bridge} 
\label{sec:mc_spec}

We extracted \nustar\ and \xmm\ EPIC-PN source spectra of MC1 and the Bridge using the same regions quoted in \citet{Ponti2010}, 
as indicated by the green  
regions in Figure \ref{fig:nustar_10-20keV}. \xmm\ observations 0694641101 (35.5~ks total exposure) and  
0694640401 (43.9~ks total exposure) were used for MC1 and the Bridge respectively since the sources are not intercepted by detector 
gaps and the signal-to-noise 
ratio is highest in these observations. This allows us to extend our energy band to 10~keV for the \xmm\ spectra, 
while the background dominates above $\sim8$~keV in the other observations. 
We selected appropriate \nustar\ observations 
and focal plane module data that cover the full extent of the clouds that are free from 
high background counts. 
We extracted MC1 source spectra from FPMA data of one Sgr A* observation (ObsID: 30001002004) 
and three mini-survey observations (ObsID: 40010001002, 40010002001 and 40010004001), for a total exposure time of 125.7 ksec.  
We extracted the Bridge spectra from FPMA data of three \nustar\ mini-survey 
observation (ObsID: 0010003001, 0010004001 and 40010006001) with a total exposure time of 71.5 ksec. 
Although there are two bright regions in the Bridge \citep[the so-called Br1 and Br2 region in][]{Capelli2012}, 
separate spectral analysis of each region does not yield sufficient photon statistics.
Since we are not certain whether SLB or focused diffuse emission dominates as the background of 
these regions, we applied both the conventional and off-source background subtraction methods described
in the Appendix. We found that the final results were not significantly different  
between the two methods because the contribution of SLB and focused diffuse emission is similar in these molecular cloud regions.  

We fit the joint \xmm\ and \nustar\ spectra with XSPEC models based on the 
XRN and LECR scenarios. 
For all spectral models considered here, we applied {\verb=tbabs=}*{\verb=[apec + apec + cloud_model]=} where {\verb=cloud_model=} 
represents one of the X-ray spectral models that 
are intrinsic to the GCMCs and described in the subsequent sections. In either case, the common 
model components are {\verb=tbabs=} and two {\verb=apec=} models representing foreground (galactic) absorption, $kT\sim1$ 
and $kT\sim8$~keV thermal components in the GC, respectively. 
We linked all the fit parameters between the \xmm\ and \nustar\ spectra except the flux normalizations for the two thermal ({\tt apec}) 
model components since background spectra mostly composed of the two thermal components were extracted differently
for the \xmm\ and \nustar\ spectra. Hereafter, we present the best-fit flux normalizations of the two thermal components from the \xmm\ spectral fitting. 

For the LECR scenario, we fit a self-consistent X-ray spectral model available in XSPEC for both the
 LECR electron and proton cases, by taking into account both X-ray continuum and fluorescent
line components calculated from the energy loss of cosmic-rays penerating 
into a slab-like cloud of neutral gas at a constant rate \citep{tatischeff2012}. 
Since the observed year-scale time variability of Fe K$\alpha$ line flux in the Sgr A clouds
rules out the LECR proton scenario, we fit an absorbed LECR electron model ({\verb=tbabs*lecre=}) as the intrinsic cloud model 
{\verb=cloud_model=}. 
\footnote{A similar model was 
used to fit X-ray spectra of
Sgr B2 clouds \citep{Terrier2010, Zhang2015}. The photo-absorption term
takes into acount intrinsic absorption in the cloud with a characteristic column density $N_{\rm H}$. Although Fe K-shell
electrons are ionized by cosmic-rays coming from an external source,
continuum X-rays emitted via non-thermal bremsstrahlung can undergo photo-absorption before escaping from the cloud. We also set $N_{\rm H}=0$ for the opposite case where continuum X-rays are emitted
near the surface of the cloud, in which case most of them are not absorbed in the cloud.
Although this is not a self-consistent treatment of the intrinsic absorption in the cloud, the two cases should bound the problem where 
the radiative transfer of continuum X-ray photons is not considered in the LECR models.}  
Following \citet{tatischeff2012} and \citet{Krivonos2014}, we fixed the path length of cosmic-ray electrons to
$\Lambda = 5\times10^{24}$~cm$^{-2}$ since we find that the fitting results are insensitive to $\Lambda$.
In all cases, the LECRe models do not fit the \xmm\ and \nustar\ spectra for MC1 and the Bridge as well as the XRN models,
yielding $\chi^2_\nu = 1.2$-1.4 (Figure~\ref{fig:spec_mc1} and \ref{fig:spec_bridge}).
The spectral fitting requires unreasonably high metallicity $Z \approx 4$ in order for the LECRe model to
account for the strong Fe~K$\alpha$ line.
Therefore, we conclude that the LECRe models are not consistent with the X-ray spectra of the
two Sgr A clouds.

\subsection{Spectral fitting results with MYTorus model}
\label{sec:fit_reflection}

Hereafter we present spectral fitting results primarily with MYTorus model, which is the only Monte-Carlo based XRN 
model that is available in XSPEC with finite cloud column density \citep{Murphy2009, Yaqoob2012}. Unlike other XRN models that have been 
applied to GCMC X-ray data, the MYTorus model can determine the cloud and primary X-ray source parameters self-consistently.  
Indeed, we find that the MYTorus model yields better spectral fits than the other XRN models as shown in this section.  
The Appendix fully describes the MYTorus model application to the GCMC X-ray data and compare it with other widely-used XRN models.   
For comparison between the different models and also with the previous results, we present the fit results using
an ad hoc XRN model {\verb= tbabs*(powerlaw + gauss + gauss) =} 
and a slab geometry model {\verb=reflionx =} with infinite optical depth \citep{Magdziarz1995, Nandra2007, Ross2005}. 
Other slab geometry models such as {\verb=pexmon=} yield similar results.
In the ad hoc XRN model, we fixed the line energy and width of (weak) Fe K$\beta$ line to 7.06~keV
and 0.01~keV, respectively and linked its flux normalization to that of Fe K$\alpha$ line
multiplied by 0.15, i.e. the ratio of the K$\alpha$
and K$\beta$ line fluorescence yields \citep{Murakami2001}. 

The MYTorus model includes three components, namely the transmitted continuum (MYTZ), scattered continuum (MYTS) and
Fe fluorescent emission lines (MYTL), in a range of equatorial
hydrogen column density through the tube of the torus
$N_{\rm H} = 10^{22}-10^{25}$~cm$^{-2}$, power-law photon index $\Gamma = 1.4$-2.6
and incident angle (between an observer and the symmetry axis of the torus) $\theta_{\rm obs} = 0-90^\circ$. See Figure
\ref{fig:mc_geometry} in the 
Appendix for the
geometry of the MYTorus model in comparison with the conventional geometry used in many publications on GCMCs.
Note that $\theta_{\rm obs} = 0^\circ$ and $90^\circ$
correspond to a face-on and edge-on observing view, respectively.
Since we observe only the reflected X-ray emission from GC molecular clouds, we adopted two additive XSPEC models for
scattered continuum (MYTS) and Fe fluorescent lines (MYTL):   \\

\noindent {\verb= atable{mytorus_scatteredH500_v00.fits} + =} \\
{\verb= atable{mytl_V000010pEp040H500_v00.fits} =} \\

\noindent in a ``coupled'' mode where the same primary X-ray spectrum  is
input for the both components.
We selected the MYTS and MYTL tables with a power-law model with the highest energy cut-off at $E=500$ keV.
Following the MYTorus manual \footnote{http://mytorus.com/mytorus-instructions.html},
we selected the MYTL data table with an energy offset of +40 [eV] since the best-fit Fe K$\alpha$ line
centroids with a Gaussian line profile are
6.44 [keV] probably due to slightly ionized Fe in the clouds and/or instrumentral energy offset \citep[note that
\nustar\ has a systematic uncertainty of 40~eV
near Fe emission lines; ][]{Madsen2015}. We bound the incident angle to
$\theta_{\rm obs} \le 60^\circ$ since we find that the MYTorus model is valid to
fit the X-ray spectra of GCMCs in this range (Appendix).

\begin{figure*}[t]
\includegraphics[height=0.5\linewidth, angle=270]{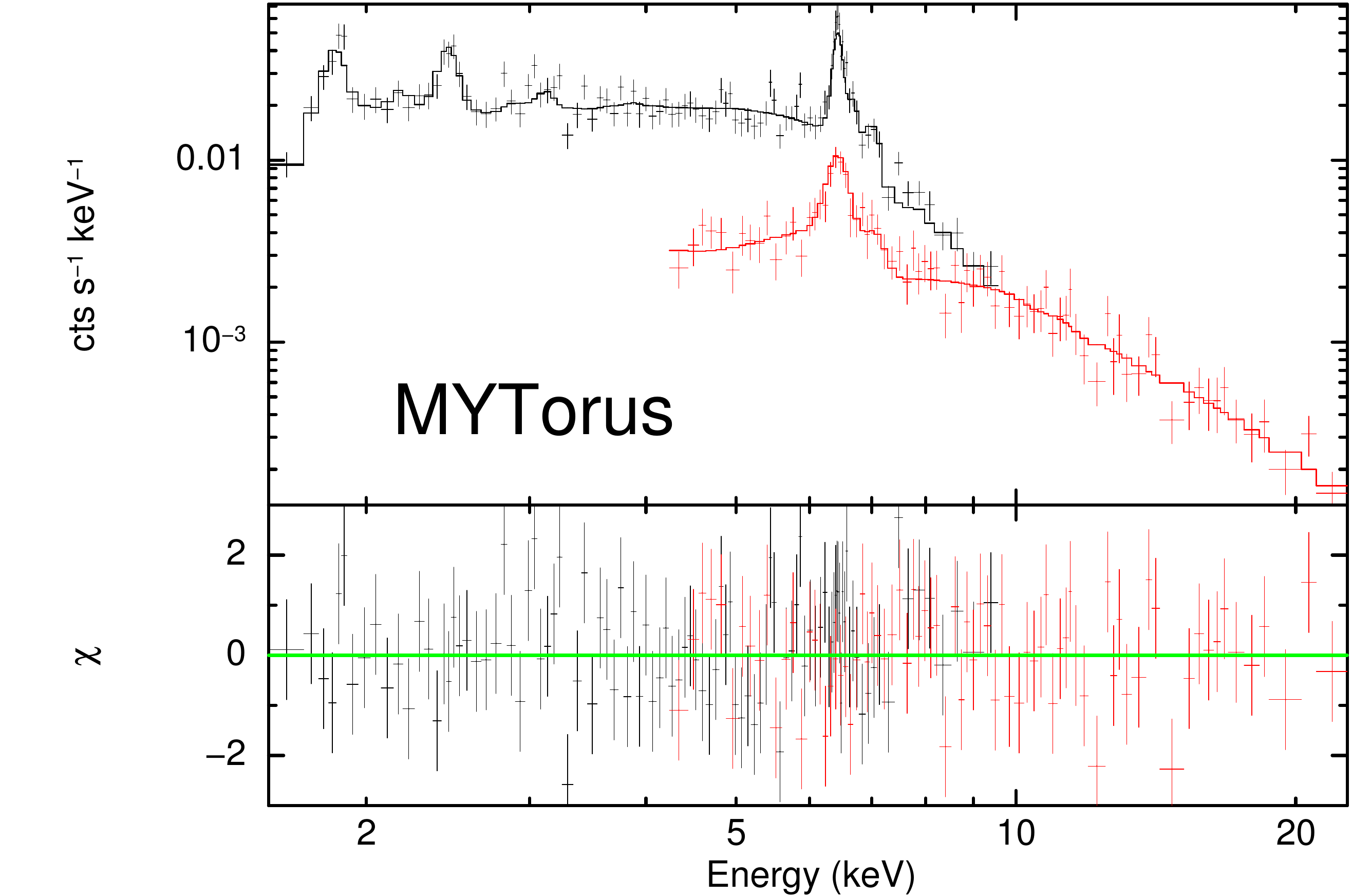}
\includegraphics[height=0.5\linewidth, angle=270]{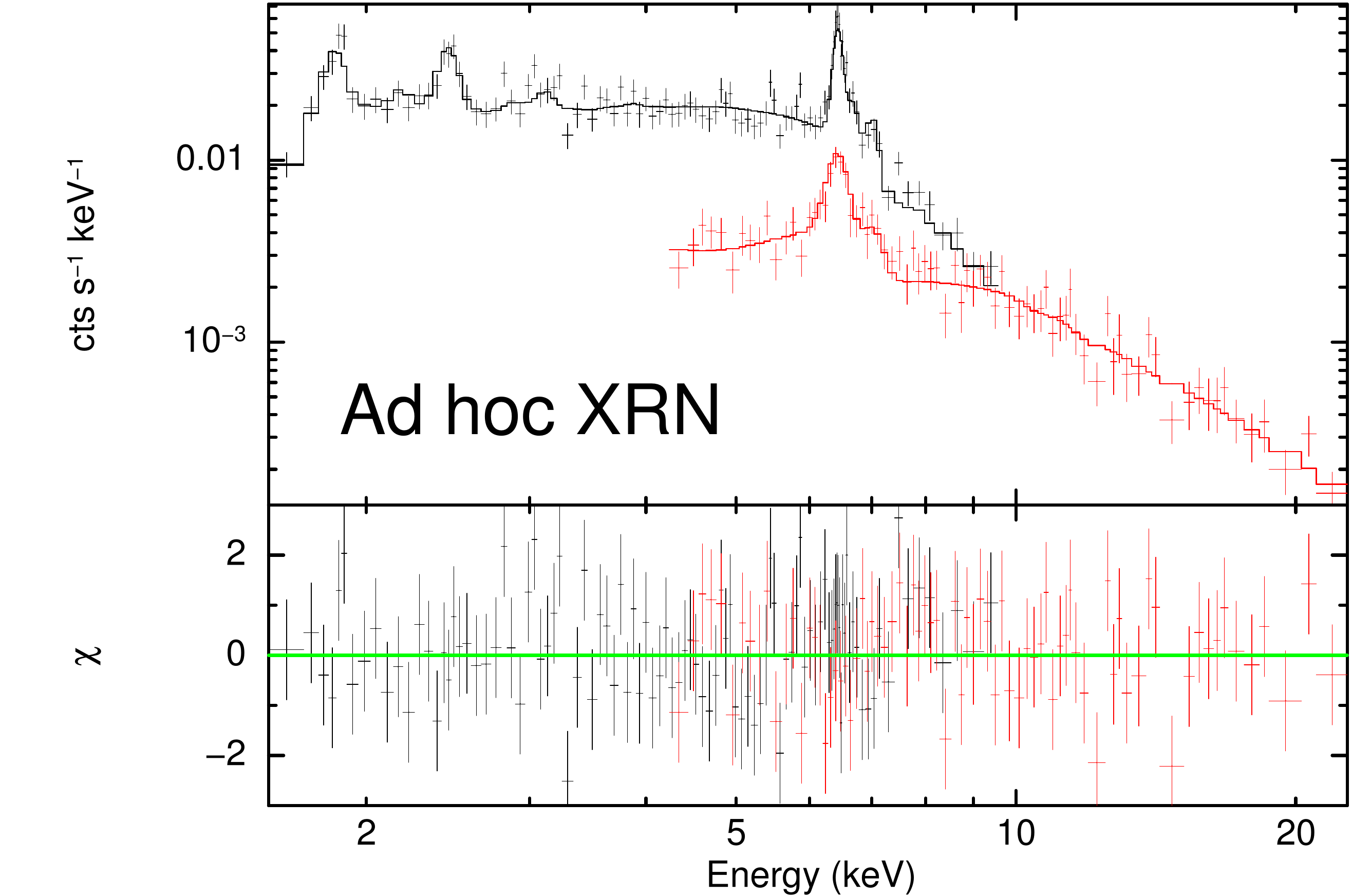} 
\includegraphics[height=0.5\linewidth, angle=270]{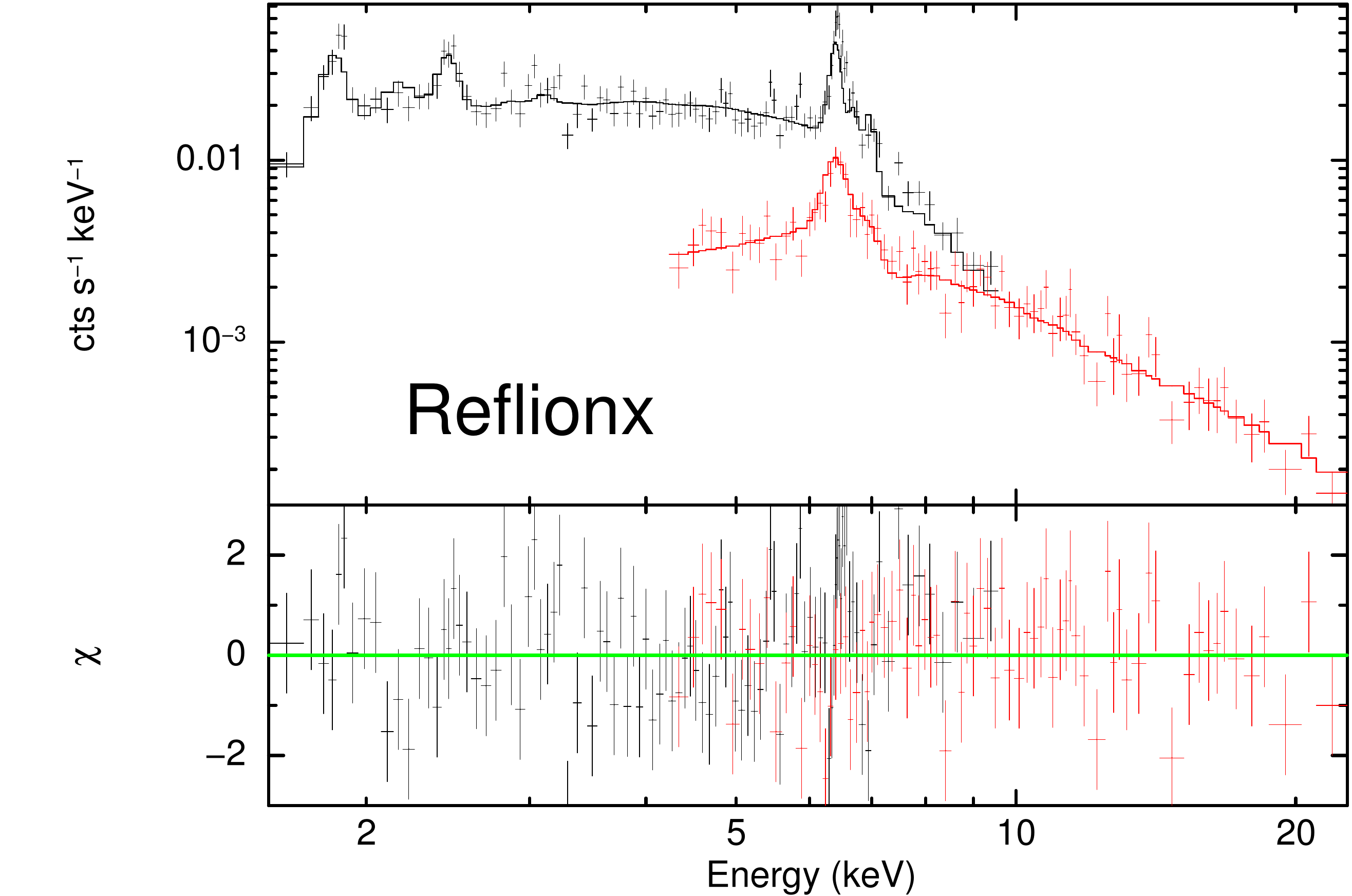} 
\includegraphics[height=0.5\linewidth, angle=270]{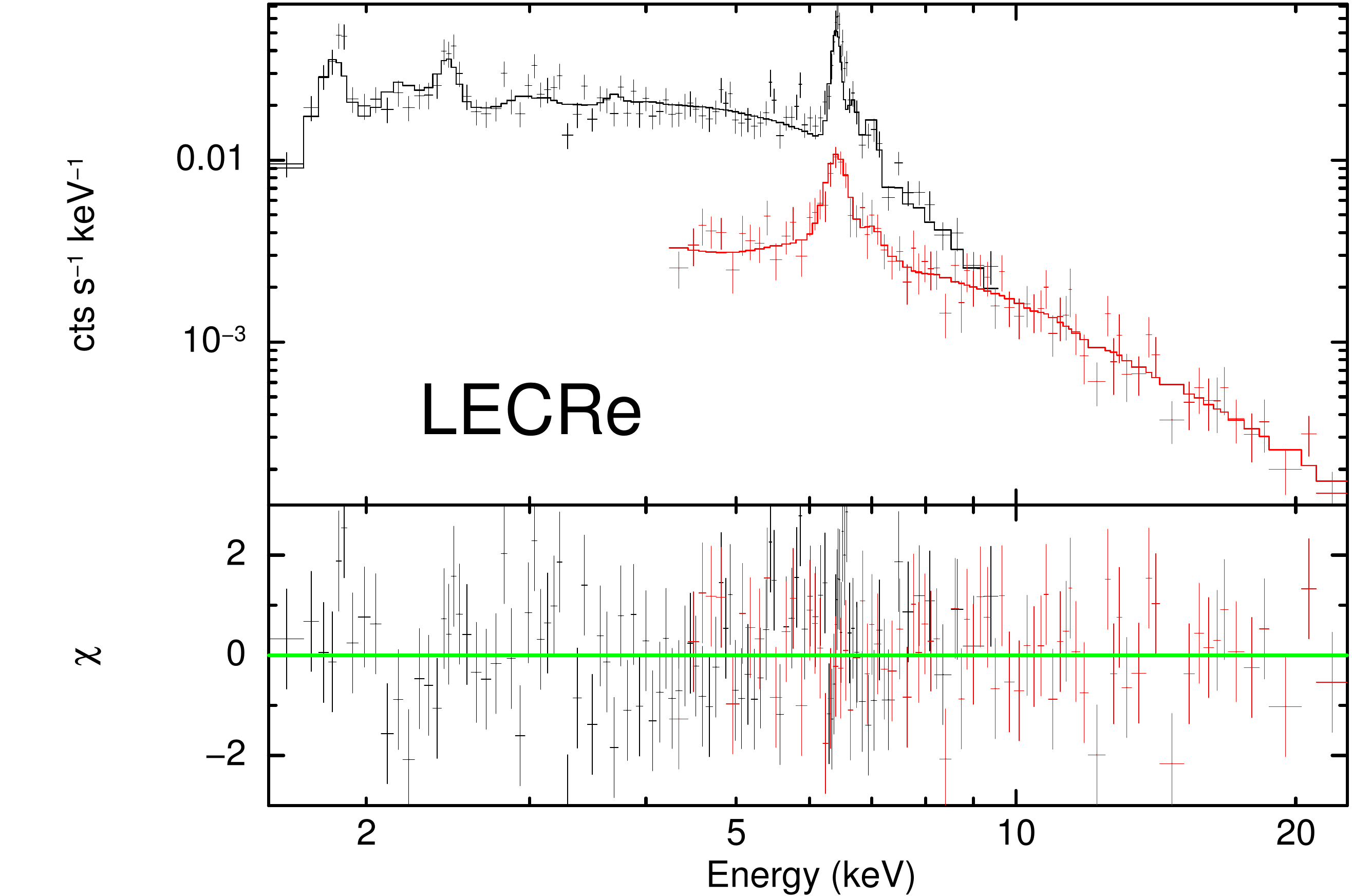} 
\caption{1.5-20 keV \xmm\ (black) and \nustar\ (red) spectra of MC1 
fit with the MYTorus (upper left), ad hoc XRN (upper right), reflionx (lower left) and LECR electron model (lower right). }
\label{fig:spec_mc1}
\end{figure*}

\begin{figure*}[t]
\includegraphics[height=0.5\linewidth, angle=270]{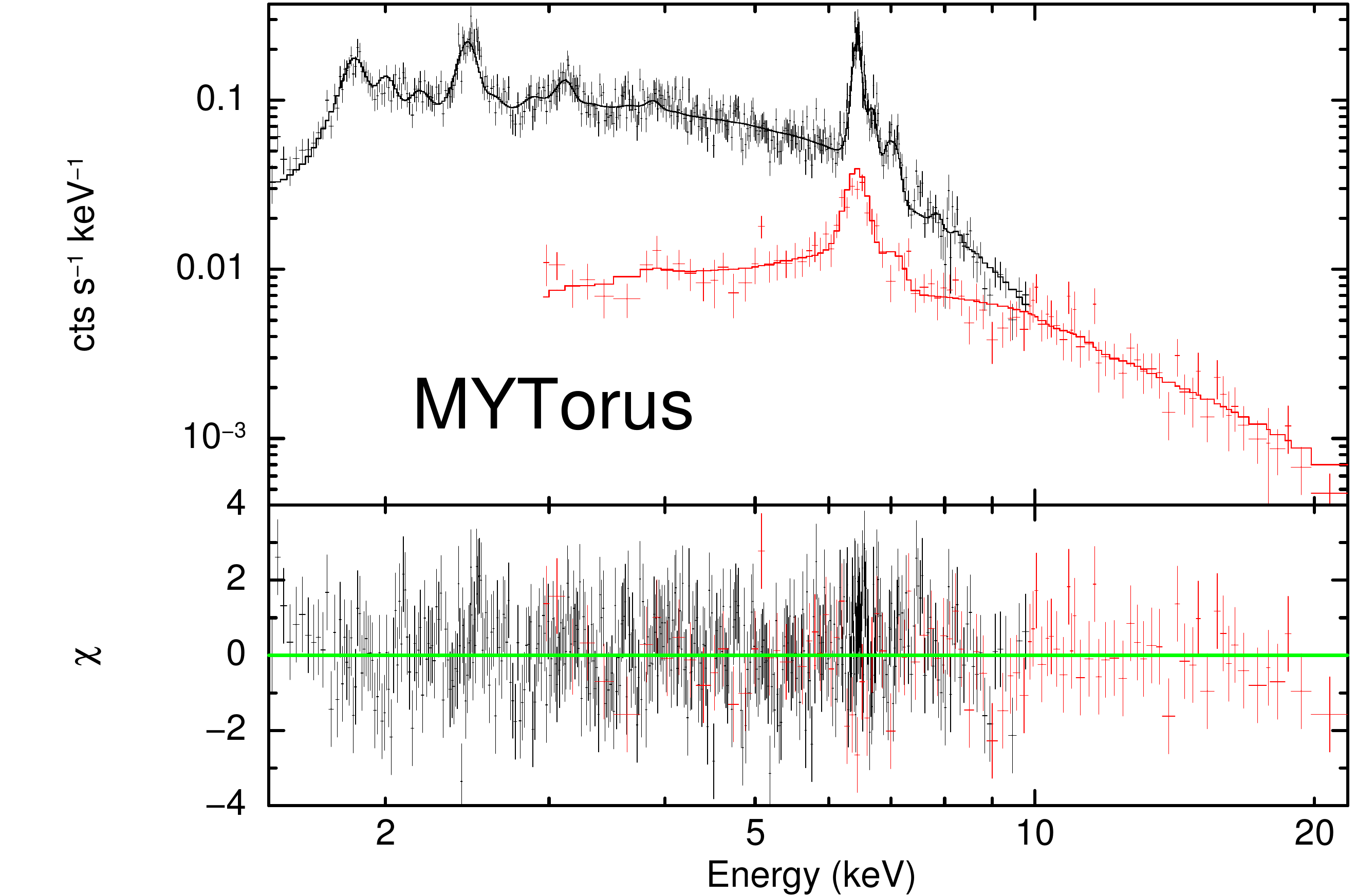}
\includegraphics[height=0.5\linewidth, angle=270]{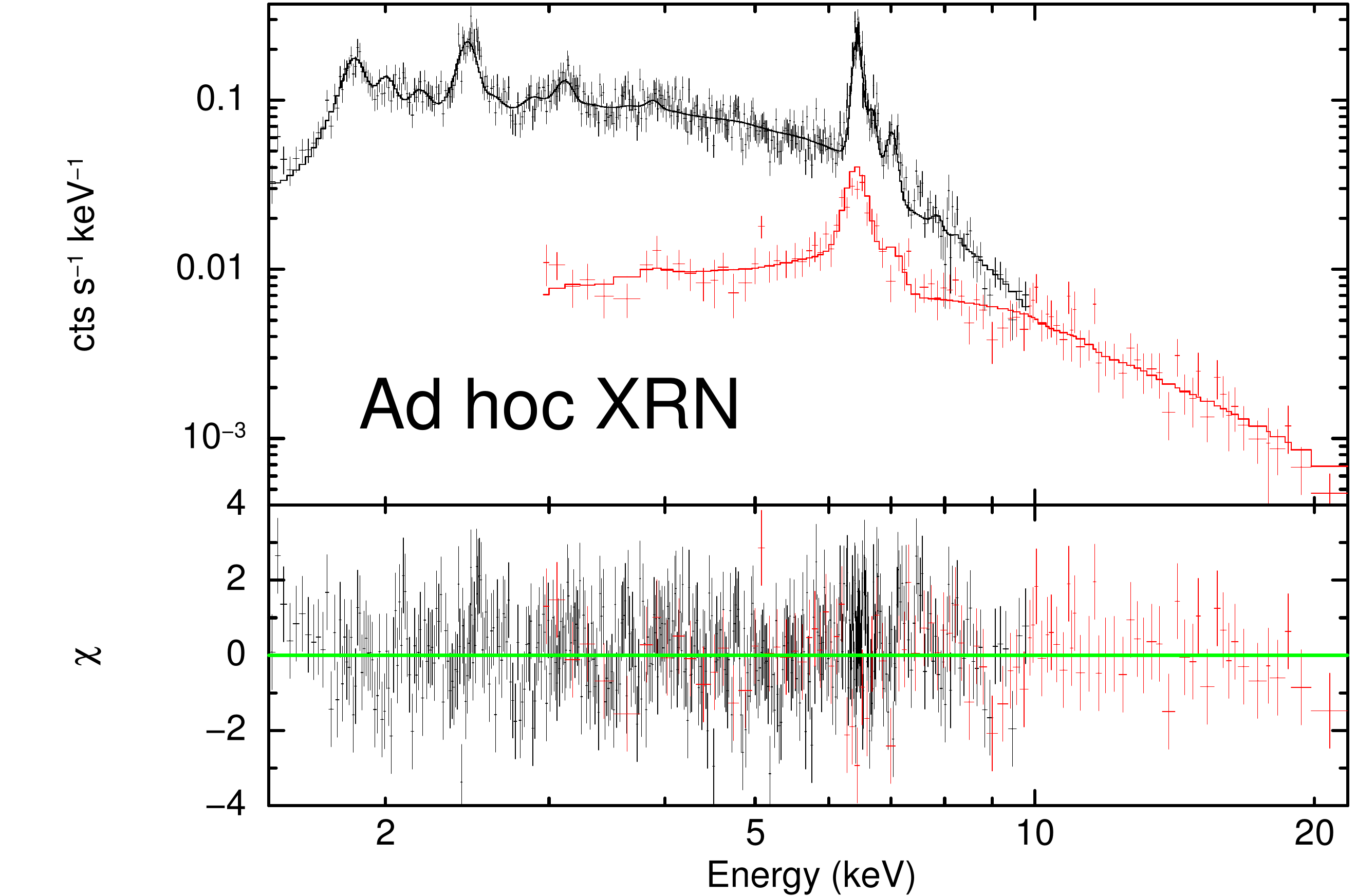} 
\includegraphics[height=0.5\linewidth, angle=270]{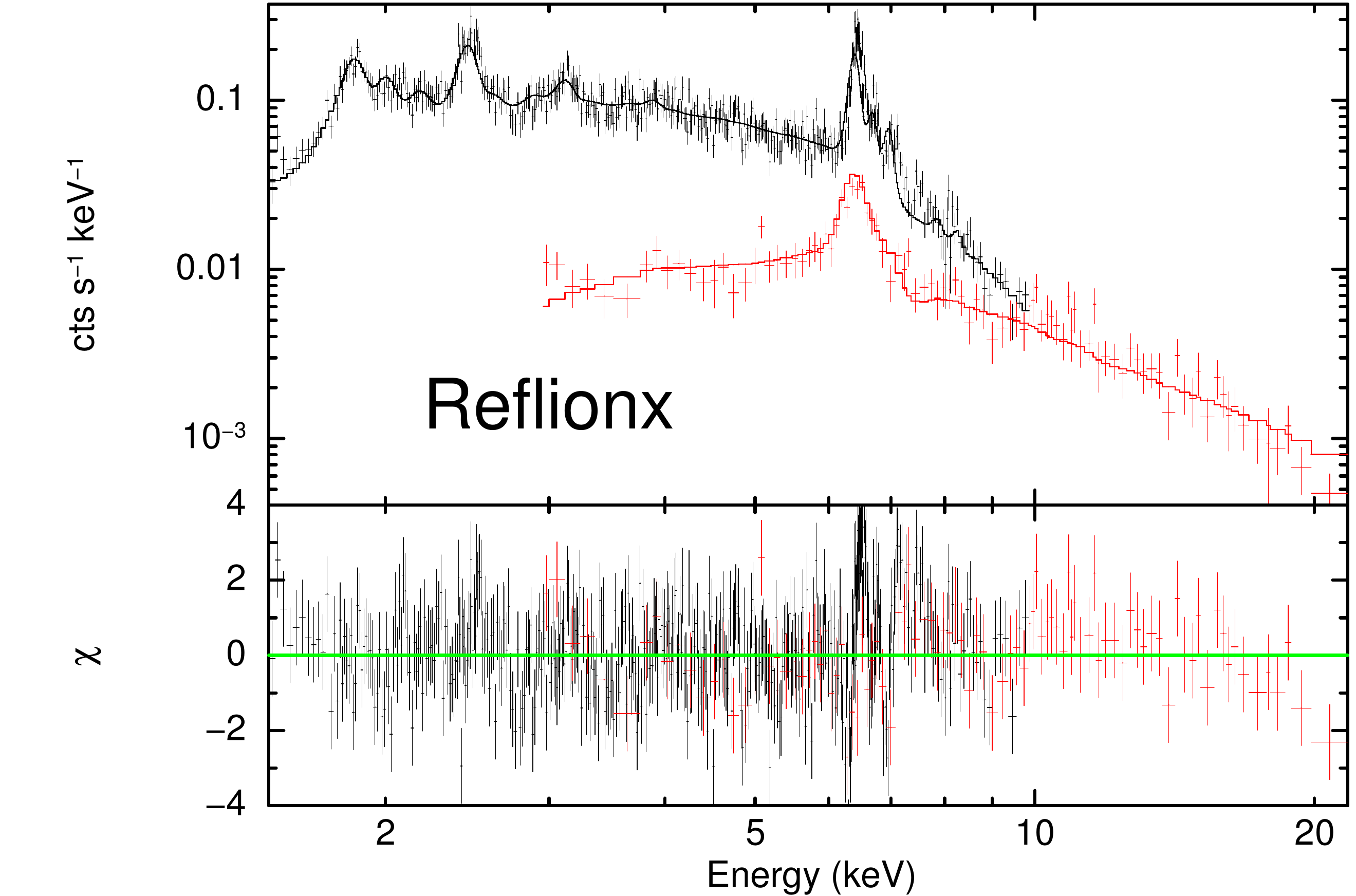}
\includegraphics[height=0.5\linewidth, angle=270]{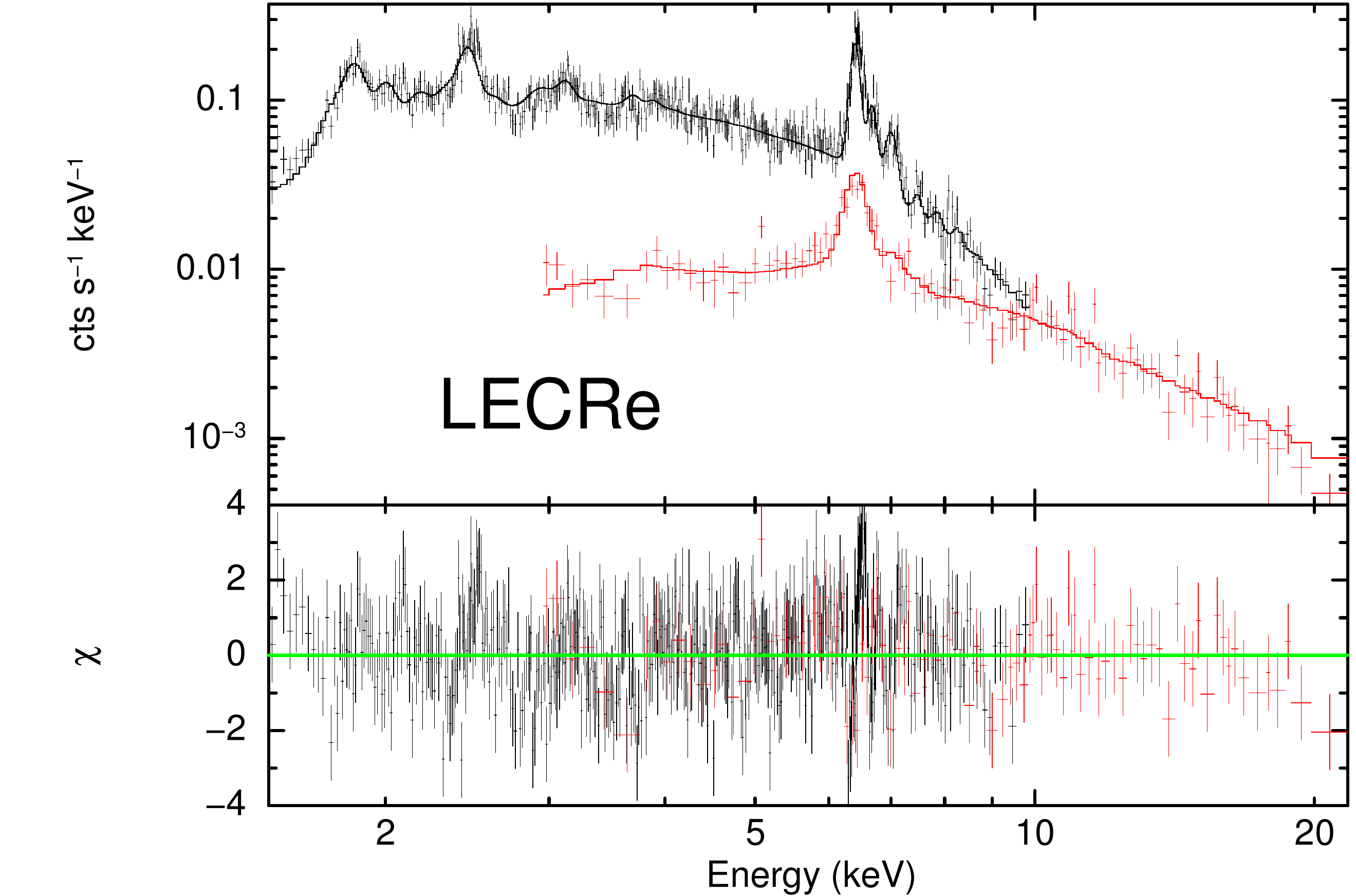}
\caption{1.5-20 keV \xmm\ (black) and \nustar\ (red) spectra of the Bridge 
fit with the MYTorus (upper left), ad hoc XRN (upper right), reflionx (lower left) and LECR electron model (lower right). }
\label{fig:spec_bridge}
\end{figure*}

The MYTorus model fits the joint \xmm\ and \nustar\ spectra of MC1 and 
the Bridge well, yielding $\chi^2_\nu/{\rm dof} = 1.01/170$ (MC1) and 1.13/524 (the Bridge), with 
all parameters well constrained 
(Figure \ref{fig:spec_mc1}, \ref{fig:spec_bridge} and Table \ref{tab:mc_spectra}). 
We found that  the intrinsic absorption and power-law continuum were accurately measured
only by the joint fitting of
\xmm\ and \nustar\ spectra, as a result of combining high-resolution Fe line spectroscopy from \xmm\ with broad-band 
X-ray spectroscopy from \nustar. 
The two thermal components 
have $kT_1 \sim1$ and $kT_2 \sim8$ keV and are consistent with the previous measurements 
in this region \citep{MunoDiffuse2004, Koyama2007}. 
Although the abundance for the lower $kT_1 \sim1$~keV temperature component is poorly constrained, we find that 
it does not affect the XRN model parameters. Although the ad hoc XRN model yields similar fit quality with 
$\chi^2_\nu/{\rm dof} = 1.01/168$ (MC1) and 1.16/522 (the Bridge), 
the MYTorus model has fewer fit parameters due to its self-consistency - the power-law index and flux normalization are 
linked between the scattered continuum (MYTS) and the fluorescent line component (MYTL). 
The {\verb=reflionx =} models do not fit MC1 and the Bridge 
spectra well with $\chi^2_\nu/{\rm dof} = 1.40/170$ and 1.60/525, respectively. 
When we fit the MC1 spectra with the {\verb=reflionx =} model, we fixed 
the plasma temperature of the second thermal component to 8~keV. Otherwise, the spectral fitting yields unreasonable parameters 
such as $kT_2 \sim 20$~keV and Fe abundance higher than 10 for the {\verb=reflionx =} model. 

\begin{deluxetable*}{lcccccc}
\tablecaption{{\it NuSTAR} + \xmm\ spectral fitting results of MC1 and the Bridge using the three XRN models}
\tablewidth{0pt}
\tablecolumns{7}
\tablehead { \colhead{} & \multicolumn{3}{c}{MC1} & \multicolumn{3}{c}{Bridge} \\  
\hline \\
\colhead{Parameters}  & \colhead{Ad hoc XRN} & \colhead{reflionx} & \colhead{MYTorus} & \colhead{Ad hoc XRN} & \colhead{reflionx} & \colhead{MYTorus}   }
\startdata
$N^f_{\rm H} [10^{22}$\,cm$^{-2}$]    & $7.1_{-0.6}^{+0.7}$      &  $8.3_{-0.6}^{+0.5}$  & $7.1\pm0.7$  & $6.1\pm0.2$  & $5.9\pm0.2$ & $6.1\pm0.2$ \\
$kT_{1}$ [keV] &   $0.62_{-0.07}^{+0.09}$      & $0.43\pm0.04$  &  $0.59_{-0.04}^{+0.1}$ & $0.90\pm0.03$  & $0.91\pm0.02$ &  $0.91\pm0.03$ \\
Abundance $Z_{1}$ &  $3.0_{-1.7}^{+2.0}$ & $5.0_{-2.9}$  &  $5.0_{-3.3}$  & $2.9_{-0.9}^{+2.1}$  & $5.0_{-0.7}$ & $2.8_{-0.3}^{+1.8}$ \\ 
$norm_{1}$ &  $5.5_{-2.8}^{+10.4} \times 10^{-3}$ & $1.4_{-0.6}^{+2.1}\times10^{-2}$ &  $3.7_{-0.2}^{+1.0}\times10^{-3}$  & $8.8_{-3.8}^{+4.2}\times10^{-3}$  & $4.2_{-0.4}^{+0.8}\times10^{-3}$ & $8.9_{-3.4}^{+3.9}\times10^{-3}$\\
$kT_{2}$ [keV] &   $8.1_{-0.5}^{+1.0}$      &  8 (fixed) &  $9.1_{-1.5}^{+1.6}$   &   $6.5_{-0.6}^{+0.8}$  & $10.6_{-0.6}^{+0.8}$ & $7.5_{-0.7}^{+0.3}$ \\
Abundance $Z_{2}$ & $0.6\pm0.2$ &  $0.7_{-0.1}^{+0.2}$  &    $0.7\pm0.2$  &  $0.5\pm0.1$   & $0.77_{-0.07}^{+0.08}$ &  $0.75_{-0.07}^{+0.15}$  \\
$norm_{2}$ &  $(9.3\pm0.4)\times10^{-4}$ & $(1.1\pm0.2)\times10^{-3}$ & $9.3_{-0.1}^{+0.2}\times10^{-4}$  & $(3.4\pm0.4)\times10^{-3}$  & $(3.8\pm0.2)\times10^{-3}$ &  $(2.7\pm0.3)\times10^{-3}$\\
\\
$N_{\rm H} [10^{23}$\,cm$^{-2}$]    &   $2.1\pm0.6$        &  ---   & $2.3_{-0.6}^{+1.0}$   &   $1.5\pm4$  & --- & $1.5_{-0.3}^{+0.6}$ \\
PL photon index ($\Gamma$)                     & $2.20\pm0.15$    &  $2.95_{-0.16}^{+0.14}$ &  $2.11_{-0.14}^{+0.23}$  & $1.81\pm0.11$  & $2.29_{-1.7}^{+1.6}$ & $1.81\pm0.10$\\
PL norm \tablenotemark{a}     &  $(1.5\pm0.5)\times10^{-3}$  & $7.3_{-3.3}^{+5.8}\times10^{-5} $& $1.8_{-0.4}^{+0.6}\times10^{-2}$  & $9.4\times10^{-4}$  & $2.2_{-0.9}^{+1.2}\times10^{-5}$  &  $(3.8\pm0.5)\times10^{-2}$  \\
Fe K$\alpha$ energy [keV]      & $6.444\pm0.008$   & --- & ---    & $6.439\pm0.003$   & --- & ---  \\
Fe K$\alpha$ flux \tablenotemark{b} &   $1.6\pm0.1$ & ---  & ---   &  $5.6\pm0.2$   & --- & --- \\
Fe K$\alpha$ EW [keV] &    $0.93\pm0.12$       &  ---    &  ---  &    $1.38\pm0.14$  & --- & --- \\
Fe abundance & --- & $1.4\pm0.2$ & 1 (fixed)   & --- & $3.8\pm0.6$ & 1 (fixed)  \\ 
Inclination angle [$^\circ$] & --- & --- & $60_{-23}$ \tablenotemark{c}   & --- & ---  & $4.5^{+15.3}_{-4.5}$ \\
\\
$\chi_\nu^2$ (dof)                  & 1.01  (168)              &  1.40 (170)  & 1.01 (168)  & 1.16 (522)  & 1.60 (525) & 1.13 (524)      
\enddata
\tablecomments {The errors are $68\%$ confidence level. Fe K$\beta$ line parameters are not listed since they are either fixed or 
linked to Fe K$\alpha$ parameters. $N^{f}_{\rm H}$ and $N_{\rm H}$ refer to the best-fit hydrogen column density for 
the foreground and 
intrinsic absorption term in the X-ray reflection models defined in \S~\ref{sec:fit_reflection}. }
\tablenotetext{a}{Flux normalization [photons\,cm$^{-2}$\,s$^{-1}$\,keV$^{-1}$] at 1 keV. The flux normalizations are defined 
differently in the three XRN models. For example, the ad hoc XRN model refers to the observed X-ray flux, while the MYTorus model refers to the incident X-ray source flux. Therefore, their best-fit values cannot be simply compared with each other. }
\tablenotetext{b}{Flux unit is $10^{-5}$ ph\,cm$^{-2}$\,s$^{-1}$.}
\tablenotetext{c}{We set the upper bound of $\theta_{\rm obs}$ to $60^\circ$ since it is the valid range for the MYTorus model to approximate the spectrum of a quasi-spherical molecular cloud (Appendix B1).}
\label{tab:mc_spectra}
\end{deluxetable*}

\paragraph{Intrinsic column density} 

Our joint \xmm\ and \nustar\ analysis using the MYTorus model measured the equatorial hydrogen column 
density (with 90\% c.l. errors) with $N_{\rm H} = 2.3_{-0.7}^{+1.2}\times10^{23}$ (MC1) and $1.5_{-0.4}^{+0.8}\times10^{23}$~cm$^{-2}$ (the Bridge). The ad hoc XRN model yields similar $N_{\rm H}$ 
values in good agreement with the results of \citet{Capelli2012}, who applied a similar ad hoc XRN model. Although one cannot simply compare the $N_{\rm H}$ values from the 
 MYTorus model and ad hoc XRN model (which has no geometry defined for $N_{\rm H}$), our simulation 
shows that spectral fitting with the ad hoc XRN model ``measures'' an $N_{\rm H}$ that deviates from the geometrical $N_{\rm H}$ of the MYTorus model by a factor 
of $\sim2$, at $N_{\rm H}\sim10^{23}$~cm$^{-2}$ (Appendix).   
In comparison with the measurements of \citet{Ponti2010} (MC1: $N_{\rm H} \sim4\times10^{22}$ and the Bridge: $9\times10^{22}$~cm$^{-2}$) 
based on the CS line intensity map \citep{Tsuboi1999, Ponti2010}, our $N_{\rm H}$ value for MC1 is higher by a factor of $\sim5$ while our result 
for the Bridge is close to their value. However, \citet{Ponti2010} and  \citet{Capelli2012} pointed out the difficulty with 
constraining $N_{\rm H}$ using molecular emission lines. For example, CS \citep{Amo2009} and 
H$^{13}$CO$^+$ \citep{Handa2006} emission line measurements deduced 
nearly two orders of magnitude different $N_{\rm H}$ values for another Sgr A cloud, G0.11$-$0.11. 
Still, the measured $N_{\rm H}$ yields 
the Thomson depth of $\tau_T \sim 0.1$, indicating that these clouds are optically thin. Therefore, it is more accurate to apply XRN models properly 
suited for optically thin cases rather than the 
slab geometry models which assume infinite column density.  

\paragraph{Power-law index}

The power-law photon indices (with 68\% c.l. errors) of the primary X-ray source 
are well constrained at $\Gamma = 2.11_{-0.14}^{+0.23}$ (MC1) and $1.81\pm0.10$  
(the Bridge) by the MYTorus model. The systematic errors associated with the angular dependence of the MYTorus model are 
smaller than the statistical errors for a measurement of $\Gamma$ at $N_{\rm H}=10^{23}$~cm$^{-2}$ (Appendix). Therefore, our results indicate that 
MC1 and the Bridge have consistent photon indices of the primary X-ray source at $\sim2$-$\sigma$ level.   
The measured photon indices are both softer than those of \citet{Ponti2010}: $\Gamma = 0.8^{+0.4}_{-0.5}$ for MC1 and 
$\Gamma = 1.0^{+1.0}_{-0.3}$ for the Bridge, based on 
\xmm-only spectral analysis over a narrower band between 4 and 8 keV. 
The ad hoc XRN model measures similar photon indices to the MYTorus model since the clouds are optically thin and 
the primary X-ray spectrum shape is not significantly 
perturbed by photo-absorption and Compton scattering. 
The {\verb=reflionx =} model yields softer photon indices ($\Gamma = 3.0$ and 2.3 for MC1 and the Bridge, respectively). Due to the infinite column 
density assumed in 
the {\verb=reflionx =} model, low energy photons are overly absorbed thus requiring a softer power-law 
photon index to fit the X-ray spectra as similarly observed in \nustar\ spectral analysis of the Arches cluster and Sgr B2 \citep{Krivonos2014, Zhang2015}.  

\paragraph{Fe K$\alpha$ fluorescent line} 

Both the flux and the EW of the Fe K$\alpha$ fluorescent line have often been used to track the time evolution of GC molecular clouds. Among our spectral models, 
only the ad hoc XRN model can provide Fe K$\alpha$ line parameters separately since the Fe fluorescent lines and scattered 
continuum are coupled in the self-consistent models, which are not parameterized to easily provide Fe K fluorescent line fluxes or EWs in XSPEC.  
Using the fit results with the ad hoc XRN model, we calculated the Fe K$\alpha$ line EW with respect to the power-law continuum, 
which is the only component instrinsic to the                     
clouds and thus can be compared to the predictions from the XRN and LECR models (\S \ref{sec:mc}). For comparison with other results, it is crucial to specify 
which X-ray continuum component is used to calculate the Fe~K$\alpha$ line EW.  
Our EW values, $0.93\pm0.12$ (MC1) and $1.38\pm0.14$ keV (the Bridge), are larger than 
those of \citet{Ponti2010}: 0.68 (MC1) and 0.75 keV (the Bridge), while the best-fit K$\alpha$ line flux 
normalizations are consistent between our work and \citet{Ponti2010}.                              
Note that the Fe K$\alpha$ line flux from the entire MC1 cloud has stayed nearly constant for years,               
 although \chandra\ found different time variations across the cloud \citep{Clavel2013}.           
The discrepancy in Fe K$\alpha$ line EWs is likely due to the fact that \xmm\ continuum spectra are heavily contaminated 
by diffuse thermal emission, thus the continuum level                           
is enhanced compared to the intrinsic non-thermal emission from the cloud. Indeed, using the Fe K$\alpha$ line flux normalization 
and the power-law continuum parameters from Table 3 in \citet{Ponti2010},                                                       
we obtain EW = 0.83 keV (MC1) and 1.16 keV (the Bridge) - they are similar to our measurements.                                   
Our Fe K$\alpha$ line EW for MC1 is also consistent with \citet{Capelli2012}, who measured EW $= 0.9\pm0.1$ keV.      

\paragraph{Inclination angle} 

The inclination angle is constrained to $\theta_{\rm obs} = 4.5^{\circ +15}_{-4.5}$ for the Bridge, while $\theta_{\rm obs} = 60^\circ_{-23}$ is less constrained for MC1 likely because the overall X-ray reflection 
spectrum is rather insensitive to $\theta_{\rm obs}$ at $N_{\rm H}\la10^{24}$~cm$^{-2}$ (Appendix) and 
the MC1 data have poorer photon statistics than the Bridge data. 
While it is tempting to suggest the Bridge with the best-fit $\theta_{\rm obs} \approx 0^\circ$ is located close to the 
projection plane of the primary source, we cannot uniquely infer line-of-sight location of the cloud based 
on the measured inclination angle and also we cannot estimate systematic errors on 
$\theta_{\rm obs}$ in the MYTorus model (Appendix). A precise measurement of the cloud line-of-sight location  
should be performed with an improved XRN model implementing more realistic geometry for the Sgr A clouds in the future. 

%%%%%%%%%%%%%%%%%%%%%%%%%%%%%%%%%%%%%%%%%%%%%%%

\subsection{Implications for the primary source illuminating the Sgr A clouds}
\label{sec:mc_imply}

Table \ref{tab:mc_summary} summarizes the observed and derived parameters from MC1 and the Bridge using 
the self-consistent MYTorus model as well as the known geometrical parameters. 
For comparison with the Sgr A clouds, we adopted the Sgr B2 results from the 2003-2004 \xmm\ 
and \integral\ observations \citep{Terrier2010} soon after both the hard X-ray
continuum and Fe K$\alpha$ line fluxes started decaying in 2000, therefore they can determine the primary X-ray source spectra 
more accurately than the \nustar\ observation in 2012. 
\citet{Terrier2010} fit 2-100 keV spectra of a $r=4.5$\amin\ circular region centered at the core of the cloud using a self-consistent XRN model, yielding the power-law index 
$\Gamma = 2.0\pm0.2$ and 1-100 keV luminosity $L_X = 1.1\times10^{39} $~erg\,$^{-1}$ 
of an illuminating X-ray source  (Table~\ref{tab:mc_summary}). \footnote{The projected distance of 100 pc between Sgr A* and Sgr B2 was assumed for $L_X$. 
The intrinsic column density $N_H=6.8\pm0.5$~cm$^{-2}$ was also 
measured using an adhoc XRN model {\tt wabs*(apec+gaus+gaus+wabs*pegpw)} \citep{Terrier2010}. 
Since the hydrogen column density distribution is highly non-uniform in the Sgr B2 region 
\citep{Etxaluze2013}, $N_{\rm H}$ measured by X-ray spectral fitting may vary 
in the range of $\sim10^{23}$-$10^{25}$~cm$^{-2}$ depending on a choice of the 
extraction region \citep{Zhang2015}.} 
Similar results ($\Gamma=1.8\pm0.2$ and $L_{X}\sim10^{39}$~erg\,s$^{-1}$) were obtained by \citet{Revnivtsev2004} who analyzed {\it ASCA}, {\it GRANAT} and {\it INTEGRAL} data from a larger region ($r=6.5$\amin) in Sgr B2 using a self-consistent XRN model.  

MC1, the Bridge and Sgr B2 cloud have the consistent primary power-law indice $\Gamma\sim2$.  
Since $\theta_{\rm obs}$ cannot be uniquely associated with the light-of-sight location of a cloud, the MYTorus model can give  
a lower bound of the illuminating source luminosity $L_X$ in which case the cloud and the primary X-ray source are located in the 
same projection plane. Using the best-fit primary X-ray fluxes at $\theta_{\rm obs} = 0^\circ$, we determined $L_X \ge 
1.1\times10^{38}$ (MC1) and $0.9\times10^{38}$~erg\,s$^{-1}$ (the Bridge). Here, we assumed that the illuminating X-ray source 
is located at Sgr A* and we rescaled the primary X-ray flux following the recipe in the Appendix.     
For reference, using the observed Fe K$\alpha$ line flux (which is also subject to Fe abundance uncertainty), 
\citet{Clavel2013} inferred higher X-ray luminosity of $5\times10^{38}$~erg\,s$^{-1}$ 
for MC1 since their estimate was based on the lower cloud column
density ($N_{\rm H} = 4\times10^{22}$~cm$^{-2}$) determined from the CS molecular line measurements 
\citep{Ponti2010} contrary to our direct measurements of $N_{\rm H}$ by fitting the broad-band X-ray spectra.

\begin{deluxetable*}{lccc}
\tablecaption{Comparison of molecular cloud and primary source parameters between MC1, the Bridge and Sgr B2 core using self-consistent XRN models}
\tablewidth{0pt}
\tablecolumns{4}
\tablehead { \colhead{Parameters}  & \colhead{MC1} & \colhead{Bridge} & \colhead{Sgr B2}}
\startdata
Cloud angular size $S$ [arcmin$^{2}$]  &  2.1      &  8.5  & 64  \\
Projected distance from Sgr A* [pc] &  $\sim12$  &   $\sim20$  & $\sim100$ \\
Equatorial column density $N_{\rm H} [10^{23}$\,cm$^{-2}$]    &   $2.3_{-0.6}^{+1.0}$        &   $1.5_{-0.3}^{+0.6}$  & $6.8\pm0.5$ \\
PL photon index ($\Gamma$)                     & $2.11_{-0.14}^{+0.23}$    & $1.81\pm0.10$  & $2.0\pm0.2$ \\
$L_X$ [erg\,s$^{-1}$] (2-20 keV) & $\ge 1.1\times10^{38}$ \tablenotemark{a}   & $\ge 0.9\times10^{38}$ \tablenotemark{a} & $1.0_{-0.5}^{+0.8}\times10^{39}$ \tablenotemark{b}
\enddata
\tablecomments{The errors are 68\% confidence level for MC1 and the Bridge, while the error confidence level for the Sgr B2 results was not specified in \citet{Terrier2010}.}
\tablenotetext{a}{The lower bound of $L_X$ was determined from the best-fit parameters at $\theta_{\rm obs} = 0^\circ$ where the cloud is 
located in the same projection plane of Sgr A*.}
\tablenotetext{b}{The errors are associated with the line-of-sight distance measurement of $130\pm60$~pc by \citet{Reid2009}. The X-ray luminosity quoted in the 1-100~keV band \citep{Terrier2010} was converted to the 2-20~keV band to match with our results for the Sgr A clouds.}
\label{tab:mc_summary}
\end{deluxetable*}

Given that MC1 and the Bridge require $L_X$ as low as $\sim 10^{38}$ erg\,s$^{-1}$, it is possible that an 
outbursting X-ray transient could have illuminated these clouds. 
Previously, \chandra\ found short temporal evolution of two Fe K$\alpha$ features in 
G0.11$-$0.11, and their Fe K$\alpha$ emission was attributed to reflection of an outburst of 
Sgr A* or an X-ray binary with a few year duration \citep{Muno2007}. Also, \citet{Capelli2012} proposed that one of the Sgr A  
clouds emitting an Fe K$\alpha$ line could be illuminated by the nearby X-ray transient 
XMMU~J174554.4$-$285456.  
Within the inner 10\amin\ of the GC, about a dozen X-ray transients have been detected, with their maximum 2-10 keV luminosities ranging from $\sim1\times10^{34}$ to $\sim7\times10^{38}$ erg\,s$^{-1}$ 
\citep{Muno2005, Degenaar2012}. Only 1A~1742$-$289 had its maximum outburst 
X-ray luminosity ($7\times10^{38}$ erg\,s$^{-1}$) exceed the inferred $L_X$ for MC1 and the Bridge. The only outburst from 1A~1742-289,  
observed in 1975, decayed rapidly over a few 
months \citep{Branduardi1976}, which is far shorter than the time variation of Fe~K$\alpha$ line 
flux  observed from MC1 and the Bridge \citep{Clavel2013}. 
Similarly, none of the other X-ray transients in the GC had persistent outbursts  
over a long enough period ($\ga 10$ years) to illuminate MC1 and the Bridge at the observed flux levels. In general, it is extremely rare for a bright 
outburst with $L_X \ga 10^{36}$ erg\,s$^{-1}$ to last for a few years \citep{Chen1997}. Therefore, we rule out the known 
X-ray transients in the GC as primary sources for MC1 and the Bridge. Alternatively, an undetected X-ray transient with X-ray outburst 
luminosity $L_X \ga 10^{37}$ erg\,s$^{-1}$ and $\sim10$ year burst duration, such as the black 
hole binary GRS~1915$+$105 \citep{Fender2004}, could be a primary source. 
However, \citet{Clavel2013} found this scenario implausible since it requires unrealistic cloud distribution around the GC to account for the 
observed Fe K$\alpha$ flux variation in the Sgr A clouds. 

As a result, Sgr A* is the most likely illuminating source for MC1 and the Bridge.
This is supported by the fact that the measured power-law indices ($\Gamma=1.8$-2.1) for Sgr A clouds as well as 
$\Gamma=2.0$ for Sgr B2 \citep{Terrier2010} are consistent with
those from the current Sgr A* flares \citep{Baganoff2001, Nowak2012, Degenaar2013, Neilsen2013, Barriere2014}, and
low-luminosity AGNs typically with $\Gamma\sim1.9$ \citep{Reeves2000}.
Several studies based on Fe K$\alpha$ line and X-ray continuum flux measurements
suggest that Sgr B2 was illuminated by a giant Sgr A* flare with
$L_X \sim 10^{39}$~erg\,s$^{-1}$ about 100 years ago \citep{Koyama1996, Murakami2001, Terrier2010, Ponti2010, Capelli2012}.
Based on the different temporal variations of Fe K$\alpha$ line emission from various GCMCs, 
\citet{Capelli2012} and \citet{Clavel2013} claimed 
that Sgr A* flaring activity in the past hundred years had multiple distinct periods with vastly
different flaring powers before declining to the current flaring state with $L_X \la 5\times10^{35}$ erg\,s$^{-1}$ \citep{Nowak2012}.
Given the error bars in $\Gamma$ and the lower bounds of $L_X$ in Table \ref{tab:mc_summary}, our analysis shows that 
the primary X-ray spectra are consistent between MC1, the Bridge and Sgr B2, thus it is still 
inconclusive whether these clouds were  
illuminated by different Sgr A* flares in the past or not. 
Continuing long-term monitoring of the 
Sgr A clouds by \nustar, with improved XRN models and photon statistics, will be able to constrain Sgr A* flaring activity (e.g., number of giant Sgr A* flares, their 
X-ray luminosities, and durations) over the last few hundred years more tightly.

%%%%%%%%%%%%%%%%%%%%%%%%%%%%%%%%%%%%%%%%%%%%%%%%%%%%%%%%%%%%%%%%%%%%%%%%%%

\section{The central 10 pc around Sgr A*} 
\label{sec:10pc}

The central 10 pc around Sgr A* is a highly crowded region with an extremely rich variety of radio, IR, soft X-ray, GeV and TeV sources. 
However, the \nustar\ 
view of the GC above $\sim20$ keV exhibits only two 
hard X-ray features --- a point-like feature coincident with \pwncen\ and the CHXE (Figure \ref{fig:central_20-40keV}). 
In the gamma-ray band, HESS detected a single source, \hesssrc\ at 
RA = $17^h45^m39^s.6$ and DEC = $-29^\circ00'22''$ (J2000) \citep{Acero2010}, that is spatially consistent with both Sgr A* and \pwncen. 
In this section, we investigate a connection between the hard X-ray sources and the HESS source as well as their emission mechanisms, based on our spectral 
analysis results using \nustar, \xmm\ and 
\chandra\ data. A multi-wavelength SED analysis is discussed in the subsequent section 
to elucidate the TeV emission mechanisms.

\subsection{Joint \nustar\ and \xmm\ spectral analysis}
\label{sec:10pc_spec}

We extracted \nustar\ spectra from a circular region of radius 40\asec\ around the \chandra\ position of \pwncen\ \citep{Wang2006}, 
at RA= $17^h45^m39^s.80$ and DEC = $-29^\circ00'19''.9$ (J2000). This radius was chosen to maximize the significance of the highest-energy spectral bins and 
minimize contamination from diffuse thermal emission given the \nustar\ PSF (e.g., HPD$\sim60$\asec).  
The \nustar\ background spectra were extracted from a region from the same detector chip but excluding Sgr A East, the 
``Plume'' \citep{Park2005}, and the molecular clouds \citep{Ponti2010}. We extended the energy band for spectral fitting 
to 50 keV, above which the internal detector background dominates. 

To better constrain the low-energy components of this spectrum, 
we used EPIC-PN data from the two \xmm\ observations (ObsID: 0694640301 and 0694641101) carried out in 2012 
for which Sgr A* was placed near the center of the FOV 
(See Table \ref{tab:obslog}). 
X-ray spectroscopy with \xmm\ EPIC instruments constrains the Galactic column 
density better than \nustar\ by fitting the spectrum below 3 keV, and it resolves 
 Fe lines at 6.7 keV (He-like) and 6.9 keV (H-like), thus measuring the plasma temperature 
accurately. 
We extracted an \xmm\ spectrum from the same region as used for the \nustar\ analysis. We generated response files and background spectra  
following the procedured described in \S \ref{sec:spectra}. We used the 2-8 keV band for fitting the \xmm\ spectra. 

\begin{figure*}
\centerline{
\includegraphics[height=0.65\linewidth, angle=270]{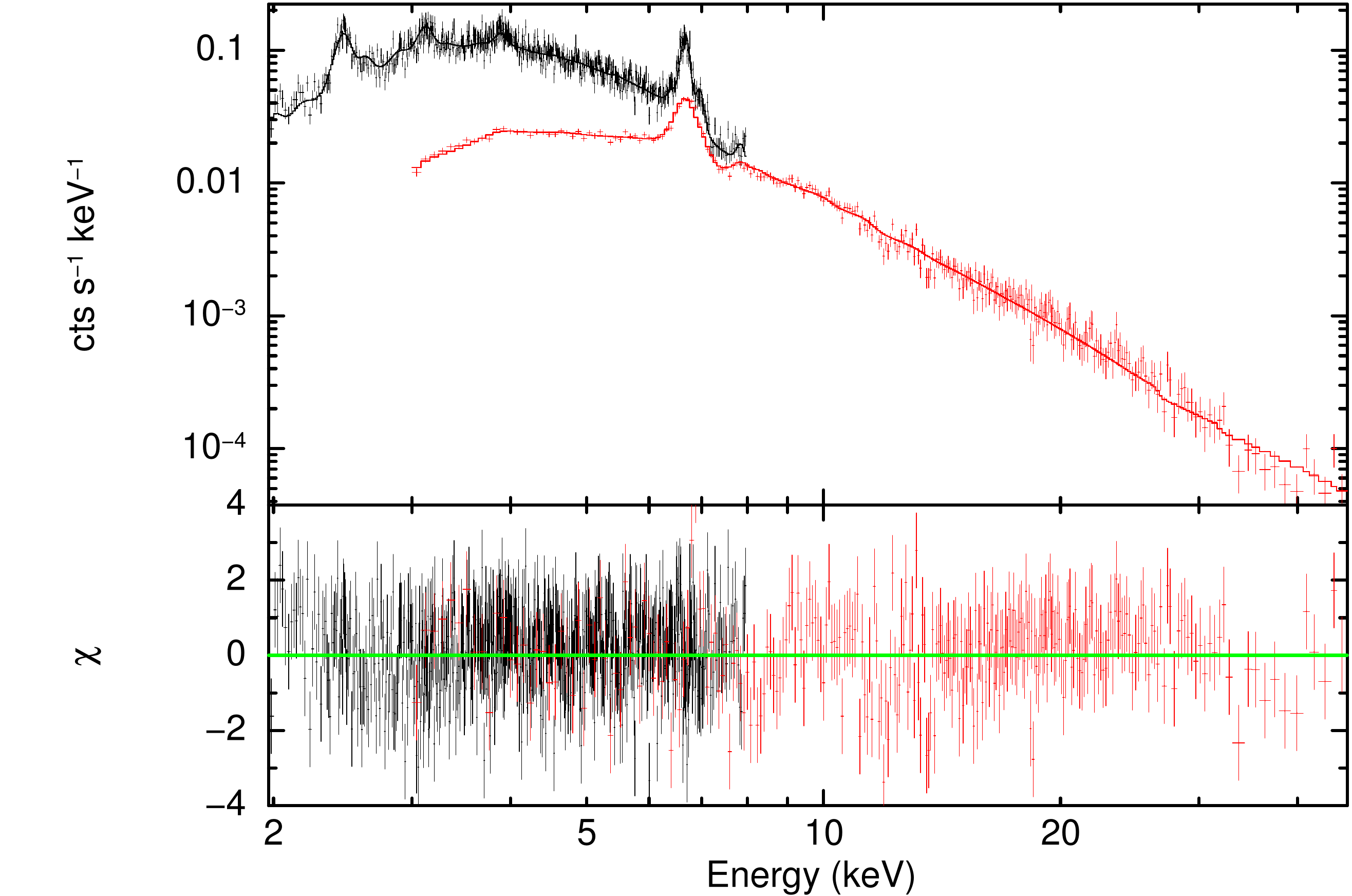}
}
\caption{2-50 keV \xmm\ EPIC-PN (black) and \nustar\ (red) spectra of the central $r=40$\asec\ circular region around \pwncen. 
The 2--8 keV spectrum is constructed from
\xmm\ EPIC-PN data, while the 3-50 keV spectrum is constructed from \nustar\ data. The model used is an absorbed two-temperature thermal
plasma plus a power-law and a Gaussian line at 6.4 keV.}
\label{fig:spec_40arcsec}
\end{figure*}

To fit the 2-50~keV spectrum of \xmm\ and \nustar, we used the model 
{\verb=const=} * {\verb=tbabs=} * {\verb=(apec + apec + pegpwrlw + gauss)=}. 
To account for the different overall normalization between \nustar\ and \xmm, a constant relative normalization factor was allowed 
to vary. The abundances of the two thermal components were fit freely 
within 
the range measured by the previous analysis \citep{Baganoff2003, Sakano2004}. 
The \nustar\ and \xmm\ spectra of the central 40\asec\ region 
are shown in Figure \ref{fig:spec_40arcsec} and the best-fit 
parameters are listed in Table \ref{tab:spectable}. 
The lower-temperature component, at $kT = 1.2$~keV, corresponds to a combination of the 
thermal emission from stellar winds in the central 10\asec\ region \citep{Baganoff2003} and the
low-temperature component of Sgr A East \citep{Sakano2004}.
The higher-temperature component, at $kT = 6.7$~keV, is consistent with the high-temperature 
component of Sgr A East in the region near Sgr A* \citep{Sakano2004}.
The power-law component has a best-fit photon index $\Gamma = 1.5\pm0.2$ and a 
20-40~keV flux of $F_X = (2.3\pm0.1)\times10^{-12}$~erg\,cm$^{-2}$\,s$^{-1}$. 

\begin{deluxetable*}{lc}
\tablecaption{\xmm\ and \nustar\ spectral analysis results of the central $r=40''$ region around \pwncen}
\tablewidth{0pt}
\tablecolumns{2}
\tablehead { \colhead{Parameters}  & \colhead{Best-fit values}}
\startdata
$N_{\rm H} [10^{22}$\,cm$^{-2}$]    & $17\pm3$      \\
$kT_{1}$ [keV] & $1.17\pm0.03$       \\
Abundance $Z_{1}$ &  $1.9_{-0.2}^{+0.3}$        \\
$norm_{1}$ & $(3.7\pm0.4)\times10^{-2}$   \\
$kT_{2}$ [keV] &  $6.7\pm0.4$       \\
Abundance $Z_{2}$ &   $1.6_{-0.2}^{+0.4}$        \\
$norm_{2}$ &  $4.5_{-0.2}^{+0.4}\times10^{-3}$  \\
Fe K$\alpha$ equivalent width [eV] &  $26\pm5$   \\
PL photon index ($\Gamma$)  &  $1.52_{-0.16}^{+0.19}$    \\
PL flux (20--40 keV)\tablenotemark{a} & $(2.3\pm0.1)\times10^{-12}$        \\
$\chi_\nu^2$ [dof]                  &  1.13 (846)
\enddata
\tablecomments {The energy band is 2-50 keV. The errors are 68\% confidence level. The overall flux normalization between the \nustar\ 
and \xmm\ spectra is $1.3\pm0.1$. }
\tablenotetext{a}{The flux unit is $10^{-12}$ \eflux.}

\label{tab:spectable}
\end{deluxetable*}

\subsection{Can \pwncen\ and the CHXE account for the 20-40 keV emission in the central 10 pc region?}

In order to assess the presence of a hard X-ray extension of \pwncen\ above 10 keV, we performed spatially-resolved spectral analysis using the \chandra\ data.  
\chandra\ measured a photon index of $\Gamma = 1.94_{-0.14}^{+0.17}$ for \pwncen, but with a spectral softening from the pulsar head to tail ranging from $\Gamma\sim$1.3-3.0 \citep{Wang2006}. The high-energy spectrum of this object is then most accurately modeled as a summation of spectra with different photon indices rather than a simple extrapolation of 
the best-fit photon index. We refit the spectrum extracted from the same region used by \citet{Wang2006} using a more extensive set of \chandra\ data.
This yields a photon index of $\Gamma = 1.8\pm0.1$, 
consistent with the previous measurements. We then divide the \citet{Wang2006} region into three sub-regions of equivalent areas, yielding $\Gamma_1 = 1.5\pm0.1$, 
$\Gamma_2 = 1.7\pm0.1$ and $\Gamma_3 = 2.6\pm0.1$, listed in order from the head toward the tail. 
Then, we constructed a composite spectral model for \pwncen\ from a set of \chandra\ fluxes and photon indices from the three segmented regions in the filament, and extrapolated it to the hard X-ray band.  The composite model gives a 20-40 keV flux of $0.97_{-0.09}^{+0.22}\times10^{-12}$ \eflux.  

\citet{Perez2015} showed the south-west and north-east region symmetrically located inside 
the CHXE ellipse have identical hard X-ray spectra, described equivalently well by either a power-law with $\Gamma \approx 1.6$ or thermal bremsstrahlung with   
$kT \approx 55$~keV. By repeating the same spectral analysis with the latest \nustar\ pipeline version, we determined the 20-40 keV flux of 
$0.56\times10^{-12}$ \eflux\ in the southwest region of the CHXE. 
Using the spatial model of the CHXE presented in \S\ref{sec:image_20-40keV}, 
we calculate the 20-40 keV flux of the CHXE to be $(1.08\pm0.14)\times10^{-12}$ \eflux\ in the central 40\asec\ region around Sgr A*, assuming 
that the CHXE has a power-law spectrum with $\Gamma=1.6$ throughout its entire region.

The sum of the estimated \pwncen\ and CHXE flux ($2.1_{-0.3}^{+0.5}\times10^{-12}$ \eflux) matches with the observed 20-40 keV flux in the central 40\asec\ region ($2.3\pm0.1\times10^{-12}$ \eflux) within the error bars. 
In addition, the spectral model consisting of a hard X-ray extension of \pwncen\ and the CHXE emission reproduces the 20-40 keV 
\nustar\ spectrum since the measured PL index of $\sim1.5$ is similar to those of \pwncen\ and the CHXE. 
This result confirms that 20-40 keV emission in the central 40\asec\ region is predominantly due to the 
CHXE and \pwncen. Our imaging analysis in \S \ref{sec:image_40-79keV} shows \pwncen\ is more prominent above 40 keV likely because 
\pwncen\ has a slightly harder X-ray spectrum and is more compact than the CHXE. 
While other X-ray sources may contribute to the hard X-ray emission in the central 40\asec\ region, our error analysis indicates their 
contribution should be less than $2\times10^{-13}$ \eflux\ in the 20-40 keV band. This upper limit will be useful in constraining models 
of X-ray and particle emission in the central parsec region around Sgr A*. 

\subsection{Connection with the TeV source \hesssrc}

Our imaging and spectral analysis shows that above 20 keV hard X-ray emission in the central 10 pc region is composed of 
the CHXE and \pwncen. 
No gamma-ray emission is expected from the CHXE since it is likely an unresolved 
population of massive magnetic CVs, while only a rare subclass of HMXBs are known 
to emit TeV gamma-rays \citep{Dubus2013}. Thus, it leaves only \pwncen\ as a hard X-ray 
counterpart candidate for \hesssrc. 

Previously, any models proposing that leptons are emitted from Sgr A* or its vicinity within a few pc 
have had  great difficulties with explaining 
the large extent of the diffuse hard X-ray source
\intsrc\ 
since the synchrotron cooling time of $\ga10$ TeV electrons emitting hard X-ray
photons is as short as $\sim10$ years \citep{Neronov2005, Hinton2007}.
Now that \nustar\ has revealed the compact hard X-ray emission above 40 keV is centered around \pwncen, this
``cooling time'' problem associated with \intsrc\ no longer exists.

In order to explore whether \pwncen\ alone can account for the GC TeV emission spectroscopically, we developed a one-zone PWN model following 
\citet{Zhang2008} and used it to fit the broad-band SED 
data in the central parsec region. Our model inputs are PWN age, magnetic field strength 
at present, a broken power-law spectrum for electron injection [$F(E_e)\sim E_e^{-p1}$ at $E_e \le E_{\rm break}$ and $E_e^{-p2}$ at $E_e > E_{\rm break}$] with the lower and upper energy limits, the 
radiation density in the IR, optical and UV bands, as well as pulsar spin-down parameters. 
For the SED data, we adopt the X-ray spectrum of \pwncen\ from our \chandra\ and \nustar\ 
spectral analysis and the TeV spectrum of \hesssrc\ from \citet{Aharonian2009}. 
The radio non-detection with an upper limit of $5\times10^{-17}$ \eflux\ from 6~cm observations 
\citep{Hinton2007} is also taken into account in our analysis. 
Hereafter, we do not consider the {\it Fermi} GeV source \gevsrc\ at
RA= $17^h45^m41^s.6$ and DEC = $-28^\circ58'43''$ (J2000) since it lies outside the error circle  
of \hesssrc\ \citep{Nolan2012, Yusef2013}.
\gevsrc\ may well be associated with the aformentioned X-ray filament \fila\ \citep{Nynka2014NF}. 

\begin{figure*}
\centerline{
\includegraphics[height=0.45\linewidth]{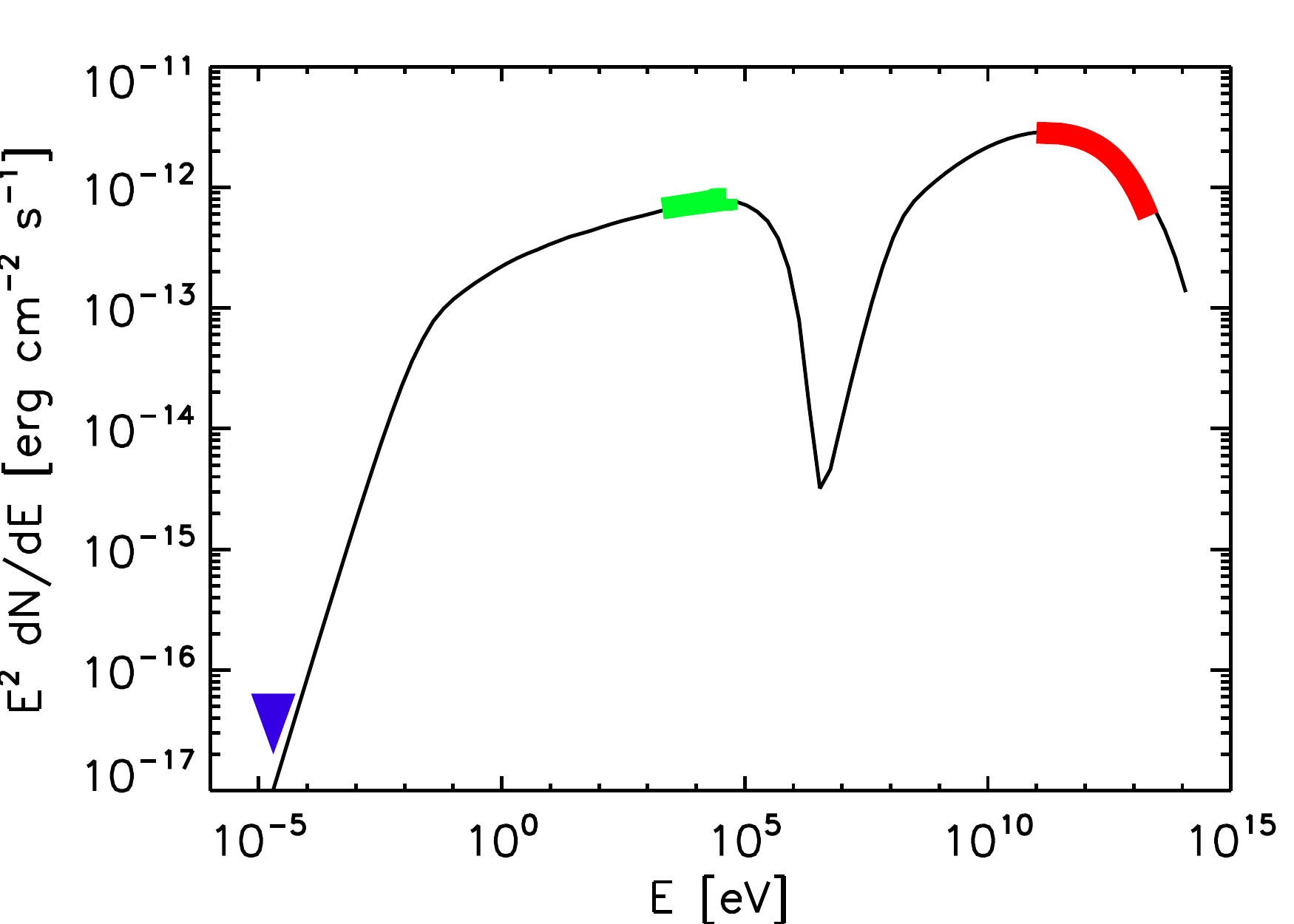}
}
\caption{A one-zone PWN model fit to the broad-band SED data including the 6cm radio flux upper limit (blue arrow), \pwncen\ X-ray spectrum (green) and 0.1-10 TeV HESS spectrum of \hesssrc\ (red).\\\\}

\label{fig:sed}
\end{figure*}

Figure \ref{fig:sed} shows the best-fit SED model along with the radio, X-ray and TeV data. 
Following \citet{Hinton2007}, we assumed \pwncen\ is a $10^4$ year-old PWN with a spin-down power of 
$5\times10^{35}$~erg\,s$^{-1}$ and its magnetic field strength is 300 $\mu$G at present. We find that these parameters fit the SED 
data reasonably well. In the central 
parsec region, the radiation density can be as high as $\sim5\times10^3$~eV\,cm$^{-3}$ 
\citep{Davidson1992}. We adopted the FIR, optical and UV radiation density from \citet{Hinton2007} who quoted the original work of 
\citet{Davidson1992}. 
We found that 
a broken power-law electron spectrum with $p_1 = 1.8$, $p_2 = 2.0$ and $E_{\rm break} = 50$~TeV  
represents the 
shape of the X-ray and TeV spectrum well. These electron injection parameters are typical of young PWNe \citep{Zhang2008}. 
The maximum electron energy was set to 200 TeV to account for the energy 
cut-off at 4~TeV in the gamma-ray spectrum \citep{Aharonian2009}. 
A low energy 
cut-off at $E_e \ga 0.5$~TeV was required so that the model is consistent with the non-detection 
of \pwncen\ in the radio band. Alternatively, \citet{Hinton2007} proposed fast electron diffusion to account for the lack of a radio 
counterpart. Both the upper and lower limit in the electron injection spectrum are similar to those of \citet{Hinton2007} who analyzed the same SED data except using the X-ray spectrum of the {\it INTEGRAL} source \intsrc. 
In conclusion, we find that \pwncen\ alone, likely a $\sim10^{4}$ year-old PWN with nominal electron injection parameters, can 
account for the broad-band SED of the central parsec region including \hesssrc. 
It is noteworthy that the other distinct TeV gamma-ray source within a degree from the GC is 
associated with another 
young PWN in SNR G0.9+0.1 \citep{Aharonian2005}.

As an alternative leptonic scenario for the GC TeV emission, stellar wind shocks from massive stars in the central 
stellar cluster can efficiently accelerate electrons \citep{Quataert2005}, or Sgr A* itself can 
eject high-energy 
electrons \citep{Kusunose2012}. Due to the fast electron cooling time either by synchrotron radiation or inverse Compton scattering in the GC, hard X-ray emission should be localized around an electron acceleration site. While these emission mechanisms may be in action, 
our \nustar\ analysis indicates that their contribution is insignificant since the hard X-ray emission above 40 keV is spatially and spectroscopically 
consistent with \pwncen. Therefore, the PWN candidate \pwncen\ seems to be the most plausible 
hard X-ray counterpart of the TeV source \hesssrc.

The hadronic scenario may be plausible as well based on the fact that diffuse TeV emission is 
spatially well correlated with molecular clouds in the GC \citep{Aharonian2006c}. One potential source of ejecting high energy protons is Sgr A* 
\citep{Aharonian2006c, Ballantyne2007, Dogiel2009a, Ballantyne2011}. Among a number of hadronic models proposed for the gamma-ray emission in the GC, there has been no specific 
prediction for X-ray spectra in the central parsec region. Rather, they have been focused on larger degree-size regions over which 
protons can propagate without losing kinetic energies significantly 
\citep{Dogiel2009b}. Similar to the LECR model, either non-thermal bremsstrahlang or synchrotron is expected to be the primary X-ray emission mechanism 
via secondary electrons produced by hadronic interactions between 
high energy protons and molecular clouds \citep{Dogiel2009a, Dogiel2009b, Gabici2009}. 
Such X-ray emission must be most prominent at the location of molecular clouds or high-density gas in the vicinity of Sgr A*, potentially with Fe K$\alpha$ line emission at 6.4 keV \citep{Dogiel2009c}. 
In addition, X-ray emission in the hadronic scenario should have a larger extent than in 
the leptonic scenario since protons have significantly longer cooling times than electrons in the 
GC where both magnetic field and 
radiation density are high. To the contrary, hard X-ray emission above 40 keV is highly concentrated around \pwncen, and \chandra\ did not detect strong Fe K$\alpha$ emission within 10 pc around Sgr A*, including the circumnuclear disk \citep{Baganoff2003}. Therefore, the hadronic scenario is unlikely to be a 
major contributor for the hard X-ray and TeV emission in the central parsec region.

%%%%%%%%%%%%%%%%%%%%%%%%%%%%%%%%%%%%%%%%%%%%%%%%%%%%%%%%%%%%%%%%%%%%%%%%%%%%%%%%%%%%%%%%%%%%%%%%%

\section{Summary}
\label{sec:summary}

The initial 450 ksec phase of the \nustar\ GC program, with its high-resolution 
imaging and spectroscopic capability from 3 to 79 keV, has made unique contributions to understanding high-energy phenomena in the crowded GC region as listed below. 

\begin{itemize}

\item[1.]  {\it NuSTAR} resolved the \integral\ source \intsrc\ into 
non-thermal X-ray filaments, molecular clouds, point sources and the 
previously-unknown central hard X-ray emission (CHXE) above 20 keV.   

\item[2.] The X-ray emission from Sgr A East is thermal with $kT\sim$~1-6 keV with no evidence of non-thermal emission, and is consistent with the previous soft X-ray observations \citep{Maeda2002, Sakano2004}. 

\item[3.] In the 20-40 keV band, \nustar\ discovered hard X-ray emission (CHXE) 
centered on Sgr A*. The CHXE is elongated along the Galactic Plane with an elliptical extent of $\sim8$~pc (Galactic longitude)  
and $\sim4$~pc (Galactic latitude) \citep{Perez2015}. The most likely explanation for the CHXE is an unresolved 
population of massive magnetic CVs (largely intermediate polars) with $M_{WD} \sim 0.9 M_{\odot}$ \citep{Hailey2015}.  

\item[4.]  {\it NuSTAR} detected four non-thermal X-ray filaments (\aeknot, \fila, \filc\ and \pwncen) 
above 10 keV. The origin of non-thermal X-ray filaments may 
be heterogeneous and associated with different emission mechanisms such as magnetic flux tubes trapping TeV 
electrons \citep{Zhang2014}, SNR-cloud interaction and PWNe \citep{Nynka2014NF}. 

\item[5.] For the first time, \nustar\ resolved hard X-ray emission from the Sgr A clouds above 10~keV 
and unambiguously detected hard X-ray continuum emission from MC1 and the 
Bridge. Hard X-ray continuum emission is spatially correlated with Fe K$\alpha$ line 
emission (EW $\sim1$~keV) from these clouds. 
We fit the Monte-Carlo based MYTorus model to the \xmm\ + \nustar\ spectra of MC1 and the Bridge, and determined their intrinsic column densities ($N_{\rm H}\sim10^{23}$~cm$^{-2}$) and the primary X-ray spectra with $\Gamma\sim2$ self-consistently. 
We set a firm lower bound for X-ray luminosity of Sgr A* flares illuminating  MC1 and the Bridge to $L_X \ga 10^{38}$~erg\,s$^{-1}$. 
It is still unclear whether the Sgr A and Sgr B clouds were illuminated by different Sgr A* flares in the past. 

\item[6.] A point-like hard X-ray source observed in the 20--60 keV band is identified as the PWN 
candidate \pwncen, $9''$ away from Sgr A*. The hard X-ray emission in the central 10 pc 
region is predominantly composed of two sources, \pwncen\ and the CHXE. 

\item[7.] In the central 10 pc around Sgr A*, \pwncen\ is the primary hard 
X-ray feature that is expected to emit TeV gamma-rays via inverse Compton scattering of 
IR, optical and UV photons. Our 
SED study suggests that \pwncen\ is the hard X-ray counterpart of the persistent TeV source \hesssrc, thus strongly favoring 
the leptonic origin of 
the TeV emission at the very center of our galaxy. 

\end{itemize}

Follow-up deep observations by \nustar\ will lead to further spectral identification of X-ray filaments and point sources. Monitoring time variation of 
the GC molecular clouds jointly by \nustar\ and \xmm\ will elucidate their X-ray 
emission mechanism and probe the primary illuminating source or Sgr A* flaring activity in the past. Starting from April 2015, the 
\nustar\ Legacy program will follow up some of the hard X-ray sources discussed in this paper with deeper exposures.

\acknowledgements

This work was supported under NASA Contract No. NNG08FD60C, and made use of data from the \nustar\ mission, a project led by the California Institute of 
Technology, managed by the Jet Propulsion Laboratory, and funded by the National Aeronautics and Space Administration. We thank the \nustar\ Operations, 
Software and Calibration teams for support with the execution and analysis of these observations. This research has made use of the \nustar\ Data Analysis 
Software (NuSTARDAS) jointly developed by the ASI Science Data Center (ASDC, Italy) and the California Institute of Technology (USA). 
R.~Krivonos acknowledges support from Russian Science Foundation through grant 14-22-00271.
G.~Ponti acknowledges support via an EU Marie Curie IntraEuropean fellowship under contract no. FP-PEOPLE-2012-IEF-331095, the Bundesministerium  f{\"u}r Wirtschaft und Technologie/DeutschesZentrum f{\"u}̈r Luf-und R fahrt (BMWI/DLR, FKZ 50 OR 1408) and the Max Planck Society.
F.E.~Bauer acknowledges support from CONICYT-Chile (Basal-CATA PFB-06/2007, FONDECYT 1141218, ``EMBIGGEN'' Anillo ACT1101), and the Ministry of Economy, Development, and Tourism's Millennium Science
Initiative through grant IC120009, awarded to The Millennium Institute of Astrophysics, MAS. 
S.~Zhang is supported by NASA Headquarters under the NASA Earth and Space Science Fellowship Program - Grant ``NNX13AM31''. 
D.~Barret acknowledges support from the French Space Agency (CNES). We thank Tahir Yaqoob for useful discussions on the MYTorus model. 

\appendix

\section{\nustar\ background in the Galactic Center observation}
\label{sec:bkg}

{\it NuSTAR} imaging and spectral analysis of GC sources is challenging due to the high background level and to its complex multiple components.
The \nustar\ background is generally characterized by four different components as outlined below. More detailed discussion on the CXB and internal background 
can be found in \citet{Wik2014}. 

\begin{itemize}
\item[1.] Focused diffuse background (2-bounce background photons): diffuse background photons in the FOV are reflected twice by the optics and focused 
on the detector plane.
\item[2.] Ghost-rays (1-bounce background photons): Photons from outside the FOV are reflected once by the optics and
reach the detector plane. Ghost-ray photons from a bright persistent source or X-ray transient can be
significant, with a visible pattern
in the \nustar\ image.
Although some observations of the GC and Norma field have been severely affected by ghost-ray background \citep{Bodaghee2014}, it is not important in the 
\nustar\ mini-survey and Sgr A* observations.
\item[3.] Stray-light or aperture background (0-bounce background photons):
Photons from any X-ray source at $\sim1$-$5^\circ$ away from the telescope
pointing vector, that are not blocked by the aperture stop, illuminate the detector plane. Stray-light background (SLB) is not uniform over the detector plane, and it is not
identical between the two focal plane modules. The location of SLB is sensitive to the position angle (PA) of the telescope.
\item[4.] Internal detector background (cosmic-ray induced background photons): atmospheric albedo and activation components with several emission lines
in the 20-40 keV band. Above $\sim40$ keV, this component usually becomes more important than the other background
components.
\end{itemize}

\begin{figure}[t]
\centerline{
\includegraphics[height=0.5\linewidth]{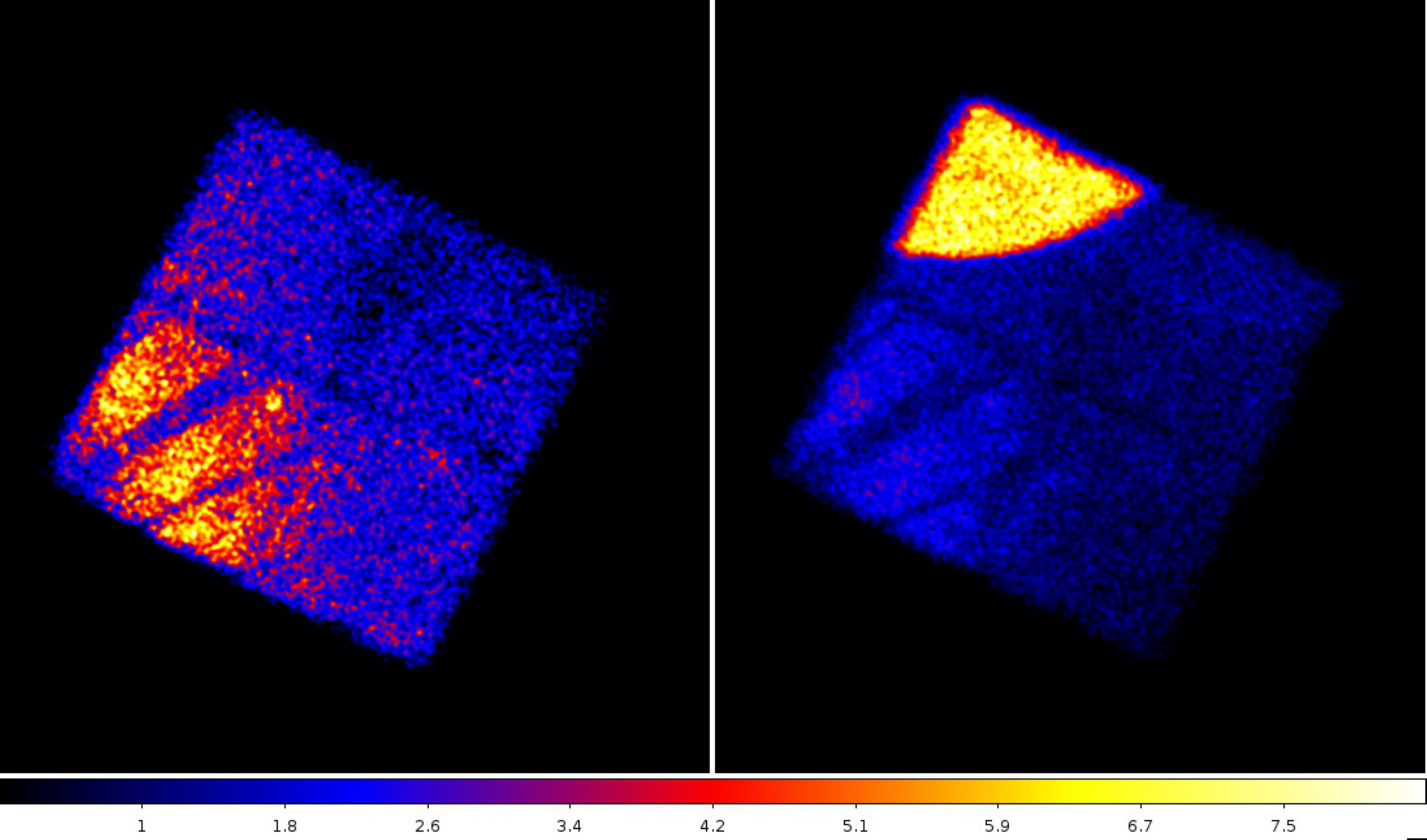}
}
\caption{\nustar\ FPMA (left) and FPMB (right) image in the sky coordinates from \nustar\ observation 40032010001 pointing at RA = 
266.0754$^\circ$ and DEC = -29.2988$^\circ$ (J2000), with PA = $332^\circ$. }
\label{fig:bkg_image}
\end{figure}

Figure \ref{fig:bkg_image} shows example \nustar\ images (FPMA and FPMB images from ObsID: 40032010001) exhibiting both ghost-ray background and SLB. The 
radiating pattern in the lower-left corner of both the FPMA and FPMB image is due to the ghost-ray background photons from the bright persistent 
LMXB 1A1742$-$294 at RA= $17^h46^m05^s.201$ and DEC = $-29^\circ30'53.3''$ (J2000) \citep{Wijnands2006}. On the other hand, the bright region in the upper-left corner of the FPMA 
image is due to the SLB from the bright X-ray source GX~3+1 \citep{Seifina2012}. 

\subsection{Stray-light background removal from bright X-ray point sources}

In some observations, SLB from a point source brighter than $\sim 10^{-11}$ \eflux\ can be  easily 
visible in raw \nustar\ images (Figure \ref{fig:bkg_image}). 
For a given  position angle (PA) and a list of nearby bright point sources, we
can exactly predict the SLB pattern on the detector plane. 
Using a  code developed specifically for calculating the SLB pattern from a  
point source with known position, we can generate a bad-pixel map for each detector module and observation, and then filter out events and exclude exposure map 
in regions of high SLB. This is one of the \nustar\ data filtering processes discussed in \S \ref{sec:obs}, before proceeding to further 
imaging and spectral analysis. 

\subsection{Background spectrum subtraction}

\begin{figure}[t]
\centerline{
\includegraphics[height=0.4\linewidth]{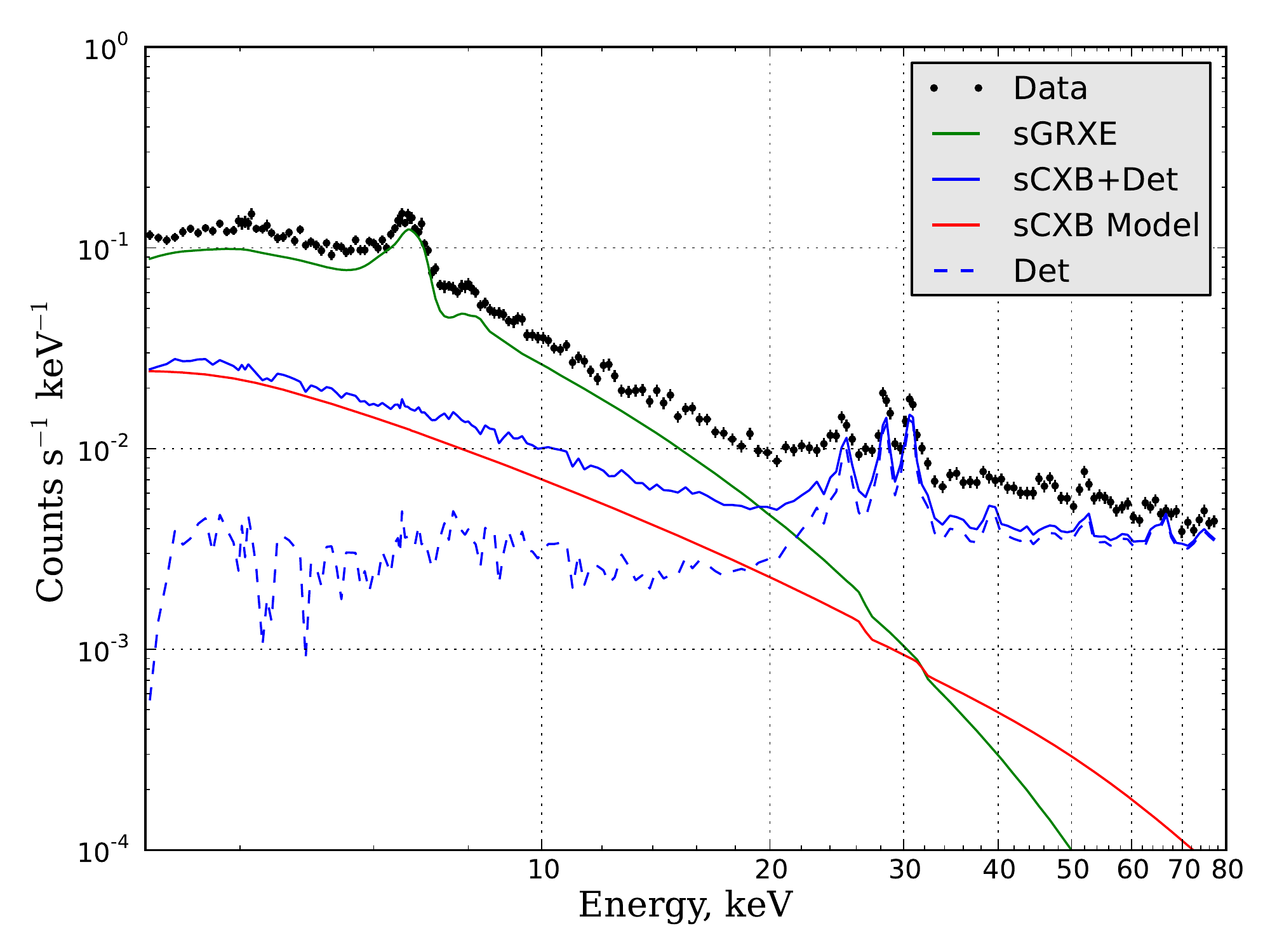} 
}
\caption{\nustar\ background spectrum taken from a region (ObsID: 40032001002) where there is no known X-ray source,
overlaid with the CXB components and internal background.}
\label{fig:bkg_spec}
\end{figure}

For all GC sources discussed in this paper, the primary background component below $\sim40$ keV is focused diffuse
background and SLB, while the instrumental background dominates above $\sim40$
keV. For example, Figure \ref{fig:bkg_spec} shows a \nustar\ background spectrum extracted from a region free from point sources 
and molecular clouds in one of the 
\nustar\ mini-survey observations (ObsID 40032001002).
As shown in Figure \ref{fig:bkg_spec}, SLB from the Galactic Ridge X-ray 
emission (sGRXE) is usually dominant over that of cosmic X-ray background (sCXB). This background spectrum is typical to \nustar\ 
GC observations. Focused CXB and GRXE components are not shown in the figure since their
fluxes are lower than their stray-light components by an order of magnitude. Both the CXB and internal background are modeled 
by the {\it nulyses} or {\it nuskybkg} software package using the high-latitude \nustar\ data \citep{Wik2014}.  
The sGRXE spectrum is well represented by an absorbed thermal spectrum (APEC model in XSPEC)  
with $kT\sim12$ keV and an Fe K$\alpha$ emission line at 6.4 keV. 

The background count rate per detector area [cm$^{-2}$] extracted from relatively source-free regions in the \nustar\ mini-survey observations varies between $2.1\times10^{-2}$ and
$6.3\times10^{-2}$~cts\,sec$^{-1}$\,cm$^{-2}$ in the 3--20 keV band. 
After subtracting the model count rates from the sCXB component and internal detector background
\citep{Wik2014}, the 3--20 keV sGRXE count rate ranges from $1.4\times10^{-2}$ (ObsID: 40010003001) to $5.5\times10^{-2}$~cts\,sec$^{-1}$\,cm$^{-2}$ 
(ObsID: 40010001002) with the 
mean count rate of $4\times10^{-2}$~cts\,sec$^{-1}$\,cm$^{-2}$. The sGRXE component accounts for $\sim70-90$\% of the total background count rate in 
the 3-20 keV band. Both sGRXE count rate and its fraction to the overall background varies between different observations, position angles and detector modules. 

The relative significance between the focused diffuse background and the SLB varies between different regions in the \nustar\ image.
When focused diffuse emission is dominant, one can extract a background
spectrum from a region away from the source on the same detector chip over which the instrumental background is
uniform \citep{Harrison2013}. When SLB is dominant, we extract a background spectrum using the same
{\it detector} region used for extracting a source spectrum from another nearby
observation with a similar position angle (off-source background subtraction) \citep{Krivonos2014}.
For instance, the Sgr A* complex has significant GC thermal emission so the former conventional background subtraction should be applied.
On the other hand, the off-source background subtraction is more appropriate for the Arches cluster where the
SLB is more significant than the focused GC diffuse emission \citep{Krivonos2014}.

While the sCXB component can be accurately modeled using \nustar\ extragalactic deep survey data \citep{Wik2014}, there is no 
reliable background model for the sGRXE component due to the complex and unknown spatial distribution of the GRXE. 
Instead, as guidance, we used the sGRXE count rates determined from the \nustar\ mini-survey data to estimate whether the sGRXE is dominant over other background components for a given source region.
For some sources, the situation is ``mixed'' where both the focused diffuse background and SLB 
have similar count rates. In this case, we applied both background subtraction methods to bound
the problem. \\

\subsection{Comparison of X-ray reflection models applied to GC molecular clouds} 

In general, the X-ray reflection spectrum from a GC cloud is composed of scattered continuum, Fe fluorescent
lines and photo-absorption edges.
A popular XRN model for GCMCs is an absorbed power-law continuum with two Gaussian emission lines for the
Fe K$\alpha$ and K$\beta$ lines respectively: {\verb= tbabs*(powerlaw + gauss + gauss) =} where {\verb=tbabs=} model, sometimes replaced by {\verb=wabs=}, represents intrinsic absorption in the cloud.
This model assumes that Fe fluorescent photons come from the center of a cloud, while another form
{\verb= tbabs*powerlaw + gauss + gauss =} assumes that Fe fluorescent photons come from the surface and are
therefore not subject to photo-absorption in the cloud. In practice, there is almost no difference between these two cases
unless the cloud column density is extremely high.
However, this ad hoc XRN model lacks self-consistency since photo-absorption, scattered continuum  and fluorescent lines are
decoupled and fit separately. The scattered continuum is represented by a single
power-law model assuming that the primary X-ray spectrum shape is unperturbed by Compton scattering.
This assumtion is valid only for low energy photons where Compton scattering is negligible, and when a cloud is optically
thin ($N_{\rm H} \ll 10^{24}$~cm$^{-2}$).

In the other extreme case, slab geometry models such as {\verb= pexrav, pexmon=} and {\verb= reflionx =} calculate X-ray
reflection
spectra self-consistently from a slab with infinite optical depth \citep{Magdziarz1995, Nandra2007, Ross2005}.
\citet{Ponti2010} applied the {\verb=pexrav=} model to Sgr A molecular clouds, while {\verb=reflionx =} has been
used to fit X-ray spectra of the Arches cluster \citep{Krivonos2014} and the Sgr B2 cloud \citep{Zhang2015}.
However, the major drawback of these slab geometry models is that they are applicable only for Compton-thick clouds
($\tau_T \gg 1 $
or $N_{\rm H}\gg 10^{24}$~cm$^{-2}$), and they do not allow for a measurement of the column density.
Both our spectral analysis and
independent $N_{\rm H}$ measurements suggest that the Sgr A clouds are optically thin with $N_{\rm H} \sim 10^{23}$~cm$^{-2}$.

None of the above models can determine the intrinsic column density of a cloud and the primary X-ray 
spectrum self-consistently. 
As demonstrated by recent X-ray studies of Compton-thick AGN, Monte-Carlo simulation is the only viable approach to build a 
self-consistent X-ray reflection model \citep{Murphy2009}. 
In the past, \citet{Sunyaev1998}, \citet{Revnivtsev2004} and \citet{Odaka2011} studied X-ray morphology and spectra of GCMCs 
using Monte-Carlo based X-ray reflection models. These models explored some limited parameter space primarily for Sgr B2, 
but they are not implemented in XSPEC for spectral fitting. 
At present, the MYTorus model is the only X-ray reflection model that is available in XSPEC that can measure the intrinsic 
column density 
self-consistently for GCMCs \citep{Murphy2009, Yaqoob2012}. The MYTorus model employs Monte-Carlo simulation of reprocessing 
X-ray photons 
from a toroidal reprocessor, and it enables ``real-time'' spectral fitting in XSPEC using  
tabulated Green's function data. 
The other X-ray reflection models available in XSPEC either assume infinite column density or do not separate a reflected 
component  \citep{Brightman2011}. 
Although the MYTorus model was originally developed to study Compton-thick AGNs with a toroidal X-ray reflector, 
we find that it is applicable to X-ray spectral analysis of GCMC data with some restrictions as shown in the next section.  
Table~\ref{tab:mc_models} compares the three XRN models used in our
analysis, their assumptions, limitations and valid parameter ranges.

\begin{deluxetable*}{lccc}
\tablecaption{Comparison of three X-ray reflection spectral models applied to GC molecular clouds}
\tablewidth{0pt}
\tablecolumns{4}
\tablehead { \colhead{Model}  & \colhead{Ad hoc XRN} & \colhead{Slab geometry} & \colhead{MYTorus}}
\startdata
Geometry  &   undefined  &  semi-infinite slab   & torus  \\
XSPEC model & {\verb= tbabs*(powerlaw + gauss + gauss) =}   & {\verb= pexrav, pexmon, reflionx =}   &  {\verb= MYTS + MYTL =} \\
Column density &  absorption only    &  infinite column density  &  self-consistent measurement \\
Primary X-ray PL index &  same as the best-fit PL index   &  model output   &  model output    \\
Primary X-ray source flux  &   adjustment by Thomson depth   &  model output scaled by solid angle  & model output scaled by solid angle \\
Fe abundance &     unspecified        &   variable  &   fixed to solar \\
Valid parameter range & no self-consistency & only for optically thick cloud  & $\theta_{\rm obs} \la 60^\circ$ and $N_{\rm H} \la 10^{24}$~cm$^{-2}$
\enddata
%\tablecomments{ }                                                                                                               
%\tablenotetext{a}{ }                                                                                                            
\label{tab:mc_models}
\end{deluxetable*}

\subsection{Applicability of MYTorus model to X-ray spectroscopy of GC molecular clouds}

\begin{figure}[t]
\centerline{
\includegraphics[height=0.5\linewidth]{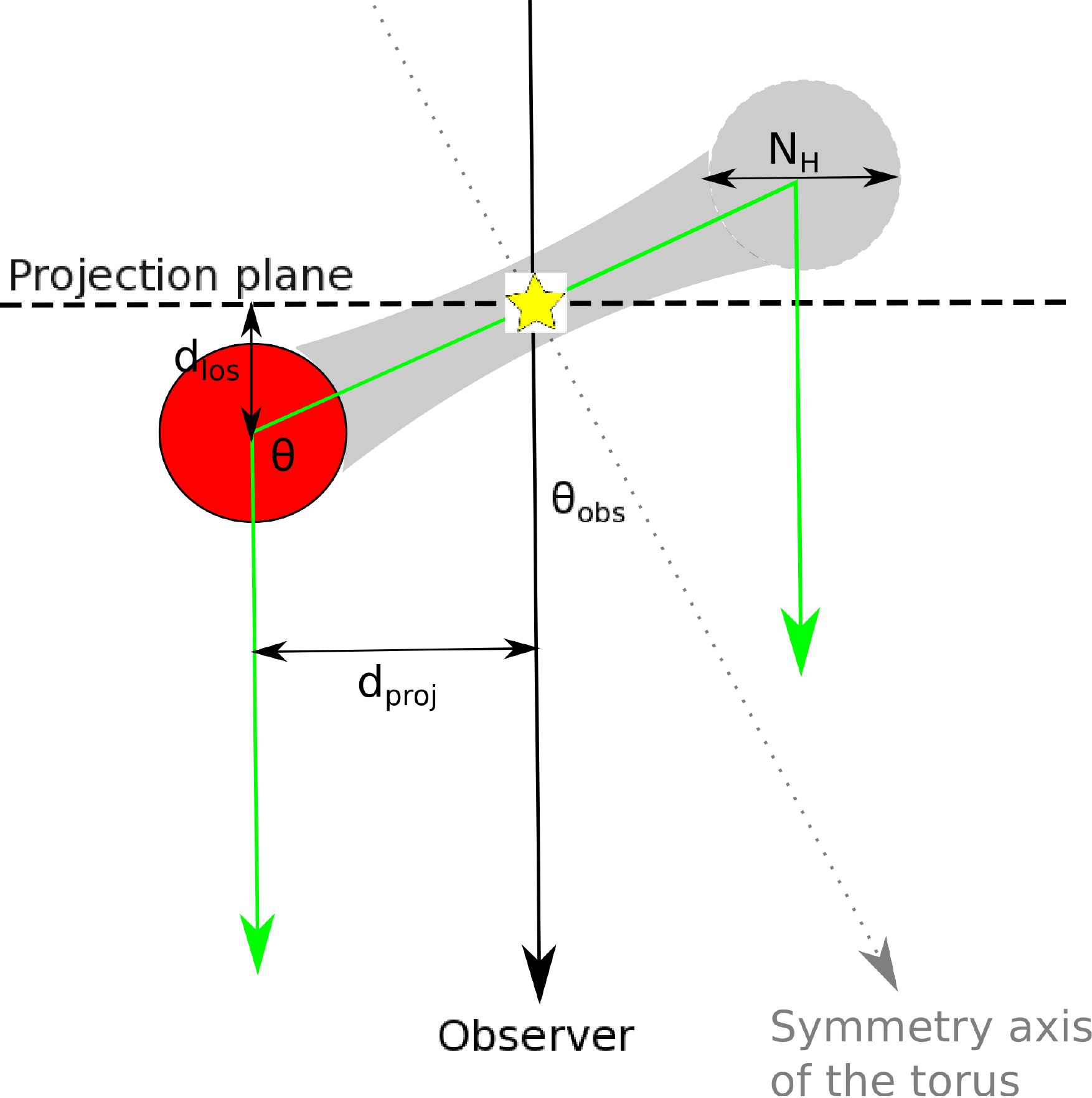} 
}
\caption{The geometry of a cloud (bright red circle) along with the observer's LOS  
  (solid vertical line) and the projection plane (dashed horizontal line) of the X-ray source (yellow star). 
$d_{\rm proj}$ is the projected distance between the cloud and the source seen by the observer at the bottom, 
while $d_{\rm los}$ is the LOS distance of the cloud measured from the projection plane. 
A virtual torus for the MYTorus model is indicated by grey area.   
The inclination angle $\theta_{\rm obs}$ 
is between the observer's LOS and the symmetry axis of the torus (dotted grey line). The equatorial column density $N_{\rm H}$ is defined over the minor diameter of the torus. Photons from the X-ray source are scattered off the cloud into the observer's LOS at an angle $\theta$ (X-ray photon's paths are indicated by green lines). $\theta_{\rm obs}=0^\circ$ (face-on view) corresponds to 
$\theta = 90^\circ$ ($d_{\rm los} = 0$) where the cloud is in the projection plane of  the X-ray source. \\} 
\label{fig:mc_geometry}
\end{figure}

Since the MYTorus model was developed primarily for studying X-ray reflection spectra of Compton-thick AGN, it assumes a 
torus with 
completely neutral material and uniform density \citep{Murphy2009}. 
The MYTorus model covers a range of the equatorial column density $N_{\rm H} = 10^{22}$-$10^{25}$~cm$^{-2}$ and power-law 
photon indices 
$\Gamma = 1.4$-2.6. 
Note that the MYTorus model defines the inclination angle between an observer's line of sight (LOS) and the symmetry axis 
of the torus ($\theta_{\rm obs}$), while most publications on GCMCs use a 
scattering angle ($\theta$) of illuminating photons off the cloud to the observer \citep{Capelli2012}. 
See Figure \ref{fig:mc_geometry} for the geometry 
of a cloud and the torus as well as the definition of the incident and scattering angle. 
 A face-on viewing case for the MYTorus model ($\theta_{\rm obs} = 0^\circ$) corresponds to the 
scattering angle $\theta=90^\circ$ when a cloud is in the same projected plane 
as the primary X-ray source. 
For each of the three key assumptions associated with the MYTorus model, below we investigate the valid parameter 
space where the model is applicable to analyze X-ray reflection spectra of molecular clouds in general. 

\begin{figure}[t]
\centerline{
\includegraphics[height=0.4\linewidth]{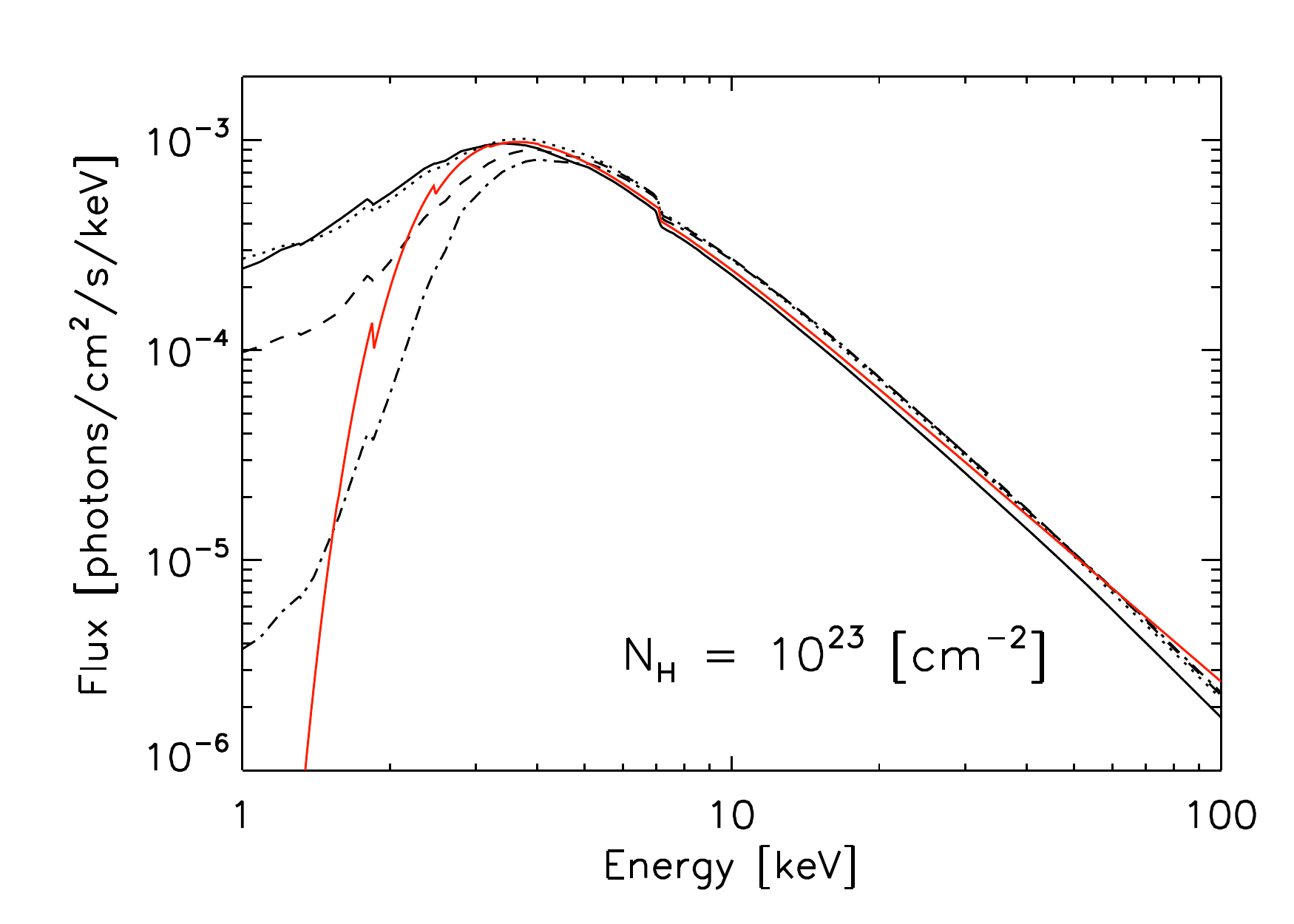}
\includegraphics[height=0.4\linewidth]{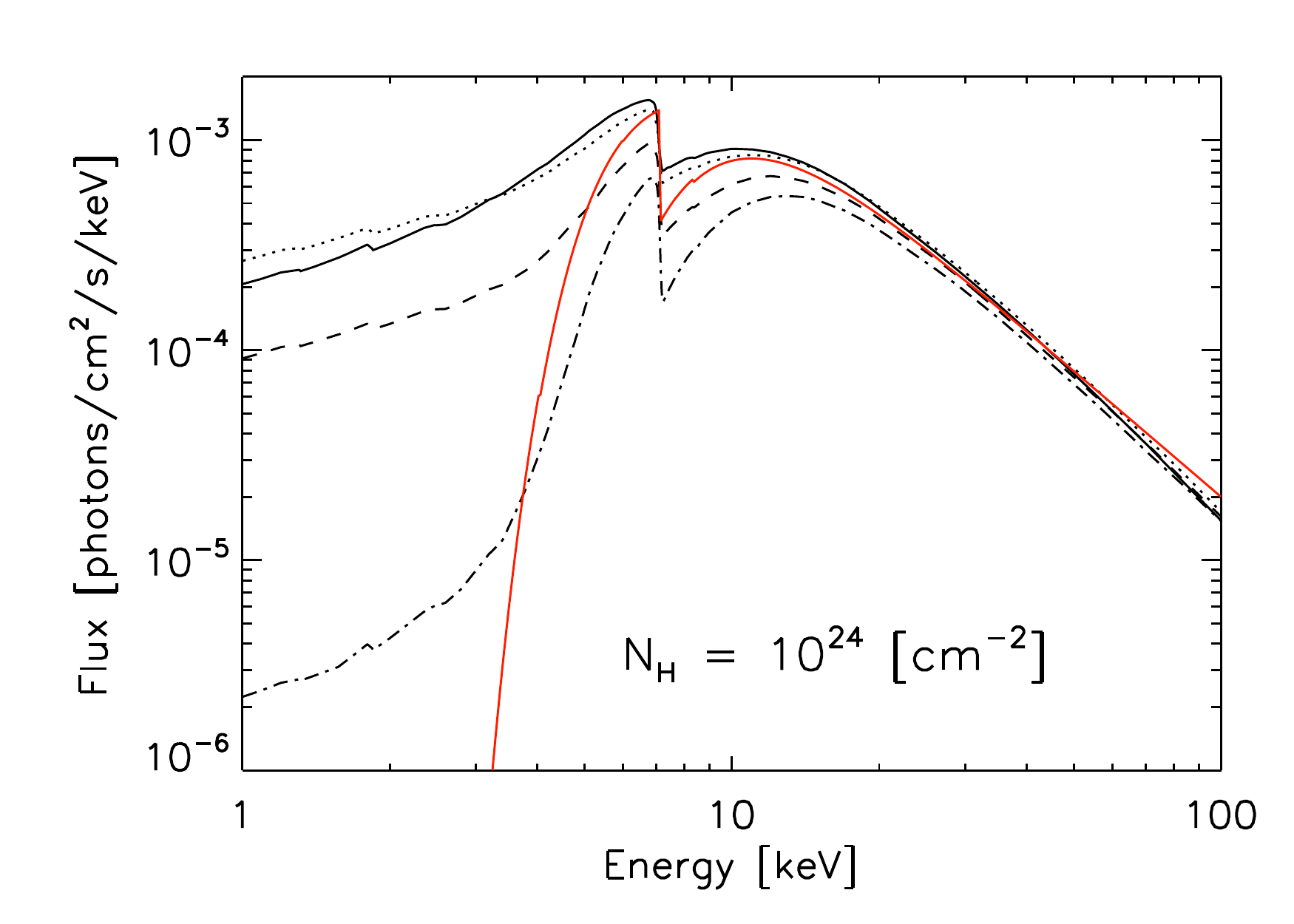}
}
\caption{MYTorus model spectra at $\theta_{\rm obs} = $ $0^\circ$ (solid), $60^\circ$ (dotted),
$75^\circ$ (dashed)  and 90$^\circ$ (dotted-dashed) for the equatorial column density $N_{\rm H} = 10^{23}$ (left) and
$10^{24}$~cm$^{-2}$ (right). In all cases, we assumed an input power-law spectrum with $\Gamma=2$. We used the same flux normalization for all
the MYTorus model spectra. Note that the model spectra show strong angular
dependence at $\theta_{\rm obs} > 60^\circ$ below $\sim3$~keV ($N_{\rm H} = 10^{23}$~cm$^{-2}$) and $\sim 10$~keV
($N_{\rm H} = 10^{24}$~cm$^{-2}$). For comparison, we plot an absorbed power-law model ({\tt tbabs*powerlaw}, red solid lines) with the
mean column density of $(\pi/4)N_{\rm H}$ over all lines-of-sight through the torus \citep{Murphy2009}, power-law index of $\Gamma=2$ and
flux normalization roughly adjusted to the MYTorus model spectra in high energy band. \\}
\label{fig:mytorus}
\end{figure}

\begin{itemize} 

\item The reflector geometry is toroidal. 

We explored a large range of $\theta_{\rm obs}$ and $N_{\rm H}$ to investigate the validity of MYTorus model application 
to a quasi-spherical cloud. To begin with, a face-on case ($\theta_{\rm obs} = 0^\circ$) provides an accurate solution for a cloud since the 
axial symmetry is preserved for X-ray photon reflection with respect to a distant observer. We can obtain reflected X-ray flux 
from the cloud (which is a part of the virtual torus, i.e. the red circle in Figure~\ref{fig:mc_geometry}) by scaling the best-fit primary X-ray flux by a 
solid angle ratio of the torus (fixed to $\Omega/4\pi = 0.5$) and the cloud. In this way, we 'collect' X-rays reflected from the cloud only and 'abandon'  
X-rays reflected from the rest of the torus (the grey area in Figure~\ref{fig:mc_geometry}). $N_{\rm H}$ and $\Gamma$ 
remain the same regardless of the cloud geometry.  
As $\theta_{\rm obs}$ deviates from the face-on case, different azimuthal parts around the torus 
can scatter X-rays at different angles  therefore the MYTorus model spectrum may show 
some variation with 
the inclination angle $\theta_{\rm obs}$ and become inaccurate for a quasi-spherical cloud.      

However, we find that the scattered 
continuum component (MYTS) does not vary with $\theta_{\rm obs}$ strongly as long as $\theta_{\rm obs} \la 60^\circ$ and the cloud is 
optically thin ($N_{\rm H} \la 10^{24}$~cm$^{-2}$). Figure \ref{fig:mytorus} shows MYTS model spectra at various inclination angles 
for $N_{\rm H} = 10^{23}$ and $10^{24}$~cm$^{-2}$.  
There are two reasons for the strong angular dependence at $\theta_{\rm obs} \ga 60^\circ$.  
First, since the half-opening angle of the torus is fixed to $60^\circ$, some back-scattered X-ray photons from one side of 
the torus can hit the other side thus they are subject to further absorption before reaching an observer at $\theta_{\rm obs} \ge 60^\circ$. This is 
peculiar to the assumed torus geometry of the MYTorus model. Second, multiple scattering can induce some angular dependence of X-ray reflection 
spectra but it is negligible at $N_{\rm H} \la 10^{24}$~cm$^{-2}$. \citet{Odaka2011}, who simulated X-ray reflection spectra for 
the Sgr B2 cloud (with a spherical shape assumed), found that the scattered X-ray continuum spectrum and morphology do not depend on the location of the cloud and incident angle significantly when the cloud is not Compton thick. Moreover, the scattered X-ray flux is proportional to the total mass (or solid angle for a given $N_{\rm H}$) of a cloud regardless of its shape, if it is optically thin \citep{Cramphorn2002}. 

To quantify the (in)sensitivity to $\theta_{\rm obs}$ or assumed geometry, we made simulated MYTS spectra for $\theta_{\rm obs}=60^\circ$ at 
$N_{\rm H} = 10^{23}$-$10^{25}$~cm$^{-2}$, then we fit the simulated
spectra with the MYTS model with $\theta_{\rm obs}$ fixed at $0^\circ$. We measured the deviation of $N_{\rm H}$, $\Gamma$ and normalization from their input values, and adopted them as systematic errors associated with the MYTorus model. At $N_{\rm H} = 10^{23}$~cm$^{-2}$, $N_{\rm H}, \Gamma$ and 
normalization deviate from the input values by $\sim10$\%, 1\% and 7\%, respectively. At $N_{\rm H} =10^{24}$~cm$^{-2}$, the deviation increases to $\sim25$\%, 3\% and 10\%. 

We also compared an absorbed power-law model ({\verb=tbabs*powerlaw=}) with the MYTorus model spectra in Figure~\ref{fig:mytorus}. We 
adopted the mean column density of $(\pi/4)N_{\rm H}$ over 
all LOS through the torus \citep{Murphy2009}, power-law index of $\Gamma=2$ and flux normalization roughly adjusted to the 
MYTorus model spectra in high energy band where photo-absorption is negligible. Note that the ad hoc XRN models predict significantly lower 
X-ray fluxes than the MYTorus models in low energy band. This is due to the fact that the ad hoc XRN model uses a single absorption term 
with a charateristic column density 
whereas photo-absorption takes place in various locations in the cloud with different optical depths. Based on simulation, 
we find that the ad hoc 
XRN model yields a column density lower than the equatorial column density ($N_{\rm H}$) from the MYTorus model by a factor 
of $\sim$2-3 in the range of $N_{\rm H}=10^{23}$-$10^{24}$~cm$^{-2}$. 

Our primary goal is to determine the primary X-ray spectrum (e.g., Sgr A* flares) 
 therefore we set $N_{\rm H} \la 10^{24}$~cm$^{-2}$ as a valid range of MYTorus model application to molecular clouds because otherwise the systematic errors for $\Gamma$ and normalization become larger than our statistical errors.  
We conclude that the reflected X-ray spectrum model in the \nustar\ energy band (3-79 keV) is not sensitive to the geometry of a reflector 
as long as $\theta_{\rm obs} \la 60^\circ$ and $N_{\rm H} \la 10^{24}$~cm$^{-2}$. In this range, similarly to the face-on case, 
the incident X-ray flux for a given cloud can be
obtained by scaling the best-fit X-ray flux from the MYTorus fit by the solid angle ratio of a cloud 
(typically $\Omega/4\pi \sim 10^{-2}$ for GCMCs) and the torus of the MYTorus model ($\Omega/4\pi = 0.5$) with $\la 10$\% errors. 
The best-fit power-law index from the MYTorus model fit can be adopted as that of the primary X-ray source 
with $\la 3$\% errors. On the other hand, a relation between $\theta_{\rm obs}$ and LOS distance of a cloud as well as its 
systematic errors cannot be well established. 
A modified version of the MYTorus model for more realistic cloud 
geometry (e.g. sphere) is under development and it will be used to determine the location of the GC molecular clouds and their primary X-ray 
source spectra more precisely without the restrictions described above.

\item The torus density profile is uniform. 

Since we extract X-ray spectra from the entire cloud and collect all 
X-ray photons reflected toward us, we assume that any effects due to the non-uniformity of the density profile will be negligible at $N_{\rm H} \la 10^{24}$~cm$^{-2}$ as multiple scattering is 
insignificant at these column densities. Also, the reflected X-ray flux is proportional to the total mass of a cloud if it is optically thin \citep{Cramphorn2002}. Thus, the primary X-ray flux will not be affected by different density profiles. 
It is, however, more important to take into account non-uniform density profile for X-ray morphology studies of GCMCs \citep{Sunyaev1998, Odaka2011}.

\item Fe abundance is fixed to solar. 

Non-solar Fe abundance primarily affects Fe fluorescent lines at 6.4 and 7.0 keV and the Fe K absorption edge at 7.1 keV. While an unknown Fe abundance 
adds some uncertainty when one attempts to determine the primary X-ray luminosity solely from the Fe K$\alpha$ line EW,  
broadband X-ray spectroscopy with \nustar\ extends to $E\ga10$ keV where the contribution of Compton scattering dominates over that of Fe fluorescent lines or photo-absorption. 
We confirmed that the fit parameters did not vary 
significantly when we fit the MYTS (scattered continuum) model only to the \xmm\ + \nustar\ spectra of MC1 and the Bridge without 6-10 keV energy bins where the Fe fluorescent lines and the K-edge are prominent. A new self-consistent XRN model based on the MYTorus model will implement 
data tables for different Fe abundances in the range of $Z_{Fe}=0.5$-3 (private communication with T. Yaqoob).

\end{itemize}

\bibliography{NuSTAR_GC}

\begin{thebibliography}{}
\expandafter\ifx\csname natexlab\endcsname\relax\def\natexlab#1{#1}\fi

\bibitem[{{Acero} {et~al.}(2010){Acero}, {Aharonian}, {Akhperjanian}, Anton,
  Barres~de Almeida, Bazer-Bachi, Becherini, Behera, Bernlöhr, Bochow,
  Boisson, Bolmont, Borrel, Braun, Brucker, Brun, Brun, Bühler, Bulik,
  Büsching, Boutelier, Chadwick, Charbonnier, Chaves, Cheesebrough, Conrad,
  Chounet, Clapson, Coignet, Dalton, Daniel, Davids, Degrange, Deil, Dickinson,
  Djannati-Ataï, Domainko, Drury, Dubois, Dubus, Dyks, Dyrda, Egberts, Eger,
  Espigat, Fallon, Farnier, Fegan, Feinstein, Fiasson, Förster, Fontaine,
  Füßling, Gabici, Gallant, Gérard, Gerbig, Giebels, Glicenstein, Glück,
  Goret, Göring, Hauser, Heinz, Heinzelmann, Henri, Hermann, Hinton, Hoffmann,
  Hofmann, Hofverberg, Holleran, Hoppe, Horns, Jacholkowska, de~Jager, Jahn,
  Jung, Katarzyński, Katz, Kaufmann, Kerschhaggl, Khangulyan, Khélifi, Keogh,
  Klochkov, Kluźniak, Kneiske, Komin, Kosack, Kossakowski, Lamanna, Lenain,
  Lohse, Marandon, Martineau-Huynh, Marcowith, Masbou, Maurin, McComb, Medina,
  Méhault, Moderski, Moulin, Naumann-Godo, de~Naurois, Nedbal, Nekrassov,
  Nicholas, Niemiec, Nolan, Ohm, Olive, de~Oña~Wilhelmi, Orford, Ostrowski,
  Panter, Arribas, Pedaletti, Pelletier, Petrucci, Pita, Pühlhofer, Punch,
  Quirrenbach, Raubenheimer, Raue, Rayner, Reimer, Renaud, Rieger, Ripken, Rob,
  Rosier-Lees, Rowell, Rudak, Rulten, Ruppel, Ryde, Sahakian, Santangelo,
  Schlickeiser, Schöck, Schönwald, Schwanke, Schwarzburg, Schwemmer, Shalchi,
  Sikora, Skilton, Sol, Stawarz, Steenkamp, Stegmann, Stinzing, Superina,
  Sushch, Szostek, Tam, Tavernet, Terrier, Tibolla, Tluczykont, van Eldik,
  Vasileiadis, Venter, Venter, Vialle, Vincent, Vivier, Völk, Volpe, Wagner,
  Ward, Zdziarski, \& Zech}]{Acero2010}
{Acero}, F., {Aharonian}, F., {Akhperjanian}, A.~G., {et~al.} 2010, Monthly
  Notices of the Royal Astronomical Society, 402, 1877

\bibitem[{{Aharonian} \& {Neronov}(2005)}]{Aharonian2005b}
{Aharonian}, F., \& {Neronov}, A. 2005, \apss, 300, 255

\bibitem[{{Aharonian} {et~al.}(2004){Aharonian}, {Akhperjanian}, {Aye},
  {Bazer-Bachi}, {Beilicke}, {Benbow}, {Berge}, {Berghaus}, {Bernl{\"o}hr},
  {Bolz}, {Boisson}, {Borgmeier}, {Breitling}, {Brown}, {Bussons Gordo},
  {Chadwick}, {Chitnis}, {Chounet}, {Cornils}, {Costamante}, {Degrange},
  {Djannati-Ata{\"i}}, {O'C.~Drury}, {Ergin}, {Espigat}, {Feinstein}, {Fleury},
  {Fontaine}, {Funk}, {Gallant}, {Giebels}, {Gillessen}, {Goret}, {Guy},
  {Hadjichristidis}, {Hauser}, {Heinzelmann}, {Henri}, {Hermann}, {Hinton},
  {Hofmann}, {Holleran}, {Horns}, {de Jager}, {Jung}, {Kh{\'e}lifi}, {Komin},
  {Konopelko}, {Latham}, {Le Gallou}, {Lemoine}, {Lemi{\`e}re}, {Leroy},
  {Lohse}, {Marcowith}, {Masterson}, {McComb}, {de Naurois}, {Nolan},
  {Noutsos}, {Orford}, {Osborne}, {Ouchrif}, {Panter}, {Pelletier}, {Pita},
  {Pohl}, {P{\"u}hlhofer}, {Punch}, {Raubenheimer}, {Raue}, {Raux}, {Rayner},
  {Redondo}, {Reimer}, {Reimer}, {Ripken}, {Rivoal}, {Rob}, {Rolland},
  {Rowell}, {Sahakian}, {Saug{\'e}}, {Schlenker}, {Schlickeiser}, {Schuster},
  {Schwanke}, {Siewert}, {Sol}, {Steenkamp}, {Stegmann}, {Tavernet},
  {Th{\'e}oret}, {Tluczykont}, {van der Walt}, {Vasileiadis}, {Vincent},
  {Visser}, {V{\"o}lk}, \& {Wagner}}]{Aharonian2004}
{Aharonian}, F., {Akhperjanian}, A.~G., {Aye}, K.-M., {et~al.} 2004, \aap, 425,
  L13

\bibitem[{{Aharonian} {et~al.}(2005){Aharonian}, {Akhperjanian}, {Aye},
  {Bazer-Bachi}, {Beilicke}, {Benbow}, {Berge}, {Berghaus}, {Bernl{\"o}hr},
  {Boisson}, {Bolz}, {Borgmeier}, {Braun}, {Breitling}, {Brown}, {Bussons
  Gordo}, {Chadwick}, {Chounet}, {Cornils}, {Costamante}, {Degrange},
  {Djannati-Ata{\"i}}, {O'C.~Drury}, {Dubus}, {Ergin}, {Espigat}, {Feinstein},
  {Fleury}, {Fontaine}, {Funk}, {Gallant}, {Giebels}, {Gillessen}, {Goret},
  {Hadjichristidis}, {Hauser}, {Heinzelmann}, {Henri}, {Hermann}, {Hinton},
  {Hofmann}, {Holleran}, {Horns}, {de Jager}, {Jung}, {Kh{\'e}lifi}, {Komin},
  {Konopelko}, {Latham}, {Le Gallou}, {Lemi{\`e}re}, {Lemoine}, {Leroy},
  {Lohse}, {Marcowith}, {Masterson}, {McComb}, {de Naurois}, {Nolan},
  {Noutsos}, {Orford}, {Osborne}, {Ouchrif}, {Panter}, {Pelletier}, {Pita},
  {P{\"u}hlhofer}, {Punch}, {Raubenheimer}, {Raue}, {Raux}, {Rayner},
  {Redondo}, {Reimer}, {Reimer}, {Ripken}, {Rob}, {Rolland}, {Rowell},
  {Sahakian}, {Saug{\'e}}, {Schlenker}, {Schlickeiser}, {Schuster}, {Schwanke},
  {Siewert}, {Sol}, {Steenkamp}, {Stegmann}, {Tavernet}, {Terrier},
  {Th{\'e}oret}, {Tluczykont}, {Vasileiadis}, {Venter}, {Vincent}, {Visser},
  {V{\"o}lk}, \& {Wagner}}]{Aharonian2005}
---. 2005, \aap, 432, L25

\bibitem[{{Aharonian} {et~al.}(2006){Aharonian}, {Akhperjanian}, {Bazer-Bachi},
  {Beilicke}, {Benbow}, {Berge}, {Bernl{\"o}hr}, {Boisson}, {Bolz}, {Borrel},
  {Braun}, {Breitling}, {Brown}, {Chadwick}, {Chounet}, {Cornils},
  {Costamante}, {Degrange}, {Dickinson}, {Djannati-Ata{\"i}}, {Drury}, {Dubus},
  {Emmanoulopoulos}, {Espigat}, {Feinstein}, {Fontaine}, {Fuchs}, {Funk},
  {Gallant}, {Giebels}, {Gillessen}, {Glicenstein}, {Goret}, {Hadjichristidis},
  {Hauser}, {Hauser}, {Heinzelmann}, {Henri}, {Hermann}, {Hinton}, {Hofmann},
  {Holleran}, {Horns}, {Jacholkowska}, {de Jager}, {Kh{\'e}lifi}, {Klages},
  {Komin}, {Konopelko}, {Latham}, {Le Gallou}, {Lemi{\`e}re},
  {Lemoine-Goumard}, {Leroy}, {Lohse}, {Marcowith}, {Martin},
  {Martineau-Huynh}, {Masterson}, {McComb}, {de Naurois}, {Nolan}, {Noutsos},
  {Orford}, {Osborne}, {Ouchrif}, {Panter}, {Pelletier}, {Pita},
  {P{\"u}hlhofer}, {Punch}, {Raubenheimer}, {Raue}, {Raux}, {Rayner}, {Reimer},
  {Reimer}, {Ripken}, {Rob}, {Rolland}, {Rowell}, {Sahakian}, {Saug{\'e}},
  {Schlenker}, {Schlickeiser}, {Schuster}, {Schwanke}, {Siewert}, {Sol},
  {Spangler}, {Steenkamp}, {Stegmann}, {Tavernet}, {Terrier}, {Th{\'e}oret},
  {Tluczykont}, {van Eldik}, {Vasileiadis}, {Venter}, {Vincent}, {V{\"o}lk}, \&
  {Wagner}}]{Aharonian2006c}
{Aharonian}, F., {Akhperjanian}, A.~G., {Bazer-Bachi}, A.~R., {et~al.} 2006,
  \nat, 439, 695

\bibitem[{{Aharonian} {et~al.}(2009){Aharonian}, {Akhperjanian}, {Anton, G.},
  {Barres de Almeida, U.}, {Bazer-Bachi, A. R.}, {Becherini, Y.}, {Behera, B.},
  {Bernl\"ohr, K.}, {Boisson, C.}, {Bochow, A.}, {Borrel, V.}, {Braun, I.}, \&
  {Brion, E.}}]{Aharonian2009}
{Aharonian}, F., {Akhperjanian}, A.~G., {Anton, G.}, {et~al.} 2009, A\&A, 503,
  817

\bibitem[{{Albert} {et~al.}(2006){Albert}, {Aliu}, {Anderhub}, {Antoranz},
  {Armada}, {Asensio}, {Baixeras}, {Barrio}, {Bartelt}, {Bartko}, {Bastieri},
  {Bavikadi}, {Bednarek}, {Berger}, {Bigongiari}, {Biland}, {Bisesi}, {Bock},
  {Bretz}, {Britvitch}, {Camara}, {Chilingarian}, {Ciprini}, {Coarasa},
  {Commichau}, {Contreras}, {Cortina}, {Curtef}, {Danielyan}, {Dazzi}, {De
  Angelis}, {de los Reyes}, {De Lotto}, {Domingo-Santamar{\'{\i}}a}, {Dorner},
  {Doro}, {Errando}, {Fagiolini}, {Ferenc}, {Fern{\'a}ndez}, {Firpo}, {Flix},
  {Fonseca}, {Font}, {Galante}, {Garczarczyk}, {Gaug}, {Giller}, {Goebel},
  {Hakobyan}, {Hayashida}, {Hengstebeck}, {H{\"o}hne}, {Hose}, {Jacon},
  {Kalekin}, {Kranich}, {Laille}, {Lenisa}, {Liebing}, {Lindfors}, {Longo},
  {L{\'o}pez}, {L{\'o}pez}, {Lorenz}, {Lucarelli}, {Majumdar}, {Maneva},
  {Mannheim}, {Mariotti}, {Mart{\'{\i}}nez}, {Mase}, {Mazin}, {Merck},
  {Meucci}, {Meyer}, {Miranda}, {Mirzoyan}, {Mizobuchi}, {Moralejo}, {Nilsson},
  {O{\~n}a-Wilhelmi}, {Ordu{\~n}a}, {Otte}, {Oya}, {Paneque}, {Paoletti},
  {Pasanen}, {Pascoli}, {Pauss}, {Pavel}, {Pegna}, {Peruzzo}, {Piccioli},
  {Prandini}, {Rico}, {Rhode}, {Riegel}, {Rissi}, {Robert}, {R{\"u}gamer},
  {Saggion}, {S{\'a}nchez}, {Sartori}, {Scalzotto}, {Schmitt}, {Schweizer},
  {Shayduk}, {Shinozaki}, {Shore}, {Sidro}, {Sillanp{\"a}{\"a}}, {Sobczynska},
  {Stamerra}, {Stepanian}, {Stark}, {Takalo}, {Temnikov}, {Tescaro}, {Teshima},
  {Tonello}, {Torres}, {Torres}, {Turini}, {Vankov}, {Vardanyan}, {Vitale},
  {Wagner}, {Wibig}, {Wittek}, \& {Zapatero}}]{Albert2006}
{Albert}, J., {Aliu}, E., {Anderhub}, H., {et~al.} 2006, \apjl, 638, L101

\bibitem[{{Amo-Baladr{\'o}n} {et~al.}(2009){Amo-Baladr{\'o}n},
  {Mart{\'{\i}}n-Pintado}, {Morris}, {Muno}, \&
  {Rodr{\'{\i}}guez-Fern{\'a}ndez}}]{Amo2009}
{Amo-Baladr{\'o}n}, M.~A., {Mart{\'{\i}}n-Pintado}, J., {Morris}, M.~R.,
  {Muno}, M.~P., \& {Rodr{\'{\i}}guez-Fern{\'a}ndez}, N.~J. 2009, \apj, 694,
  943

\bibitem[{{Archer} {et~al.}(2014){Archer}, {Barnacka}, {Beilicke}, {Benbow},
  {Berger}, {Bird}, {Biteau}, {Buckley}, {Bugaev}, {Byrum}, {Cardenzana},
  {Cerruti}, {Chen}, {Chen}, {Ciupik}, {Connolly}, {Cui}, {Dickinson}, {Dumm},
  {Eisch}, {Falcone}, {Federici}, {Feng}, {Finley}, {Fleischhack}, {Fortson},
  {Furniss}, {Galante}, {Griffin}, {Griffiths}, {Grube}, {Gyuk},
  {H{\aa}kansson}, {Hanna}, {Holder}, {Hughes}, {Johnson}, {Kaaret}, {Kar},
  {Kertzman}, {Khassen}, {Kieda}, {Krawczynski}, {Kumar}, {Lang}, {Maier},
  {McArthur}, {McCann}, {Meagher}, {Moriarty}, {Mukherjee}, {Nieto},
  {O'Faol{\'a}in de Bhr{\'o}ithe}, {Ong}, {Otte}, {Park}, {Perkins}, {Pohl},
  {Popkow}, {Prokoph}, {Pueschel}, {Quinn}, {Ragan}, {Rajotte}, {Reyes},
  {Reynolds}, {Richards}, {Roache}, {Sembroski}, {Shahinyan}, {Smith},
  {Staszak}, {Telezhinsky}, {Tucci}, {Tyler}, {Varlotta}, {Vincent}, {Wakely},
  {Weinstein}, {Welsing}, {Wilhelm}, {Williams}, {Zajczyk}, \&
  {Zitzer}}]{Archer2014}
{Archer}, A., {Barnacka}, A., {Beilicke}, M., {et~al.} 2014, \apj, 790, 149

\bibitem[{Arnaud(1996)}]{Arnaud1996}
Arnaud, K.~A. 1996, ASP Conference Series, 101, 17

\bibitem[{{Baganoff} {et~al.}(2001){Baganoff}, {Bautz}, {Brandt}, {Chartas},
  {Feigelson}, {Garmire}, {Maeda}, {Morris}, {Ricker}, {Townsley}, \&
  {Walter}}]{Baganoff2001}
{Baganoff}, F.~K., {Bautz}, M.~W., {Brandt}, W.~N., {et~al.} 2001, \nat, 413,
  45

\bibitem[{Baganoff {et~al.}(2003)Baganoff, Maeda, Morris, Bautz, Brandt, Cui,
  Doty, Feigelson, Garmire, Pravdo, Ricker, \& Townsley}]{Baganoff2003}
Baganoff, F.~K., Maeda, Y., Morris, M., {et~al.} 2003, The Astrophysical
  Journal, 591, 891

\bibitem[{{Ballantyne} {et~al.}(2007){Ballantyne}, {Melia}, {Liu}, \&
  {Crocker}}]{Ballantyne2007}
{Ballantyne}, D.~R., {Melia}, F., {Liu}, S., \& {Crocker}, R.~M. 2007, \apjl,
  657, L13

\bibitem[{{Ballantyne} {et~al.}(2011){Ballantyne}, {Schumann}, \&
  {Ford}}]{Ballantyne2011}
{Ballantyne}, D.~R., {Schumann}, M., \& {Ford}, B. 2011, \mnras, 410, 1521

\bibitem[{{Bamba} {et~al.}(2010){Bamba}, {Anada}, {Dotani}, {Mori}, {Yamazaki},
  {Ebisawa}, \& {Vink}}]{Bamba2010}
{Bamba}, A., {Anada}, T., {Dotani}, T., {et~al.} 2010, \apjl, 719, L116

\bibitem[{{Barri{\`e}re} {et~al.}(2014){Barri{\`e}re}, {Tomsick}, {Baganoff},
  {Boggs}, {Christensen}, {Craig}, {Dexter}, {Grefenstette}, {Hailey},
  {Harrison}, {Madsen}, {Mori}, {Stern}, {Zhang}, {Zhang}, \&
  {Zoglauer}}]{Barriere2014}
{Barri{\`e}re}, N.~M., {Tomsick}, J.~A., {Baganoff}, F.~K., {et~al.} 2014,
  \apj, 786, 46

\bibitem[{{Barri{\`e}re} {et~al.}(2015){Barri{\`e}re}, {Krivonos}, {Tomsick},
  {Bachetti}, {Boggs}, {Chakrabarty}, {Christensen}, {Craig}, {Hailey},
  {Harrison}, {Hong}, {Mori}, {Stern}, \& {Zhang}}]{Barriere2015}
{Barri{\`e}re}, N.~M., {Krivonos}, R., {Tomsick}, J.~A., {et~al.} 2015, \apj,
  799, 123

\bibitem[{B{\'e}langer {et~al.}(2006)B{\'e}langer, Goldwurm, Renaud, Terrier,
  Melia, N.~Lund, Skinner, \& Yusef-Zadeh}]{Belanger2006}
B{\'e}langer, G., Goldwurm, A., Renaud, M., {et~al.} 2006, The Astrophysical
  Journal, 636, 275

\bibitem[{{Bodaghee} {et~al.}(2014){Bodaghee}, {Tomsick}, {Krivonos}, {Stern},
  {Bauer}, {Fornasini}, {Barri{\`e}re}, {Boggs}, {Christensen}, {Craig},
  {Gotthelf}, {Hailey}, {Harrison}, {Hong}, {Mori}, \& {Zhang}}]{Bodaghee2014}
{Bodaghee}, A., {Tomsick}, J.~A., {Krivonos}, R., {et~al.} 2014, \apj, 791, 68

\bibitem[{{Branduardi} {et~al.}(1976){Branduardi}, {Ives}, {Sanford},
  {Brinkman}, \& {Maraschi}}]{Branduardi1976}
{Branduardi}, G., {Ives}, J.~C., {Sanford}, P.~W., {Brinkman}, A.~C., \&
  {Maraschi}, L. 1976, \mnras, 175, 47P

\bibitem[{{Brightman} \& {Nandra}(2011)}]{Brightman2011}
{Brightman}, M., \& {Nandra}, K. 2011, \mnras, 413, 1206

\bibitem[{{Bykov} {et~al.}(2005){Bykov}, {Bocchino}, \& {Pavlov}}]{Bykov2005}
{Bykov}, A.~M., {Bocchino}, F., \& {Pavlov}, G.~G. 2005, \apjl, 624, L41

\bibitem[{{Bykov} {et~al.}(2000){Bykov}, {Chevalier}, {Ellison}, \&
  {Uvarov}}]{Bykov2000}
{Bykov}, A.~M., {Chevalier}, R.~A., {Ellison}, D.~C., \& {Uvarov}, Y.~A. 2000,
  \apj, 538, 203

\bibitem[{{Capelli} {et~al.}(2012){Capelli}, {Warwick}, {Porquet}, {Gillessen},
  \& {Predehl}}]{Capelli2012}
{Capelli}, R., {Warwick}, R.~S., {Porquet}, D., {Gillessen}, S., \& {Predehl},
  P. 2012, \aap, 545, A35

\bibitem[{{Cembranos} {et~al.}(2013){Cembranos}, {Gammaldi}, \&
  {Maroto}}]{Cembranos2013}
{Cembranos}, J.~A.~R., {Gammaldi}, V., \& {Maroto}, A.~L. 2013, Journal of
  Cosmology and Astroparticle Physics, 4, 51

\bibitem[{{Chen} {et~al.}(1997){Chen}, {Shrader}, \& {Livio}}]{Chen1997}
{Chen}, W., {Shrader}, C.~R., \& {Livio}, M. 1997, \apj, 491, 312

\bibitem[{Chernyakova {et~al.}(2011)Chernyakova, Malyshev, Aharonian, Crocker,
  \& Jones}]{Chernyakova2011}
Chernyakova, M., Malyshev, D., Aharonian, F.~A., Crocker, R.~M., \& Jones,
  D.~I. 2011, The Astrophysical Journal, 726, 60

\bibitem[{{Christopher} {et~al.}(2005){Christopher}, {Scoville}, {Stolovy}, \&
  {Yun}}]{Christopher2005}
{Christopher}, M.~H., {Scoville}, N.~Z., {Stolovy}, S.~R., \& {Yun}, M.~S.
  2005, \apj, 622, 346

\bibitem[{{Clavel} {et~al.}(2013){Clavel}, {Terrier}, {Goldwurm}, {Morris},
  {Ponti}, {Soldi}, \& {Trap}}]{Clavel2013}
{Clavel}, M., {Terrier}, R., {Goldwurm}, A., {et~al.} 2013, \aap, 558, A32

\bibitem[{{Cramphorn} \& {Sunyaev}(2002)}]{Cramphorn2002}
{Cramphorn}, C.~K., \& {Sunyaev}, R.~A. 2002, \aap, 389, 252

\bibitem[{{Davidson} {et~al.}(1992){Davidson}, {Werner}, {Wu}, {Lester},
  {Harvey}, {Joy}, \& {Morris}}]{Davidson1992}
{Davidson}, J.~A., {Werner}, M.~W., {Wu}, X., {et~al.} 1992, \apj, 387, 189

\bibitem[{{Degenaar} {et~al.}(2013){Degenaar}, {Miller}, {Kennea}, {Gehrels},
  {Reynolds}, \& {Wijnands}}]{Degenaar2013}
{Degenaar}, N., {Miller}, J.~M., {Kennea}, J., {et~al.} 2013, \apj, 769, 155

\bibitem[{{Degenaar} {et~al.}(2012){Degenaar}, {Wijnands}, {Cackett}, {Homan},
  {in't Zand}, {Kuulkers}, {Maccarone}, \& {van der Klis}}]{Degenaar2012}
{Degenaar}, N., {Wijnands}, R., {Cackett}, E.~M., {et~al.} 2012, \aap, 545, A49

\bibitem[{{Dogiel} {et~al.}(2009{\natexlab{a}}){Dogiel}, {Cheng}, {Chernyshov},
  {Bamba}, {Ichimura}, {Inoue}, {Ko}, {Kokubun}, {Maeda}, {Mitsuda}, \&
  {Yamasaki}}]{Dogiel2009c}
{Dogiel}, V., {Cheng}, K.-S., {Chernyshov}, D., {et~al.} 2009{\natexlab{a}},
  \pasj, 61, 901

\bibitem[{{Dogiel} {et~al.}(2009{\natexlab{b}}){Dogiel}, {Chernyshov}, {Yuasa},
  {Prokhorov}, {Cheng}, {Bamba}, {Inoue}, {Ko}, {Kokubun}, {Maeda}, {Mitsuda},
  {Nakazawa}, \& {Yamasaki}}]{Dogiel2009a}
{Dogiel}, V.~A., {Chernyshov}, D.~O., {Yuasa}, T., {et~al.} 2009{\natexlab{b}},
  \pasj, 61, 1099

\bibitem[{{Dogiel} {et~al.}(2009{\natexlab{c}}){Dogiel}, {Chernyshov}, {Yuasa},
  {Cheng}, {Bamba}, {Inoue}, {Ko}, {Kokubun}, {Maeda}, {Mitsuda}, {Nakazawa},
  \& {Yamasaki}}]{Dogiel2009b}
{Dogiel}, V.~A., {Chernyshov}, D., {Yuasa}, T., {et~al.} 2009{\natexlab{c}},
  \pasj, 61, 1093

\bibitem[{{Dubus}(2013)}]{Dubus2013}
{Dubus}, G. 2013, \aapr, 21, 64

\bibitem[{{Etxaluze} {et~al.}(2013){Etxaluze}, {Goicoechea}, {Cernicharo},
  {Polehampton}, {Noriega-Crespo}, {Molinari}, {Swinyard}, {Wu}, \&
  {Bally}}]{Etxaluze2013}
{Etxaluze}, M., {Goicoechea}, J.~R., {Cernicharo}, J., {et~al.} 2013, \aap,
  556, A137

\bibitem[{{Fender} \& {Belloni}(2004)}]{Fender2004}
{Fender}, R., \& {Belloni}, T. 2004, \araa, 42, 317

\bibitem[{{Ferri{\`e}re}(2009)}]{Ferriere2009}
{Ferri{\`e}re}, K. 2009, \aap, 505, 1183

\bibitem[{Freeman {et~al.}(2001)Freeman, Doe, \& Siemiginowska}]{Freeman2001}
Freeman, P., Doe, S., \& Siemiginowska, A. 2001, Proc. SPIE, 4477, 76

\bibitem[{{Gabici} {et~al.}(2009){Gabici}, {Aharonian}, \&
  {Casanova}}]{Gabici2009}
{Gabici}, S., {Aharonian}, F.~A., \& {Casanova}, S. 2009, \mnras, 396, 1629

\bibitem[{{Gehrels} {et~al.}(2004){Gehrels}, {Chincarini}, {Giommi}, {Mason},
  {Nousek}, {Wells}, {White}, {Barthelmy}, {Burrows}, {Cominsky}, {Hurley},
  {Marshall}, {M{\'e}sz{\'a}ros}, {Roming}, {Angelini}, {Barbier}, {Belloni},
  {Campana}, {Caraveo}, {Chester}, {Citterio}, {Cline}, {Cropper}, {Cummings},
  {Dean}, {Feigelson}, {Fenimore}, {Frail}, {Fruchter}, {Garmire}, {Gendreau},
  {Ghisellini}, {Greiner}, {Hill}, {Hunsberger}, {Krimm}, {Kulkarni}, {Kumar},
  {Lebrun}, {Lloyd-Ronning}, {Markwardt}, {Mattson}, {Mushotzky}, {Norris},
  {Osborne}, {Paczynski}, {Palmer}, {Park}, {Parsons}, {Paul}, {Rees},
  {Reynolds}, {Rhoads}, {Sasseen}, {Schaefer}, {Short}, {Smale}, {Smith},
  {Stella}, {Tagliaferri}, {Takahashi}, {Tashiro}, {Townsley}, {Tueller},
  {Turner}, {Vietri}, {Voges}, {Ward}, {Willingale}, {Zerbi}, \&
  {Zhang}}]{Gehrels2004}
{Gehrels}, N., {Chincarini}, G., {Giommi}, P., {et~al.} 2004, \apj, 611, 1005

\bibitem[{{Hailey} {et~al.}(2015)}]{Hailey2015}
{Hailey}, C., {et~al.} 2015, to be submitted to ApJ

\bibitem[{{Handa} {et~al.}(2006){Handa}, {Sakano}, {Naito}, {Hiramatsu}, \&
  {Tsuboi}}]{Handa2006}
{Handa}, T., {Sakano}, M., {Naito}, S., {Hiramatsu}, M., \& {Tsuboi}, M. 2006,
  \apj, 636, 261

\bibitem[{{Harrison} {et~al.}(2013){Harrison}, {Craig}, {Christensen},
  {Hailey}, {Zhang}, {Boggs}, {Stern}, {Cook}, {Forster}, {Giommi},
  {Grefenstette}, {Kim}, {Kitaguchi}, {Koglin}, {Madsen}, {Mao}, {Miyasaka},
  {Mori}, {Perri}, {Pivovaroff}, {Puccetti}, {Rana}, {Westergaard}, {Willis},
  {Zoglauer}, {An}, {Bachetti}, {Barri{\`e}re}, {Bellm}, {Bhalerao},
  {Brejnholt}, {Fuerst}, {Liebe}, {Markwardt}, {Nynka}, {Vogel}, {Walton},
  {Wik}, {Alexander}, {Cominsky}, {Hornschemeier}, {Hornstrup}, {Kaspi},
  {Madejski}, {Matt}, {Molendi}, {Smith}, {Tomsick}, {Ajello}, {Ballantyne},
  {Balokovi{\'c}}, {Barret}, {Bauer}, {Blandford}, {Brandt}, {Brenneman},
  {Chiang}, {Chakrabarty}, {Chenevez}, {Comastri}, {Dufour}, {Elvis}, {Fabian},
  {Farrah}, {Fryer}, {Gotthelf}, {Grindlay}, {Helfand}, {Krivonos}, {Meier},
  {Miller}, {Natalucci}, {Ogle}, {Ofek}, {Ptak}, {Reynolds}, {Rigby},
  {Tagliaferri}, {Thorsett}, {Treister}, \& {Urry}}]{Harrison2013}
{Harrison}, F.~A., {Craig}, W.~W., {Christensen}, F.~E., {et~al.} 2013, \apj,
  770, 103

\bibitem[{{Heard} \& {Warwick}(2013)}]{Heard2013}
{Heard}, V., \& {Warwick}, R.~S. 2013, \mnras, 428, 3462

\bibitem[{Hinton \& Aharonian(2007)}]{Hinton2007}
Hinton, J.~A., \& Aharonian, F.~A. 2007, The Astrophysical Journal, 657, 302

\bibitem[{Hong {et~al.}(2015)}]{Hong2015}
Hong, J., {et~al.} 2015, to be submitted to ApJ

\bibitem[{{Inui} {et~al.}(2009){Inui}, {Koyama}, {Matsumoto}, \&
  {Tsuru}}]{Inui2009}
{Inui}, T., {Koyama}, K., {Matsumoto}, H., \& {Tsuru}, T.~G. 2009, \pasj, 61,
  241

\bibitem[{Johnson {et~al.}(2009)Johnson, Dong, \& Wang}]{Johnson2009}
Johnson, S.~P., Dong, H., \& Wang, Q.~D. 2009, Monthly Notices of the Royal
  Astronomical Society, 399, 1429

\bibitem[{{Jones} {et~al.}(2012){Jones}, {Burton}, {Cunningham},
  {Requena-Torres}, {Menten}, {Schilke}, {Belloche}, {Leurini},
  {Mart{\'{\i}}n-Pintado}, {Ott}, \& {Walsh}}]{Jones2012}
{Jones}, P.~A., {Burton}, M.~G., {Cunningham}, M.~R., {et~al.} 2012, \mnras,
  419, 2961

\bibitem[{{Kashyap} {et~al.}(2010){Kashyap}, {van Dyk}, {Connors}, {Freeman},
  {Siemiginowska}, {Xu}, \& {Zezas}}]{Kashyap2010}
{Kashyap}, V.~L., {van Dyk}, D.~A., {Connors}, A., {et~al.} 2010, \apj, 719,
  900

\bibitem[{{Kaspi} {et~al.}(2014){Kaspi}, {Archibald}, {Bhalerao}, {Dufour},
  {Gotthelf}, {An}, {Bachetti}, {Beloborodov}, {Boggs}, {Christensen}, {Craig},
  {Grefenstette}, {Hailey}, {Harrison}, {Kennea}, {Kouveliotou}, {Madsen},
  {Mori}, {Markwardt}, {Stern}, {Vogel}, \& {Zhang}}]{Kaspi2014}
{Kaspi}, V.~M., {Archibald}, R.~F., {Bhalerao}, V., {et~al.} 2014, \apj, 786,
  84

\bibitem[{{Koch} {et~al.}(2014){Koch}, {Bahramian}, {Heinke}, {Mori}, {Rea},
  {Degenaar}, {Haggard}, {Wijnands}, {Ponti}, {Miller}, {Yusef-Zadeh},
  {Dufour}, {Cotton}, {Baganoff}, \& {Reynolds}}]{koch2014}
{Koch}, E.~W., {Bahramian}, A., {Heinke}, C.~O., {et~al.} 2014, \mnras, 442,
  372

\bibitem[{{Koyama} {et~al.}(1989){Koyama}, {Awaki}, {Kunieda}, {Takano}, \&
  {Tawara}}]{Koyama1989}
{Koyama}, K., {Awaki}, H., {Kunieda}, H., {Takano}, S., \& {Tawara}, Y. 1989,
  \nat, 339, 603

\bibitem[{{Koyama} {et~al.}(1996){Koyama}, {Maeda}, {Sonobe}, {Takeshima},
  {Tanaka}, \& {Yamauchi}}]{Koyama1996}
{Koyama}, K., {Maeda}, Y., {Sonobe}, T., {et~al.} 1996, \pasj, 48, 249

\bibitem[{{Koyama} {et~al.}(2007){Koyama}, {Hyodo}, {Inui}, {Nakajima},
  {Matsumoto}, {Tsuru}, {Takahashi}, {Maeda}, {Yamazaki}, {Murakami},
  {Yamauchi}, {Tsuboi}, {Senda}, {Kataoka}, {Takahashi}, {Holt}, \&
  {Brown}}]{Koyama2007}
{Koyama}, K., {Hyodo}, Y., {Inui}, T., {et~al.} 2007, \pasj, 59, 245

\bibitem[{Krivonos {et~al.}(2007)Krivonos, Revnivtsev, Churazov, Sazonov,
  Grebenev, \& Sunyaev}]{Krivonos2007}
Krivonos, R., Revnivtsev, M., Churazov, E., {et~al.} 2007, Astronomy and
  Astrophysics, 463, 957

\bibitem[{{Krivonos} {et~al.}(2014){Krivonos}, {Tomsick}, {Bauer}, {Baganoff},
  {Barriere}, {Bodaghee}, {Boggs}, {Christensen}, {Craig}, {Grefenstette},
  {Hailey}, {Harrison}, {Hong}, {Madsen}, {Mori}, {Nynka}, {Stern}, \&
  {Zhang}}]{Krivonos2014}
{Krivonos}, R.~A., {Tomsick}, J.~A., {Bauer}, F.~E., {et~al.} 2014, \apj, 781,
  107

\bibitem[{{Kusunose} \& {Takahara}(2012)}]{Kusunose2012}
{Kusunose}, M., \& {Takahara}, F. 2012, \apj, 748, 34

\bibitem[{{Li} {et~al.}(2013){Li}, {Morris}, \& {Baganoff}}]{Li2013}
{Li}, Z., {Morris}, M.~R., \& {Baganoff}, F.~K. 2013, \apj, 779, 154

\bibitem[{{Linden} {et~al.}(2011){Linden}, {Hooper}, \&
  {Yusef-Zadeh}}]{Linden2011}
{Linden}, T., {Hooper}, D., \& {Yusef-Zadeh}, F. 2011, \apj, 741, 95

\bibitem[{{Lotti} {et~al.}(2015){Lotti}, {Natalucci}, {Baganoff}, {Bauer},
  {Boggs}, {Craig}, {Christensen}, {Gotthelf}, {Harrison}, {Mori}, {Nynka},
  {Stern}, {Tomsick}, \& {Zhang}}]{Lotti2015}
{Lotti}, S., {Natalucci}, L., {Baganoff}, F.~K., {et~al.} 2015, to be submitted
  to \apj

\bibitem[{{Lu} {et~al.}(2003){Lu}, {Wang}, \& {Lang}}]{Lu2003}
{Lu}, F.~J., {Wang}, Q.~D., \& {Lang}, C.~C. 2003, \aj, 126, 319

\bibitem[{{Lu} {et~al.}(2008){Lu}, {Yuan}, \& {Lou}}]{Lu2008}
{Lu}, F.~J., {Yuan}, T.~T., \& {Lou}, Y.-Q. 2008, \apj, 673, 915

\bibitem[{{Madsen} {et~al.}(2015){Madsen}, {Harrison}, {Markwardt}, {An},
  {Grefenstette}, {Bachetti}, {Miyasaka}, {Kitaguchi}, {Bhalerao},
  {Christensen}, {Craig}, {Fuerst}, {Walton}, {Hailey}, {Rana}, {Stern},
  {Westergaard}, \& {Zhang}}]{Madsen2015}
{Madsen}, K.~K., {Harrison}, F.~A., {Markwardt}, C., {et~al.} 2015, submitted
  to \apj, arXiv:1504.01672

\bibitem[{{Maeda} {et~al.}(2002){Maeda}, {Baganoff}, {Feigelson}, {Morris},
  {Bautz}, {Brandt}, {Burrows}, {Doty}, {Garmire}, {Pravdo}, {Ricker}, \&
  {Townsley}}]{Maeda2002}
{Maeda}, Y., {Baganoff}, F.~K., {Feigelson}, E.~D., {et~al.} 2002, \apj, 570,
  671

\bibitem[{{Magdziarz} \& {Zdziarski}(1995)}]{Magdziarz1995}
{Magdziarz}, P., \& {Zdziarski}, A.~A. 1995, \mnras, 273, 837

\bibitem[{{Mori} {et~al.}(2013){Mori}, {Gotthelf}, {Zhang}, {An}, {Baganoff},
  {Barri{\`e}re}, {Beloborodov}, {Boggs}, {Christensen}, {Craig}, {Dufour},
  {Grefenstette}, {Hailey}, {Harrison}, {Hong}, {Kaspi}, {Kennea}, {Madsen},
  {Markwardt}, {Nynka}, {Stern}, {Tomsick}, \& {Zhang}}]{Mori2013}
{Mori}, K., {Gotthelf}, E.~V., {Zhang}, S., {et~al.} 2013, \apjl, 770, L23

\bibitem[{Morris \& Serabyn(1996)}]{Morris1996}
Morris, M., \& Serabyn, E. 1996, Annual Review of Astronomy and Astrophysics,
  34, 645

\bibitem[{{Muno} {et~al.}(2008){Muno}, {Baganoff}, {Brandt}, {Morris}, \&
  {Starck}}]{Muno2008}
{Muno}, M.~P., {Baganoff}, F.~K., {Brandt}, W.~N., {Morris}, M.~R., \&
  {Starck}, J.-L. 2008, \apj, 673, 251

\bibitem[{{Muno} {et~al.}(2007){Muno}, {Baganoff}, {Brandt}, {Park}, \&
  {Morris}}]{Muno2007}
{Muno}, M.~P., {Baganoff}, F.~K., {Brandt}, W.~N., {Park}, S., \& {Morris},
  M.~R. 2007, \apjl, 656, L69

\bibitem[{{Muno} {et~al.}(2005){Muno}, {Pfahl}, {Baganoff}, {Brandt}, {Ghez},
  {Lu}, \& {Morris}}]{Muno2005}
{Muno}, M.~P., {Pfahl}, E., {Baganoff}, F.~K., {et~al.} 2005, \apjl, 622, L113

\bibitem[{Muno {et~al.}(2004)Muno, Baganoff, Bautz, Feigelson, Garmire, Morris,
  Park, Ricker, \& Townsley}]{MunoDiffuse2004}
Muno, M.~P., Baganoff, F.~K., Bautz, M.~W., {et~al.} 2004, The Astrophysical
  Journal, 613, 326

\bibitem[{{Muno} {et~al.}(2009){Muno}, {Bauer}, {Baganoff}, {Bandyopadhyay},
  {Bower}, {Brandt}, {Broos}, {Cotera}, {Eikenberry}, {Garmire}, {Hyman},
  {Kassim}, {Lang}, {Lazio}, {Law}, {Mauerhan}, {Morris}, {Nagata},
  {Nishiyama}, {Park}, {Ram{\`i}rez}, {Stolovy}, {Wijnands}, {Wang}, {Wang}, \&
  {Yusef-Zadeh}}]{Muno2009}
{Muno}, M.~P., {Bauer}, F.~E., {Baganoff}, F.~K., {et~al.} 2009, \apjs, 181,
  110

\bibitem[{{Murakami} {et~al.}(2001){Murakami}, {Koyama}, \&
  {Maeda}}]{Murakami2001}
{Murakami}, H., {Koyama}, K., \& {Maeda}, Y. 2001, \apj, 558, 687

\bibitem[{{Murphy} \& {Yaqoob}(2009)}]{Murphy2009}
{Murphy}, K.~D., \& {Yaqoob}, T. 2009, \mnras, 397, 1549

\bibitem[{{Nandra} {et~al.}(2007){Nandra}, {O'Neill}, {George}, \&
  {Reeves}}]{Nandra2007}
{Nandra}, K., {O'Neill}, P.~M., {George}, I.~M., \& {Reeves}, J.~N. 2007,
  \mnras, 382, 194

\bibitem[{{Neilsen} {et~al.}(2013){Neilsen}, {Nowak}, {Gammie}, {Dexter},
  {Markoff}, {Haggard}, {Nayakshin}, {Wang}, {Grosso}, {Porquet}, {Tomsick},
  {Degenaar}, {Fragile}, {Houck}, {Wijnands}, {Miller}, \&
  {Baganoff}}]{Neilsen2013}
{Neilsen}, J., {Nowak}, M.~A., {Gammie}, C., {et~al.} 2013, \apj, 774, 42

\bibitem[{{Neronov} {et~al.}(2005){Neronov}, {Chernyakova}, {Courvoisier}, \&
  {Walter}}]{Neronov2005}
{Neronov}, A., {Chernyakova}, M., {Courvoisier}, T.~J.-L., \& {Walter}, R.
  2005, ArXiv Astrophysics e-prints, astro-ph/0506437

\bibitem[{{Nobukawa} {et~al.}(2011){Nobukawa}, {Ryu}, {Tsuru}, \&
  {Koyama}}]{Nobukawa2011}
{Nobukawa}, M., {Ryu}, S.~G., {Tsuru}, T.~G., \& {Koyama}, K. 2011, \apjl, 739,
  L52

\bibitem[{{Nolan} {et~al.}(2012){Nolan}, {Abdo}, {Ackermann}, {Ajello},
  {Allafort}, {Antolini}, {Atwood}, {Axelsson}, {Baldini}, {Ballet}, \&
  et~al.}]{Nolan2012}
{Nolan}, P.~L., {Abdo}, A.~A., {Ackermann}, M., {et~al.} 2012, \apjs, 199, 31

\bibitem[{{Nowak} {et~al.}(2012){Nowak}, {Neilsen}, {Markoff}, {Baganoff},
  {Porquet}, {Grosso}, {Levin}, {Houck}, {Eckart}, {Falcke}, {Ji}, {Miller}, \&
  {Wang}}]{Nowak2012}
{Nowak}, M.~A., {Neilsen}, J., {Markoff}, S.~B., {et~al.} 2012, \apj, 759, 95

\bibitem[{{Nynka} {et~al.}(2013){Nynka}, {Hailey}, {Mori}, {Baganoff}, {Bauer},
  {Boggs}, {Craig}, {Christensen}, {Gotthelf}, {Harrison}, {Hong}, {Perez},
  {Stern}, {Zhang}, \& {Zhang}}]{Nynka2013}
{Nynka}, M., {Hailey}, C.~J., {Mori}, K., {et~al.} 2013, \apjl, 778, L31

\bibitem[{{Nynka} {et~al.}(2015){Nynka}, {Hailey}, {Reynolds}, {An},
  {Baganoff}, {Boggs}, {Christensen}, {Craig}, {Gotthelf}, {Grefenstette},
  {Harrison}, {Krivonos}, {Madsen}, {Mori}, {Perez}, {Stern}, {Wik}, {Zhang},
  \& {Zoglauer}}]{Nynka2014NF}
{Nynka}, M., {Hailey}, C.~J., {Reynolds}, S.~P., {et~al.} 2015, \apj, 800, 119

\bibitem[{{Odaka} {et~al.}(2011){Odaka}, {Aharonian}, {Watanabe}, {Tanaka},
  {Khangulyan}, \& {Takahashi}}]{Odaka2011}
{Odaka}, H., {Aharonian}, F., {Watanabe}, S., {et~al.} 2011, \apj, 740, 103

\bibitem[{Park {et~al.}(2005)Park, Muno, Baganoff, Maeda, Morris, Chartas,
  Sanwal, Burrows, \& Garmire}]{Park2005}
Park, S., Muno, M.~P., Baganoff, F.~K., {et~al.} 2005, The Astrophysical
  Journal, 631, 964

\bibitem[{{Perez} {et~al.}(2015){Perez}, {Hailey}, {Bauer}, {Krivonos}, {Mori},
  {Baganoff}, {Barri{\`e}re}, {Boggs}, {Christensen}, {Craig}, {Grefenstette},
  {Grindlay}, {Harrison}, {Hong}, {Madsen}, {Nynka}, {Stern}, {Tomsick}, {Wik},
  {Zhang}, {Zhang}, \& {Zoglauer}}]{Perez2015}
{Perez}, K., {Hailey}, C.~J., {Bauer}, F.~E., {et~al.} 2015, \nat, 520, 646

\bibitem[{{Ponti} {et~al.}(2013){Ponti}, {Morris}, {Terrier}, \&
  {Goldwurm}}]{Ponti2013}
{Ponti}, G., {Morris}, M.~R., {Terrier}, R., \& {Goldwurm}, A. 2013, in
  Advances in Solid State Physics, Vol.~34, Cosmic Rays in Star-Forming
  Environments, ed. D.~F. {Torres} \& O.~{Reimer}, 331

\bibitem[{{Ponti} {et~al.}(2010){Ponti}, {Terrier}, {Goldwurm}, {Belanger}, \&
  {Trap}}]{Ponti2010}
{Ponti}, G., {Terrier}, R., {Goldwurm}, A., {Belanger}, G., \& {Trap}, G. 2010,
  \apj, 714, 732

\bibitem[{{Ponti} {et~al.}(2014){Ponti}, {Morris}, {Clavel}, {Terrier},
  {Goldwurm}, {Soldi}, {Sturm}, {Haberl}, \& {Nandra}}]{Ponti2014}
{Ponti}, G., {Morris}, M.~R., {Clavel}, M., {et~al.} 2014, in IAU Symposium,
  Vol. 303, IAU Symposium, ed. L.~O. {Sjouwerman}, C.~C. {Lang}, \& J.~{Ott},
  333--343

\bibitem[{{Quataert} \& {Loeb}(2005)}]{Quataert2005}
{Quataert}, E., \& {Loeb}, A. 2005, \apjl, 635, L45

\bibitem[{{Reeves} \& {Turner}(2000)}]{Reeves2000}
{Reeves}, J.~N., \& {Turner}, M.~J.~L. 2000, \mnras, 316, 234

\bibitem[{{Reid}(1993)}]{Reid1993}
{Reid}, M.~J. 1993, \araa, 31, 345

\bibitem[{{Reid} {et~al.}(2009){Reid}, {Menten}, {Zheng}, {Brunthaler}, \&
  {Xu}}]{Reid2009}
{Reid}, M.~J., {Menten}, K.~M., {Zheng}, X.~W., {Brunthaler}, A., \& {Xu}, Y.
  2009, \apj, 705, 1548

\bibitem[{Reid {et~al.}(1999)Reid, Readhead, Vermeulen, \& Treuhaft}]{Reid1999}
Reid, M.~J., Readhead, A. C.~S., Vermeulen, R.~C., \& Treuhaft, R.~N. 1999, The
  Astrophysical Journal, 524, 816

\bibitem[{Revnivtsev {et~al.}(2009)Revnivtsev, Sazonov, , Churazov, Forman,
  Vikhlinin, \& Sunyaev}]{Revnivtsev2009}
Revnivtsev, M., Sazonov, S., , {et~al.} 2009, Nature, 458, 1142

\bibitem[{Revnivtsev {et~al.}(2006)Revnivtsev, Sazonov, Gilfanov, Churazov, \&
  Sunyaev}]{Revnivtsev2006}
Revnivtsev, M., Sazonov, S., Gilfanov, M., Churazov, E., \& Sunyaev, R. 2006,
  Astronomy and Astrophysics, 452, 169

\bibitem[{{Revnivtsev} {et~al.}(2004){Revnivtsev}, {Churazov}, {Sazonov},
  {Sunyaev}, {Lutovinov}, {Gilfanov}, {Vikhlinin}, {Shtykovsky}, \&
  {Pavlinsky}}]{Revnivtsev2004}
{Revnivtsev}, M.~G., {Churazov}, E.~M., {Sazonov}, S.~Y., {et~al.} 2004, \aap,
  425, L49

\bibitem[{{Ross} \& {Fabian}(2005)}]{Ross2005}
{Ross}, R.~R., \& {Fabian}, A.~C. 2005, \mnras, 358, 211

\bibitem[{Sakano {et~al.}(2003)Sakano, Warwick, Decourchelle, \&
  Predehl}]{Sakano2003}
Sakano, M., Warwick, R.~S., Decourchelle, A., \& Predehl, P. 2003, Monthly
  Notices of the Royal Astronomical Society, 430, 747

\bibitem[{Sakano {et~al.}(2004)Sakano, Warwick, Decourchelle, \&
  Predehl}]{Sakano2004}
---. 2004, Monthly Notices of the Royal Astronomical Society, 350, 129

\bibitem[{Seifina \& Titarchuk(2012)}]{Seifina2012}
Seifina, E., \& Titarchuk, L. 2012, The Astrophysical Journal, 747,
  doi:10.1088/0004-637X/747/2/99

\bibitem[{{Sunyaev} \& {Churazov}(1998)}]{Sunyaev1998}
{Sunyaev}, R., \& {Churazov}, E. 1998, \mnras, 297, 1279

\bibitem[{{Sunyaev} {et~al.}(1993){Sunyaev}, {Markevitch}, \&
  {Pavlinsky}}]{Sunyaev1993}
{Sunyaev}, R.~A., {Markevitch}, M., \& {Pavlinsky}, M. 1993, \apj, 407, 606

\bibitem[{{Tatischeff} {et~al.}(2012){Tatischeff}, {Decourchelle}, \&
  {Maurin}}]{tatischeff2012}
{Tatischeff}, V., {Decourchelle}, A., \& {Maurin}, G. 2012, \aap, 546, A88

\bibitem[{{Terrier} {et~al.}(2010){Terrier}, {Ponti}, {B{\'e}langer},
  {Decourchelle}, {Tatischeff}, {Goldwurm}, {Trap}, {Morris}, \&
  {Warwick}}]{Terrier2010}
{Terrier}, R., {Ponti}, G., {B{\'e}langer}, G., {et~al.} 2010, \apj, 719, 143

\bibitem[{{Tsuboi} {et~al.}(1999){Tsuboi}, {Handa}, \& {Ukita}}]{Tsuboi1999}
{Tsuboi}, M., {Handa}, T., \& {Ukita}, N. 1999, \apjs, 120, 1

\bibitem[{{Tsuboi} {et~al.}(2006){Tsuboi}, {Okumura}, \&
  {Miyazaki}}]{Tsuboi2006}
{Tsuboi}, M., {Okumura}, S.~K., \& {Miyazaki}, A. 2006, Journal of Physics
  Conference Series, 54, 16

\bibitem[{{Tsuchiya} {et~al.}(2004){Tsuchiya}, {Enomoto}, {Ksenofontov},
  {Mori}, {Naito}, {Asahara}, {Bicknell}, {Clay}, {Doi}, {Edwards}, {Gunji},
  {Hara}, {Hara}, {Hattori}, {Hayashi}, {Itoh}, {Kabuki}, {Kajino}, {Katagiri},
  {Kawachi}, {Kifune}, {Kubo}, {Kurihara}, {Kurosaka}, {Kushida}, {Matsubara},
  {Miyashita}, {Mizumoto}, {Moro}, {Muraishi}, {Muraki}, {Nakase}, {Nishida},
  {Nishijima}, {Ohishi}, {Okumura}, {Patterson}, {Protheroe}, {Sakamoto},
  {Sakurazawa}, {Swaby}, {Tanimori}, {Tanimura}, {Thornton}, {Tokanai},
  {Uchida}, {Watanabe}, {Yamaoka}, {Yanagita}, {Yoshida}, \&
  {Yoshikoshi}}]{Tsuchiya2004}
{Tsuchiya}, K., {Enomoto}, R., {Ksenofontov}, L.~T., {et~al.} 2004, \apjl, 606,
  L115

\bibitem[{Verner {et~al.}(1996)Verner, Ferland, Korista, \&
  Yakovlev}]{Verner1996}
Verner, D.~A., Ferland, G.~J., Korista, K.~T., \& Yakovlev, D.~G. 1996,
  Astrophysical Journal, 465, 487

\bibitem[{{Wang} {et~al.}(2002{\natexlab{a}}){Wang}, {Gotthelf}, \&
  {Lang}}]{Wang2002}
{Wang}, Q.~D., {Gotthelf}, E.~V., \& {Lang}, C.~C. 2002{\natexlab{a}}, \nat,
  415, 148

\bibitem[{{Wang} {et~al.}(2002{\natexlab{b}}){Wang}, {Lu}, \&
  {Lang}}]{Wang2002b}
{Wang}, Q.~D., {Lu}, F., \& {Lang}, C.~C. 2002{\natexlab{b}}, \apj, 581, 1148

\bibitem[{Wang {et~al.}(2006)Wang, Lu, \& Gotthelf}]{Wang2006}
Wang, Q.~D., Lu, F.~J., \& Gotthelf, E.~V. 2006, Monthly Notices of the Royal
  Astronomical Society, 367, 937

\bibitem[{{Weisskopf} {et~al.}(2007){Weisskopf}, {Wu}, {Trimble}, {O'Dell},
  {Elsner}, {Zavlin}, \& {Kouveliotou}}]{Weisskopf2007}
{Weisskopf}, M.~C., {Wu}, K., {Trimble}, V., {et~al.} 2007, \apj, 657, 1026

\bibitem[{{Wijnands} {et~al.}(2006){Wijnands}, {in't Zand}, {Rupen},
  {Maccarone}, {Homan}, {Cornelisse}, {Fender}, {Grindlay}, {van der Klis},
  {Kuulkers}, {Markwardt}, {Miller-Jones}, \& {Wang}}]{Wijnands2006}
{Wijnands}, R., {in't Zand}, J.~J.~M., {Rupen}, M., {et~al.} 2006, \aap, 449,
  1117

\bibitem[{{Wik} {et~al.}(2014){Wik}, {Hornstrup}, {Molendi}, {Madejski},
  {Harrison}, {Zoglauer}, {Grefenstette}, {Gastaldello}, {Madsen},
  {Westergaard}, {Ferreira}, {Kitaguchi}, {Pedersen}, {Boggs}, {Christensen},
  {Craig}, {Hailey}, {Stern}, \& {Zhang}}]{Wik2014}
{Wik}, D.~R., {Hornstrup}, A., {Molendi}, S., {et~al.} 2014, \apj, 792, 48

\bibitem[{Wilms {et~al.}(2000)Wilms, Allen, \& McCray}]{Wilms2000}
Wilms, J., Allen, A., \& McCray, R. 2000, The Astrophysical Journal, 542,
  914–

\bibitem[{Winkler {et~al.}(2003)Winkler, Courvoisier, Cocco, Gehrels, Gimenez,
  Grebenev, Hermsen, Mas-Hesse, Lebrun, Lund, Palumbo, Paul, Roques, Schnopper,
  Schonfelder, Sunyaev, Teegarden, Ubertini, Vedrenne, \& Dean}]{Winkler2003}
Winkler, C., Courvoisier, T. J.-L., Cocco, G.~D., {et~al.} 2003, Astronomy and
  Astrophysics, 411, L1

\bibitem[{{Worrall} {et~al.}(1982){Worrall}, {Marshall}, {Boldt}, \&
  {Swank}}]{Worrall1982}
{Worrall}, D.~M., {Marshall}, F.~E., {Boldt}, E.~A., \& {Swank}, J.~H. 1982,
  \apj, 255, 111

\bibitem[{{Yaqoob}(2012)}]{Yaqoob2012}
{Yaqoob}, T. 2012, \mnras, 423, 3360

\bibitem[{{Yuasa} {et~al.}(2012){Yuasa}, {Makishima}, \&
  {Nakazawa}}]{Yuasa2012}
{Yuasa}, T., {Makishima}, K., \& {Nakazawa}, K. 2012, \apj, 753, 129

\bibitem[{{Yusef-Zadeh} {et~al.}(2002{\natexlab{a}}){Yusef-Zadeh}, {Law}, \&
  {Wardle}}]{Yusef2002}
{Yusef-Zadeh}, F., {Law}, C., \& {Wardle}, M. 2002{\natexlab{a}}, \apjl, 568,
  L121

\bibitem[{{Yusef-Zadeh} {et~al.}(2002{\natexlab{b}}){Yusef-Zadeh}, {Law},
  {Wardle}, {Wang}, {Fruscione}, {Lang}, \& {Cotera}}]{Yusef2002b}
{Yusef-Zadeh}, F., {Law}, C., {Wardle}, M., {et~al.} 2002{\natexlab{b}}, \apj,
  570, 665

\bibitem[{{Yusef-Zadeh} \& {Morris}(1987)}]{Yusef1987}
{Yusef-Zadeh}, F., \& {Morris}, M. 1987, \apj, 322, 721

\bibitem[{{Yusef-Zadeh} {et~al.}(2007){Yusef-Zadeh}, {Muno}, {Wardle}, \&
  {Lis}}]{Yusef2007}
{Yusef-Zadeh}, F., {Muno}, M., {Wardle}, M., \& {Lis}, D.~C. 2007, \apj, 656,
  847

\bibitem[{{Yusef-Zadeh} {et~al.}(2012){Yusef-Zadeh}, {Arendt}, {Bushouse},
  {Cotton}, {Haggard}, {Pound}, {Roberts}, {Royster}, \& {Wardle}}]{Yusef2012}
{Yusef-Zadeh}, F., {Arendt}, R., {Bushouse}, H., {et~al.} 2012, \apjl, 758, L11

\bibitem[{{Yusef-Zadeh} {et~al.}(2013){Yusef-Zadeh}, {Hewitt}, {Wardle},
  {Tatischeff}, {Roberts}, {Cotton}, {Uchiyama}, {Nobukawa}, {Tsuru}, {Heinke},
  \& {Royster}}]{Yusef2013}
{Yusef-Zadeh}, F., {Hewitt}, J.~W., {Wardle}, M., {et~al.} 2013, \apj, 762, 33

\bibitem[{{Zhang} {et~al.}(2008){Zhang}, {Chen}, \& {Fang}}]{Zhang2008}
{Zhang}, L., {Chen}, S.~B., \& {Fang}, J. 2008, \apj, 676, 1210

\bibitem[{{Zhang} {et~al.}(2014){Zhang}, {Hailey}, {Baganoff}, {Bauer},
  {Boggs}, {Craig}, {Christensen}, {Gotthelf}, {Harrison}, {Mori}, {Nynka},
  {Stern}, {Tomsick}, \& {Zhang}}]{Zhang2014}
{Zhang}, S., {Hailey}, C.~J., {Baganoff}, F.~K., {et~al.} 2014, \apj, 784, 6

\bibitem[{{Zhang} {et~al.}(2015){Zhang}, {Hailey}, {Baganoff}, {Bauer},
  {Boggs}, {Craig}, {Christensen}, {Gotthelf}, {Harrison}, {Mori}, {Nynka},
  {Stern}, {Tomsick}, \& {Zhang}}]{Zhang2015}
---. 2015, submitted to \apj, arXiv:1507.08740

\end{thebibliography}

\end{document}